\documentclass[draftcls, onecolumn]{IEEEtran}
\usepackage{amsmath,amsfonts}
\usepackage{algorithmic}
\usepackage{algorithm}
\usepackage{array}
\usepackage[caption=false,font=normalsize,labelfont=sf,textfont=sf]{subfig}
\usepackage{textcomp}
\usepackage{stfloats}
\usepackage{url}
\usepackage{verbatim}
\usepackage{graphicx}
\usepackage{cite}
\usepackage{enumerate}
\usepackage{eucal}
\usepackage[hidelinks]{hyperref}
\usepackage{bm}
\usepackage{comment}
\usepackage{xcolor}
\usepackage{adjustbox}

\usepackage{tikz}
\usepackage{color, colortbl}
\usepackage{nicematrix}

\usepackage{float}

\usepackage{bigdelim}
\usepackage{arydshln}
\usepackage{mathrsfs}
\usepackage{mathtools}

\newcommand{\BB}{\mathbb{B}}

\newcommand{\JJ}{\mathpzc{J}}
\newcommand{\famJ}{\mathcal{J}}

\newcommand{\II}{\mathpzc{I}}
\newcommand{\KK}{\mathpzc{K}}

\newcommand{\F}{\mathcal{F}}

\newcommand{\Z}{\mathbb{Z}}

\newcommand{\J}{\mathcal{J}}

\newcommand{\ZZ}{\mathcal{Z}}

\newcommand{\Prob}{\text{Pr}}

\newtheorem{lemma}{Lemma}
\newtheorem{theorem}{Theorem}

\newtheorem{corollary}{Corollary}
\newtheorem{proposition}{Proposition}
\newtheorem{remark}{Remark}

\newtheorem{defn}{Definition}

\DeclareMathAlphabet{\mathpzc}{OT1}{pzc}{m}{it}

\newcommand{\neqlvl}{\eta}
\newcommand{\Var}{\text{Var}}

\newcommand{\Sizer}{{\text{size}(\vec{r})}}
\newcommand{\Sizel}{{\text{size}(\vec{l})}}
\newcommand{\Sizern}{{\text{size}(\vec{r}_N)}}
\newcommand{\Sizeln}{{\text{size}(\vec{l}_N)}}
\def\x{{\mathbf x}}

\newcommand{\nodeskew}{\mathpzc{s}}

\newcommand{\TT}{\mathcal{T}}
\renewcommand{\vec}[1]{\underline{#1}}

\begin{document}
\title{
Fast DFT Computation for Signals with Structured Support}

\author{Charantej Reddy Pochimireddy, Aditya Siripuram, Brad Osgood
\thanks{This work
was supported in part by Qualcomm Technologies through the Qualcomm Innovation
Fellowship India 2021.}
}

\markboth{Journal of \LaTeX\ Class Files,~Vol.~14, No.~8, August~2021}%
{Shell \MakeLowercase{\textit{et al.}}: A Sample Article Using IEEEtran.cls for IEEE Journals}


\maketitle
\begin{abstract}
Suppose an $N-$length signal has known frequency support of size $k$. Given sample access to this signal, how fast can we compute the DFT? The answer to this question depends on the structure of the frequency support. 

We first identify some frequency supports for which (an ideal) $O(k \log k)$ complexity is achievable, referred to as homogeneous sets. We give a generalization of radix-2 that enables $O(k\log k)$ computation of signals with homogeneous frequency support. Using homogeneous sets as building blocks, we construct more complicated support structures for which the complexity of $O(k\log k)$ is achievable. We also investigate the relationship of DFT computation with additive structure in the support and provide partial converses.
\end{abstract}

\section{Introduction}
\label{sec:intro}
The Discrete Fourier Transform (DFT) $\F$ (defined below) is a fundamental tool for analyzing discrete signals. It is employed in diverse areas of engineering and science, for example speech and image processing, medical imaging, data analysis and too many others to list. The Fast Fourier Transform (FFT) is an algorithm that computes the DFT of an $N$-length signal in $O(N \log N)$ steps,  \cite{cooley1965algorithm, good1958interaction, rader1968discrete}, by
recursively exploiting the structure of the DFT matrix. 

The importance and ubiquity of the FFT algorithm cannot be overstated. With applications involving datasets of ever-increasing size, there is a need to reduce the complexity below the $O(N \log N)$ bound provided by the traditional FFT. A key is to identify and exploit situations affecting the DFT coefficients.  For example, a widely used model considers sparsity of the DFT coefficients: one assumes that only $k<<N$ DFT coefficients are non-zero. 

We start with an even simpler assumption. For a signal $f$, let $\JJ$ denote the  \emph{frequency support} of $\F f$, consisting of the set of frequencies $n$ for which $\F f (n) \ne 0$. If the frequency support is known, is it possible to speed up the computation of the DFT? 
This work is an attempt to address this problem systematically.

As in the traditional radix-2 algorithms, we shall assume that $N$ is a power of $2$ (this can be generalized to a power of any prime). The fact that the DFT can be computed in $O(N\log N)$ 
derives from the recursive structure of the $N\times N$ DFT matrix \cite{osgood2018lectures,cooley1965algorithm, good1958interaction, rader1968discrete}. When the frequency support is known, the focus is on submatrices of the DFT, which may not inherit such recursive properties as the full DFT matrix. 

Suppose a signal has frequency support of size $k$. We can trivially give an $O(k^2)$ algorithm to find the DFT coefficients (see Section \ref{sec:preliminary_observations}) irrespective of $N$. This is much bigger than what we want, but it is then reasonable to expect that the DFT computational complexity scales only with the size of the frequency support and not with $N$. On the other hand, when the frequency support is the entire set of frequencies (i.e., $k=N$), the best-known complexity is $O(N\log N)$. With this in mind, we pose the following question, which we make precise later:

\begin{quote}
    Which structures on the frequency support, of size $k$, allow for a computational complexity of $O(k\log k)$?
\end{quote}

To the best of our knowledge, this has not been investigated. We observe that for some support structures, simple modifications of the well-known FFT algorithms can achieve the target complexity of $O(k \log k)$. However, more sophisticated techniques seem to be needed in general. 

The key structural property of the frequency support $\JJ$ that enables fast DFT computation is the number of $r$ such that
\[
j_1-j_2 = \text{ (odd number)}\times 2^{r} \text{ for some }j_1\neq j_2\in \JJ.
\]
Each such $r$ is referred to as a \emph{pivot} of $\JJ$. If the number of pivots is small compared to the size of the support, $|\JJ|$, the DFT computation achieves the target complexity of $O(k\log k)$. 

Consider the following simple technique to compute the DFT of signals supported in $\JJ$. Let $r_{\text{max}}$ be the largest pivot in $\JJ$. Then, no two elements of $\JJ$ have the same remainder when divided by $2^{r_{\text{max}}+1}$. This means that any signal (with frequency support $\JJ$) can be uniformly downsampled to $2^{r_{\text{max}}+1}$ samples, and this would not cause any aliasing in the frequency domain (details in Section \ref{sec:down-samp-cong-tree}). So by computing the $2^{r_{\text{max}}+1}$ point DFT of the downsampled signal, we should be able to recover the DFT coefficients. Since this technique involves a $2^{r_{\text{max}}+1}$ point DFT, it costs $\approx (r_{\text{max}}+1)2^{r_{\text{max}}+1}$ arithmetic operations. The central result of this work is that (by using certain non-uniform downsampling patterns) $r_{\text{max}}$ (i.e., the largest pivot) in the above expression can be replaced with the \emph{number of pivots}, which can be a significant improvement.  

We start with the frequency supports having the smallest number of pivots, referred to as \emph{homogeneous} sets. Homogeneous sets include periodic sets, sets of $2^s$ consecutive integers (for any $s$), and the trivial case when $\JJ$ is the entire frequency range. But they have some other non-trivial structures as well. 

We provide an $O(k\log k)$ algorithm to compute the DFT of signals whose frequency support is a homogeneous set. This algorithm is a generalization of the standard radix-2 algorithm. We then build more complicated support sets using homogeneous sets as building blocks. For example, by taking restricted unions of homogeneous sets, considering approximately homogeneous sets, or taking random subsets of homogeneous sets. We provide an $O(k\log k)$ algorithm for such support structures. As an application of these ideas, we also prove that a complexity of $O(k\log^2k)$  is achievable for frequency supports $\JJ$ that have additive structure, i.e., when \(|\JJ+\JJ|=|\{j+j':j,j'\in\JJ,\}|<C|\JJ|\) for some constant $C$ (that does not grow with the size of $\JJ$).

\section{Problem setup}
\label{sec:problem_setup}
While we discuss the notation and preliminaries in detail in Section \ref{sec:notation}, we provide a few definitions to set up the problem. We let $\mathbb{Z}_N$ denote the ring of integers modulo $N$, and we assume that $N=2^M$ is a power of $2$. We let $\F$ be the $N \times N$ DFT matrix, with entries $\{\exp(-2\pi i mn/N) \}$, for $m,n =0,1,2,\ldots,N-1$. Thus for any signal $f \in \mathbb{C}^N$
\[
\F f(m) = \sum_{n=0}^{N-1}f(n)e^{-2\pi i m n /N}, \text{ for }m=0,1,2,\ldots.
\]
\begin{defn}
\emph{(Bandlimited spaces)} For an index set $\JJ\subseteq \Z_N$, the subspace $\BB^\JJ$ of $\mathbb{C}^N$ consists of signals with frequency support $\JJ$:
\[
\BB^\JJ =\{f: \F f(m)=0 \text{ for every }m\notin\JJ \}.
\]
\end{defn}
Our work deals with computing the DFT of signals in bandlimited spaces. Next, we introduce the notion of families of index sets to collect all index sets with the desired property.
\begin{defn}\emph{(Families of index sets)}
\label{def:fam_idx_sets}
A (deterministic) family of index sets $\{(\famJ_{N}, k_N)\}_{N=1,2,4,8\ldots}$ is a sequence such that each $\JJ \in \famJ_{N}$ is a subset of $\Z_N$ of size $|\JJ|= k_N$; with $k_N \rightarrow \infty$. 

We also generalize this notion to allow for the constituent index sets to be generated randomly. Similar to the above, a stochastic 
family of index sets $\{(\famJ_{N}, k_N)\}_{N=1,2,8,\ldots}$ is a sequence where $\famJ_{N}$ is a probability distribution on the subsets of $\Z_N$, such that any $\JJ$ distributed according to $\famJ_N$ has a size $\Theta(k_N)$ with high probability, i.e., there exists constants $\alpha, \beta$ such that 
\[
\alpha k_N \leq |\JJ| \leq \beta k_N,
\]
with probability approaching $1$ as $N \rightarrow \infty$, and $k_N \rightarrow \infty$.
  %

For stochastic families, we write $\JJ \in \famJ_N$ to mean that $\JJ$ is distributed according to $\famJ_N$. We often drop the subscript $N$, so we refer to $k_N$ as $k$ and the family as $\famJ$ when it is clear from the context.
\end{defn}

A simple (deterministic) family of index sets is the
family consisting of all $k_N-$size subsets of $\Z_N$. Another example of a deterministic family is the family of universal sets \cite{DS-1} (see also Definition \ref{def:universal_sets}) is obtained by collecting all universal sets of size $k_N$ in $\famJ_N$. Some of our results are for singleton families: here, $\famJ_N$ contains a single index set as a member. 

The stochastic families we consider are obtained by taking random subsets of certain structured sets (see Definition \ref{def:fam_r_sets} for example).

\begin{defn}
\emph{(Structured-DFT algorithm)} Given a family of index sets $\{(\famJ_N, k_N)\}$, a structured-DFT algorithm for $\{(\famJ_N, k_N)\}$ computes, for each $N$, the DFT coefficients $\{\F f(j)\}_{j \in \JJ}$ of any $f \in \BB^\JJ,$ and  $\JJ \in \famJ_N$. The structured-DFT algorithm is given access to $\JJ$ and the entries of $f \in \BB^\JJ$.
\end{defn}
\begin{enumerate}
    \item The complexity of a structured DFT algorithm is the number of arithmetic operations required to compute the DFT coefficients: typically, this may depend on the frequency support $\JJ$.  
    For deterministic families, the complexity of the algorithm is taken
    as the worst-case complexity over $\JJ \in \famJ_N$.  For stochastic families, we say that the complexity of the algorithm is $O(g(k_N))$ if, for large enough $N$, the DFT coefficients can be computed using $c g(k_N)$ arithmetic operations (for some constant $c>0$) with probability $p_N$, and $p_N \rightarrow 1$. Thus the algorithm has a probability of failure ($1-p_N$); this is the probability (over the support $\JJ$) that the DFT coefficients are not computed fast enough (or are not computed at all).
    \item Note that we assume the Structured-DFT algorithm knows the family $\{(\famJ_N, k_N)\}$ apriori. The complexity of the algorithm is calculated from the number of arithmetic operations after being given access to $\JJ$ and the entries of $f$. We will also typically assume the indices of $\JJ$ are given to the algorithm in binary. In practice, there may be some pre-processing to extract the relevant information from the family before accepting any inputs. 

    \item We also require that the algorithm compute the DFT coefficients exactly. We assume that any finite precision-related issues are negligible. 
\end{enumerate}

Another common criterion for evaluating DFT algorithms is sample complexity. This is the number of entries of $f$ that the algorithm needs access to. The algorithms we discuss have $O(k_N)$ sample complexity for all of the families we discuss.

\begin{defn}
\emph{(Fourier computability)}  We say that a family $\{(\famJ_{N}, k_N)\}$ is Fourier-computable if there exists an algorithm for $\{(\famJ_{N}, k_N)\}$ with $O(k_N\log k_N)$ complexity. 

\end{defn} 
In this work, homogeneity (defined more comprehensively in Section \ref{sec:congruence_trees}) is the key structure that enables Fourier computability. We give a summary of our results.
\subsection{Key results}
We establish that the following families are Fourier-computable:
\begin{enumerate}[(i)]
    \item Families of homogeneous sets (Corollary \ref{cor:spectral-Fourier-computable}).
    \item Family of Approximately homogeneous sets/balanced families (Theorem \ref{thm:part-hom-Fourier-computability}), this includes the family of universal sets \cite{DS-1}, and sets in arithmetic progressions. 
    \item Restricted union of homogeneous sets (Theorem \ref{thm:UoE2}).
    \item Random subsets of homogeneous sets (Theorem \ref{thm:hom_rand_subset}).
\end{enumerate}

The key ingredient is a generalization of the radix-2 algorithm to compute the DFT of signals supported on homogeneous sets (Theorem \ref{thm:computation_Hi-DFT}), and a tree based interpretation of index sets. The results (ii)-(iv) above combine the insights gained from Theorem \ref{thm:computation_Hi-DFT} with the recursive structure of the tree-based construction. These results are not known in the literature as far as we are aware. They are proved using a structured,  non-uniform downsampling pattern: in particular, the algorithms that use uniform downsampling can be seen to fail (catastrophically) for arbitrary homogeneous sets. 

\begin{enumerate}[(i)]
 \setcounter{enumi}{4}
    \item In the other direction, we also provide an example of families (Section \ref{sec:converse_results}) for which the sampling-based framework above \emph{cannot} be used to compute the DFT in $O(k \log k)$. Thus the Fourier computability of these families remains unresolved. We also prove that the proposed algorithm always results in a complexity of $\Omega(k \log k)$ (Lemma \ref{lem:con_SAS}).
    \item We also give some preliminary results for sets with additive structure as defined in the introduction. These sets are the subject of study in additive combinatorics \cite{tao2006additive}.  We prove that the proposed algorithm achieves a complexity of $O(k\log^2k)$ for singleton families of generalized arithmetic progressions (Lemma \ref{lem:gap-complexity}) and consequently for sets with additive structure (Theorem \ref{thm:additive-structure-complexity}).
\end{enumerate}

While we assume that $N=2^M$ for ease of exposition, our results can be easily extended to the case when $N$ is a prime power.
\subsection{An example result}

Consider the following stochastic family: 
\begin{defn}
\emph{(Random subsets)} For any set $\KK\subseteq \Z_N$, we generate a random subset $\JJ$ by including each entry from $\KK$ in $\JJ$ with probability $p = k/|\KK|$, independent of the choices for all the other entries in $\KK$. We denote the resulting distribution by  $\KK^{\downarrow k}$. 
For $\JJ \in \KK^{\downarrow k} $, we note that $|\JJ|$ is binomially distributed with $E|\JJ| = k$ and $\Var |\JJ| \leq k$. Also, when $k = |\KK|,$ the generated set $\JJ =\KK$ with probability $1$.  
\end{defn}
We now have the following stochastic family of index sets.
\begin{defn}\label{def:fam_r_sets}
\emph{(Family of random subsets)}  Given sets $\KK_N \subseteq \Z_N$, $\{(\KK_N^{\downarrow k_N}, k_N)\}$ is a family of index sets for any choice of $\KK_N\subseteq \Z_N$ and $k_N \leq |\KK_N|$ such that $k_N \rightarrow \infty$ (see also Appendix \ref{proof:def_fam_r_sets}).
\end{defn}
Homogenous sets (these are sets with a small number of pivots) are known to be equivalent to spectral sets, defined below (we discuss this equivalence in detail later, in Section \ref{sec:down-samp-cong-tree}):
\begin{defn}\label{def:spectral}
\emph{Spectral sets \cite{fuglede1974commuting, tao2003fuglede, fan2016compact, siripuram2018lp}} We say that $\JJ \subseteq \Z_N$ is spectral if there exists a square unitary\footnote{By unitary we mean unitary up to scaling. For e.g. the matrix $\F$ as defined is taken to be unitary.} submatrix of $\F$ with columns indexed by $\JJ$. 
\end{defn}
Note that by definition, the set $\Z_N$ itself is a spectral set. 
As another example, for $N=2^{10}$, the set $\JJ = \{316, 384, 828, 896\}$ is a spectral set, as the Fourier submatrix with rows $\{1, 292, 641, 932\}$ and columns indexed by $\JJ$ is unitary. In general, for $N=2^M$, there exist spectral sets of sizes $2^m$ for each $0\leq m \leq M$.

We give an example result from our work: the result below makes precise the observation (iv) claimed informally earlier:

\begin{quote}
\emph{(Fourier computability for random subsets of spectral sets)} Given spectral sets $\KK_N \subseteq \Z_N$, consider the family $\{(\KK_N^{\downarrow k_N}, k_N)\}$ such that $k_N \leq |\KK_N|$ and $k_N \rightarrow \infty$. Then this family is Fourier-computable (Theorem \ref{thm:hom_rand_subset}). 
\end{quote}
The present work is a generalization of our earlier work \cite{9518104} where we gave an $O(k \log k)$ algorithm to compute the DFT of signals supported on spectral sets. Here we generalize those observations to restricted unions, random subsets and approximations of spectral sets, including generalized arithmetic progressions and additive structures; in addition to using the alternative characterization of homogeneity (using congruence trees) that aids these generalizations and different proof techniques.
\section{Related work}
\label{sec:related-work}
A closely connected problem that has been well studied in literature is the sparse-FFT problem, where the frequency support is assumed to be unknown but of size at most $k$. (Such signals are henceforth referred to as $k-$sparse signals). For the many breakthrough results on this problem, we refer the reader to \cite{gilbert2014recent}. The crucial difference in our setup is the assumption of complete knowledge of the location of the nonzero frequency coefficients. We exploit the support structure that enables fast DFT computation; as such, our results are different in form and substance from those in the sparse-FFT literature.

 Nevertheless, we may consider solving the structured-DFT problem posed in the introduction by simply ignoring the available information on the support locations and applying one of the sparse-FFT algorithms. The best-known complexity for computing the DFT of $k$-sparse signals is $O(k \log N)$, achieved by the randomized algorithm from \cite{nearlyoptimalsparse}: this is higher than our requirement of $O(k \log k)$. Randomized algorithms may have some limitations (see \cite{akavia2014deterministic} for details), and there have also been interesting works that provide deterministic algorithms for sparse-FFT \cite{akavia2014deterministic} \cite{iwen2007deterministic}. However, the complexity of these algorithms is higher than $O(k^2)$. Also note that all these algorithms recover the DFT coefficients approximately, and as discussed earlier, this work is restricted to exact DFT computation.

On the other end, seminal works in \cite{pawar-ramachandran} \cite{ghazi2013sample} provide an $O(k \log k)$ probabilistic algorithm to compute the DFT of $k$-sparse signals - these algorithms operate without the knowledge of the support, but there is a chance (depending on the actual frequency domain support) of the algorithm failing to compute the DFT coefficients. In these techniques, the signal (frequency domain) supports are modeled as random subsets of $\Z_N$. Firstly, these algorithms need to assume that $N$ has multiple prime factors, as they rely heavily on the Chinese Remainder theorem for isolating frequency coefficients. To the best of our knowledge, these approaches do not extend to the case when $N=2^M$, as in the setup of our problem. We can consider resampling a given $f \in \mathbb{C}^N$ to $N'$ samples, where $N'$ has multiple prime factors, and then apply the algorithms from \cite{pawar-ramachandran} \cite{ghazi2013sample}. However, this resampling may incur an additional computation cost, and certainly incurs an approximation cost. Secondly, the earlier techniques from \cite{pawar-ramachandran} \cite{ghazi2013sample} impose  probability distribution on the subsets of $\Z_N$ (i.e., the subsets are generated by random sampling from $\Z_N)$ and the probability of failure (i.e., DFT coefficients not getting computed fast enough) is small for supports picked with respect to this distribution.

 Our framework allows us to give $O(k \log k)$ algorithms for random subsets of certain structured sets $K \subseteq \Z_N$ as well, thus generalizing the class of distributions on subsets of $\Z_N$ for which $O(k\log k)$ complexity can be achieved. Also, we note that the time-domain downsampling  needs to be non-uniform to achieve an $O(k \log k)$ complexity for random subsets of structured sets $K$ we consider (See Section \ref{sec:problem_setup} for a summary of our results). In contrast,  the techniques from \cite{pawar-ramachandran} \cite{ghazi2013sample} use uniform downsampling patterns. Finally, we do not assume anything about the growth of $k_N$. Thus the existing works \cite{pawar-ramachandran} \cite{ghazi2013sample} do not solve the question posed in the introduction. Nevertheless, our work is heavily inspired by these algorithms. Other related works from sparse-FFT literature include \cite{lawlor2013adaptive}, which, similar to \cite{pawar-ramachandran} and \cite{ghazi2013sample}, gives a probabilistic $O(k \log k)$ algorithm. However, this algorithm assumes the time domain signal $f$ to be a function of a continuous variable (i.e., we have access to samples of $f$ at any locations, including those off-grid). As with the other algorithms discussed, this does not apply to the setup of our problem.

Apart from the sparse-FFT literature, the concept of homogeneity has been discussed extensively in the context of spectral sets, Fuglede's conjecture, and tiling (see \cite{siripuram2018lp, coven1999tiling,laba2002spectral, newman1977tesselation} for a few references); the (non-uniform) downsampling patterns we use are discussed in the signal processing literature \cite{Venk-Bresler, siripuram2019discrete} and in number theory \cite{laba2002spectral}; but to the best of our knowledge, this is the first work that ties these concepts to the computational complexity of the DFT and establishes a connection between Fourier computability and additive structure. 

\section{Notation and review of useful results}
\label{sec:notation}
In this section, we discuss the notation and review key DFT properties. Most of the notation we use is standard. Readers familiar with these properties may consider skipping to Section \ref{sec:preliminary_observations}.
\begin{enumerate}

\item The inverse DFT (IDFT) is the matrix $\F^{-1}$ 
with entries $(\F^{-1})_{mn}=(1/N)\{\exp(2\pi i mn/N) \}$, for $m,n =0,1,2,\ldots,N-1$. For any signal $f \in \mathbb{C}^N$, the IDFT of $f$, is  $\F^{-1} f$, so that \[
\F^{-1} f(m) = \frac{1}{N}\sum_{n=0}^{N-1}f(n)e^{2\pi i m n /N}, \text{ for }m=0,1,2,\ldots.
\]

\item We use $\tau$ to represent the shift operator. Namely, for signals $f:\Z_N\rightarrow \mathbb{C}$, we have $\tau f:\Z_N \rightarrow \mathbb{C}$ with $\tau f (n) = f(n-1)$. For index sets $\JJ\subseteq \Z_N$, we define $\tau \JJ$ as $\tau \JJ = \{j+1: j\in \JJ\}$. 

\item With $N=2^M$, we write $\omega_M$ for the primitive $2^M$ root of unity: $\omega_M \coloneqq \exp(-2\pi i/2^M)$. We denote by $\underline{\omega}_M^j$ the vector $\underline{\omega}_M^j \coloneqq (\omega_M^0, \omega_M^j, \omega_M^{2j}, \ldots, \omega_M^{(2^M-1)j} )$, which is  the $j^{th}$ column of the $N\times N$ DFT matrix. We drop the subscript $M$ when it is apparent from the context.

\item We denote by $\delta$ the discrete impulse/unit impulse function defined as
\[\delta(n)=\begin{cases}
1 \text{ if } n=0, \\ 0 \text{ otherwise.}
\end{cases}\]
Then $\F \delta = \vec{1}$, where $\vec{1}$ is all ones vector of the same size as $\delta$.
\item 
\emph{(Shift modulation)} $\tau^{\alpha} f$ is the signal $f$ shifted by $\alpha$, and one has 
\[\F (\tau^{\alpha} f) (n)=   e^{-2\pi i \alpha n /N} \F f (n), \text{ for }n=0,1,2,\ldots,\]
or simply \(\F (\tau^\alpha f) = \underline{\omega}^\alpha.\F f,\)
where $\vec{x}.\vec{y}$ is the element-wise product of vectors $\vec{x}$ and $\vec{y}$.

\item \emph{(Downsampling)} Sampling the signal $f$ uniformly in the time domain results in the sampled signal $f_{\downarrow 2^m}$
\[
f_{\downarrow 2^m}(k) = f(2^mk), \text{ for }k=0,1,2\ldots
\]

It is well known that the downsampling process leads to aliasing \cite{osgood2018lectures} of the frequency domain signal $\F f$:
\begin{equation}
\label{eq:DFT_down_sampling} 
\F f_{\downarrow 2^m}(n) = 2^{M-m}\sum_{r=0}^{2^m}\F f(n + r2^{M-m}).
\end{equation}

\item \emph{(Computational complexity notation)}
Given $h, g : \mathbb{Z} \rightarrow \mathbb{R}$, we say that $h(n)$ is $O(g(n))$ if $h(n) \leq c g(n)$ for some constant $c$ and all $n$ large. We say that $h(n)$ is $\Omega (g(n))$ if $h(n) \geq c g(n)$ for all $n$ large. We say that $h(n)$ is $\Theta(g(n))$ if $cg(n) \leq h(n) \leq C g(n)$ for some constants $c, C$, and all $n$ large. Similarly, we say that $h(n)$ is $o(g(n))$ if $h(n)/g(n) \rightarrow 0$. We write $h(n) = O(g(n))$ to mean $h(n)$ is $O(g(n))$.
    

\item \emph{(Computational complexity)}
\begin{itemize}
    \item The complexity of finding DFT of $N$-length signal using FFT is $c_1 N\log N$, where constant $c_1 = 3/2$.
    \item The complexity of solving a Vandermonde system \cite{parker1964inverses,gohberg1997fast} of size $N \times N$ is $c_2 N^2$ operations, where constant $c_2 = 6$.
\end{itemize}
\end{enumerate}

\section{Preliminary observations}
\label{sec:preliminary_observations}
We start by investigating some techniques to compute the DFT of a signal in $\BB^\JJ$. 
\subsection{Solving system of equations}
\label{sec:preliminary_observations_system}
The following simple method can reconstruct the DFT coefficients by constructing a system of equations involving Fourier submatrices. For this, we need the following notation. Let $\II, \JJ \subseteq \Z_N$ be index sets and $A$ an $N\times N$ matrix whose rows/columns are indexed by elements of $\Z_N$. We denote by $A(\II, \JJ)$ the submatrix of $A$ formed with columns $\JJ$ and rows $\II$. For any vector $f$, we represent by $f_\JJ$ the vector obtained by keeping only the elements indexed $\JJ$ from $f$.

Consider any signal $f \in \BB^\JJ$, say $k = |\JJ|$ and $\JJ = \{j_0, j_1,\ldots, j_{k-1}\}$. We have 
\[
f = \underbrace{\F^{-1}}_{\substack{\text{inverse }\\ \text{DFT matrix }}}\underbrace{\F f}_{ \text{DFT of }f},
\]
 Since $\F f$ is nonzero only on the locations $\JJ$ (by definition of $\BB^\JJ$), only the columns from $\F^{-1}$ indexed by $\JJ$ remain. 
 \[
 f =\F^{-1}(\Z_N,\JJ)(\F f)_\JJ.
 \]

Suppose we access $k$ entries of the vector $f$ corresponding to the locations $\II$, we get the following $k \times k$ system of equations
\begin{equation}
\label{eq:submatrix-method}
f_\II = \F^{-1}(\II,\JJ)(\F f)_\JJ.
\end{equation}
To find $(\F f)_\JJ$, we could solve the above system of equations. We can, for instance, pick $\II= \{0,1,2,\ldots,k-1\}$, in which case the $k\times k$ Fourier submatrix involved in the above system of equations becomes

\[\F^{-1}(\II,\JJ)= \frac{1}{N}
\begin{pmatrix}
1 & 1 & \ldots & 1\\
e^{2\pi j_0 /N} & e^{2\pi j_1 /N} & \ldots & e^{2\pi j_{k-1} /N} \\
\vdots&\vdots&\vdots&\vdots\\
e^{2\pi (k-1) j_0 /N} & e^{2\pi (k-1) j_1 /N} & \ldots & e^{2\pi (k-1) j_{k-1} /N}
\end{pmatrix}.
\]
Since this is a Vandermonde matrix, the system in \eqref{eq:submatrix-method} can be solved in $6k^2$ arithmetic operations \cite{parker1964inverses,gohberg1997fast}. This provides an $O(k^2)$ structured-DFT algorithm. However, this technique does not exploit the structure of the DFT submatrices beyond Vandermonde. In fact, at the extreme when $k=N$, this technique gives an $O(N^2)$ complexity, whereas $O(N\log N)$ algorithms (Radix-2 FFT, etc.) are very well known.
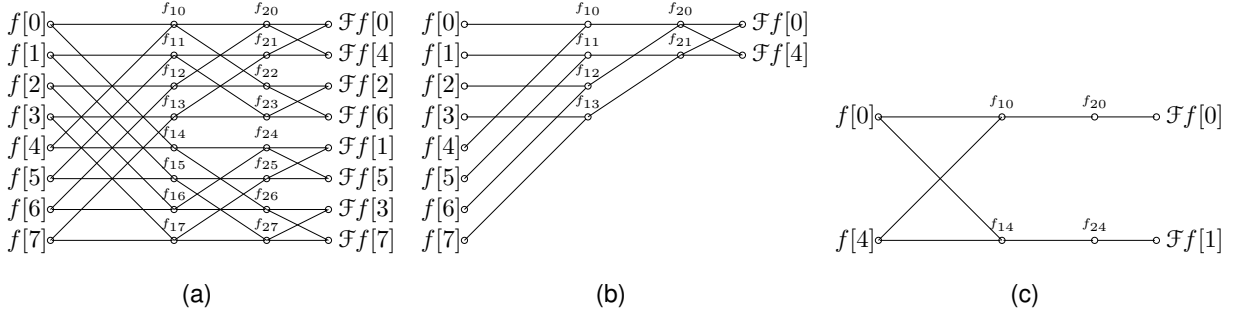
\begin{figure*}[ht]
\resizebox{\textwidth}{!}{%
\subfloat[]{
\begin{tikzpicture}[scale=0.22]

  \foreach \n in {0,...,7} {
    \draw (-1,-2*\n) node {$f[\n]$};
    \draw (0.5,-2*\n) circle(0.2)[fill=white]; 
    \draw (8.5,-2*\n) circle(0.2)[fill=white];
    \draw (8.5,-2*\n+1) node {$\scriptscriptstyle f_{1\n}$};
    \draw (14.5,-2*\n) circle(0.2)[fill=white];
    \draw (14.5,-2*\n+1) node {$\scriptscriptstyle f_{2\n}$};
    \draw (18.5,-2*\n) circle(0.2)[fill=white];
    
    \draw (0.6, -2*\n) --  (8.4, -2*\n); 
    \draw (8.6, -2*\n) -- (14.4, -2*\n);
    \draw (14.6, -2*\n) -- (18.4, -2*\n);
  }
  
 \draw (21,0) node {$\F f[0]$};
 \draw (21,-2) node {$\F f[4]$};
 \draw (21,-4) node {$\F f[2]$};
 \draw (21,-6) node {$\F f[6]$};
 \draw (21,-8) node {$\F f[1]$};
 \draw (21,-10) node {$\F f[5]$};
 \draw (21,-12) node {$\F f[3]$};
 \draw (21,-14) node {$\F f[7]$};
\foreach \n in {0,...,3} {
\draw (0.5, -2*\n) -- (8.5, -2*\n-8); 
\draw (0.5, -2*\n-8) -- (8.5, -2*\n); 
}

\foreach \n in {0,1} {
\draw (8.5, -2*\n) -- (14.5, -2*\n-4); 
\draw (8.5, -2*\n-4) -- (14.5, -2*\n); 

\draw (8.5, -2*\n-8) -- (14.5, -2*\n-12); 
\draw (8.5, -2*\n-12) -- (14.5, -2*\n-8); 
}

\foreach \n in {0,...,3} {
\draw (14.5, -4*\n) -- (18.5, -4*\n-2);
\draw (14.5, -4*\n-2) -- (18.5, -4*\n);
}

\end{tikzpicture}}
\subfloat[]{
\begin{tikzpicture}[scale=0.22]

  \foreach \n in {0,...,7} {
    \draw (-1,-2*\n) node {$f[\n]$};
    \draw (0.5,-2*\n) circle(0.2)[fill=white]; 
    
  }
  
 \draw (21,0) node {$\F f[0]$};
 \draw (21,-2) node {$\F f[4]$};
\foreach \n in {0,...,3} {
\draw (0.6, -2*\n) --  (8.4, -2*\n);

\draw (0.5, -2*\n-8) -- (8.5, -2*\n); 
\draw (8.5,-2*\n) circle(0.2)[fill=white];
\draw (8.5,-2*\n+1) node {$\scriptscriptstyle f_{1\n}$};
}

\foreach \n in {0,1} {
\draw (8.6, -2*\n) -- (14.4, -2*\n);
\draw (14.5,-2*\n) circle(0.2)[fill=white];
\draw (14.5,-2*\n+1) node {$\scriptscriptstyle f_{2\n}$};
}
\foreach \n in {0,1} {
\draw (8.5, -2*\n-4) -- (14.5, -2*\n); 

}

\foreach \n in {0,1} {
\draw (18.5,-2*\n) circle(0.2)[fill=white];
\draw (14.6, -2*\n) -- (18.4, -2*\n);
}
\foreach \n in {0} {
\draw (14.5, -4*\n) -- (18.5, -4*\n-2);
\draw (14.5, -4*\n-2) -- (18.5, -4*\n);
}

\end{tikzpicture}}
\subfloat[]{
\begin{tikzpicture}[scale=0.22]

  \foreach \n in {0,4} {
    \draw (-1,-2*\n) node {$f[\n]$};
    \draw (0.5,-2*\n) circle(0.2)[fill=white]; 
    \draw (8.5,-2*\n) circle(0.2)[fill=white];
    \draw (8.5,-2*\n+1) node {$\scriptscriptstyle f_{1\n}$};
    
    \draw (0.6, -2*\n) --  (8.4, -2*\n); 
  }
  
 \draw (21,0) node {$\F f[0]$};
 \draw (18.5,0) circle(0.2)[fill=white];
 \draw (14.6, 0) -- (18.4, 0);
 
 \draw (18.5,-8) circle(0.2)[fill=white];
\draw (21,-8) node {$\F f[1]$};
\draw (14.6, -8) -- (18.4, -8);

\foreach \n in {0} {
\draw (0.5, -2*\n) -- (8.5, -2*\n-8); 
\draw (0.5, -2*\n-8) -- (8.5, -2*\n); 
}

\draw (14.5,1) node {$\scriptscriptstyle f_{20}$};
\draw (14.5,-7) node {$\scriptscriptstyle f_{24}$};
\foreach \n in {0} {
\draw (14.5,-2*\n) circle(0.2)[fill=white];
\draw (14.5,-2*\n-8) circle(0.2)[fill=white];

\draw (8.6, -2*\n) -- (14.4, -2*\n);
\draw (8.6, -2*\n-8) -- (14.4, -2*\n-8);

}


\end{tikzpicture}}
}

\caption{Examples of trimmed radix-2 computation graphs. Fig (a) is the complete computation graph for $N=8$. For support $\JJ = \{0,1\}$ the computation graph can be trimmed to (c), and effectively computes a $2-$ point DFT. For support $\JJ = \{0,4\}$, the trimmed computation graph is (b): in this case, all the time domain samples are used, resulting in a complexity of $\Omega(N)$. }
\label{fig:trim_butterfly}
\end{figure*}
\subsection{Trimming the Radix-2 computation graph}
\label{sec:preliminary_observations_trim}
The $N-$point DFT of a signal $f$ can be computed by the Radix-2  FFT algorithm. See Fig \ref{fig:trim_butterfly}(a) for an illustration of the corresponding computation graph for $N=8$. Note that the computation graph has a recursive structure: (with reference to Fig \ref{fig:trim_butterfly}(a)) The values at the left or level $0$ (i.e., $f(0), f(1), \ldots, f(7)$ give the IDFT of the signal $\F f$, the values at the intermediate locations give the DFT of the downsampled signal. For example, the values $f_{10}, f_{11},f_{12}, f_{13}$ (at level 1) give the IDFT of the downsampled DFT \(\begin{pmatrix}
\F f(0) & \F f(2) & \F f(4) & \F f(6)
\end{pmatrix}\), the values $f_{20}, f_{21}$ (at level 2) give the IDFT of the downsampled signal \(\begin{pmatrix}
\F f(0)& \F f(4)
\end{pmatrix}\).
Also note that the DFT indices on the right are paired up when they differ by $N/2$ (i.e., $\F f [0]$ is paired with $\F f[4]$; $\F f[2]$ is paired with $\F f[6]$, etc.).

Now suppose that the frequency support is already known, then we may consider trimming the computation graph to some redundant computations. For instance, in the computation graph for $N=8$ as in Fig \ref{fig:trim_butterfly}(a), suppose the support is known to be $\JJ = \{0,1\}$, then the  values $f_{10}, f_{11},f_{12}, f_{13}$ (at level 1) referred to earlier, give the IDFT of  \(\begin{pmatrix}
\F f(0) & 0 & 0 & 0
\end{pmatrix}\): thus, all the values $f_{1j}, j=0:3$ would be equal to a multiple of $\F f(0)$. So we need to compute only one of those $4$ values, and computing any one of these four values directly gives us $\F f(0)$. Similarly, 
we need to compute only one of the four values from $f_{14}, f_{15}, f_{16}, f_{17}$. Once these two values (one from the top half of level $1$ and one from the bottom half of level $1$) are computed; we do not have to do further computation on the subsequent levels. This results in the trimmed computation graph given in Figure \ref{fig:trim_butterfly}(c). Note that in this case, the DFT coefficients $(\F f)_\JJ$ can be computed by doing a $2-$point DFT of $\begin{pmatrix}
f(0) & f(4)
\end{pmatrix}$.

We can consider generalizing this: we identify the level in the computation graph at which all the intermediate coeffecients will be equal (i.e., this is the level at which all the DFT coefficients will be isolated), and trim the computation graph up to that level. 
Note that if the support set were $\{0,4\}$ instead, it is impossible to achieve isolation at intermediate stages and the computation graph will use all the $N$ time-domain samples (see Fig \ref{fig:trim_butterfly}(b)).

So if the support $\JJ$ includes indices that differ by $N/2$, this algorithm uses all the $N$ time domain samples, and thus has a complexity of $\Omega(N)$. For the (deterministic) family of $k-$sparse signals, there are always $\JJ \in \famJ_N$ which contain, indices differing by $N/2$, and so the complexity of this algorithm for the family of $k-$ sparse signals is $\Omega(N)$. For the family $\Z_N^{\downarrow k}$, the probability that there are two indices in $\JJ\in \Z_N^{\downarrow k}$ differing by $N/2$ is 
\begin{align*}
1-\left(1- \frac{k^2}{N^2} \right)^{N/2}&=1-\left(1- \frac{k^2}{N^2} \right)^{\frac{N^2}{k^2}\frac{k^2}{2N}}\\
&\rightarrow l>0 \text{ for }k=\Omega(\sqrt{N}).
\end{align*}
Since this probability is bounded away from zero, the complexity of the algorithm is $\Omega(N)$, which is much higher than our target complexity of $O(k \log k)$. 

\begin{figure*}[ht]
\centering
\includegraphics[width=0.7\textwidth,height = 0.4\textwidth]{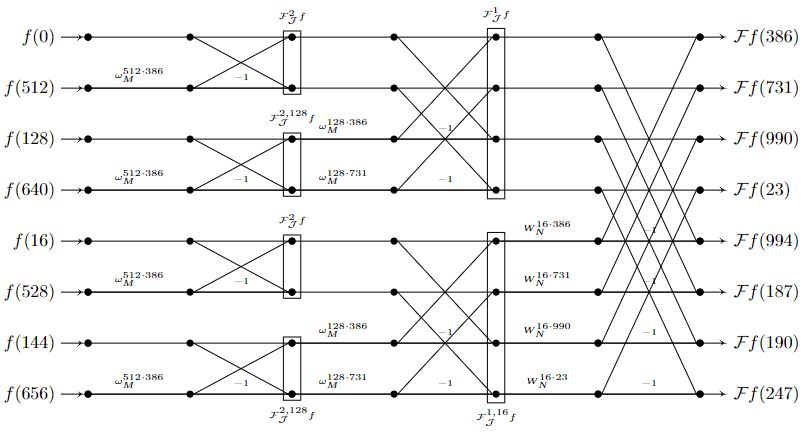}
\caption{DFT computation of $1024-$ length signal $f$ with homogeneous support set $\JJ = \{23, 187, 190, 247, 386, 731, 990, 994\}$. The congruence tree for $\JJ$ is shown in Fig \ref{fig:ex_hom_butterfly}.}
\label{fig:block-diagram_FFT}
\end{figure*}

 In the following few sections, we build towards a generalization of radix-2. This generalization is strict, in the sense that the computation graphs generated by our algorithm cannot be obtained by trimming existing radix-2 computation graphs (see Fig \ref{fig:block-diagram_FFT} for an example). Later (Section \ref{sec:Hi-DFT-to-DFT}), we discuss a simple way to combine the system of equations approach (Section \ref{sec:preliminary_observations_system}) and trimming the generalized radix-2 (similar to Section \ref{sec:preliminary_observations_trim}). For this, we first need to discuss the structure of homogeneous sets.

\section{Congruence trees}
\label{sec:congruence_trees}
In this section, we elaborate on the structure of homogeneous sets. For that (and for the rest of the results), it is helpful to formalize the notion of congruence trees for subsets of $\Z_N$.

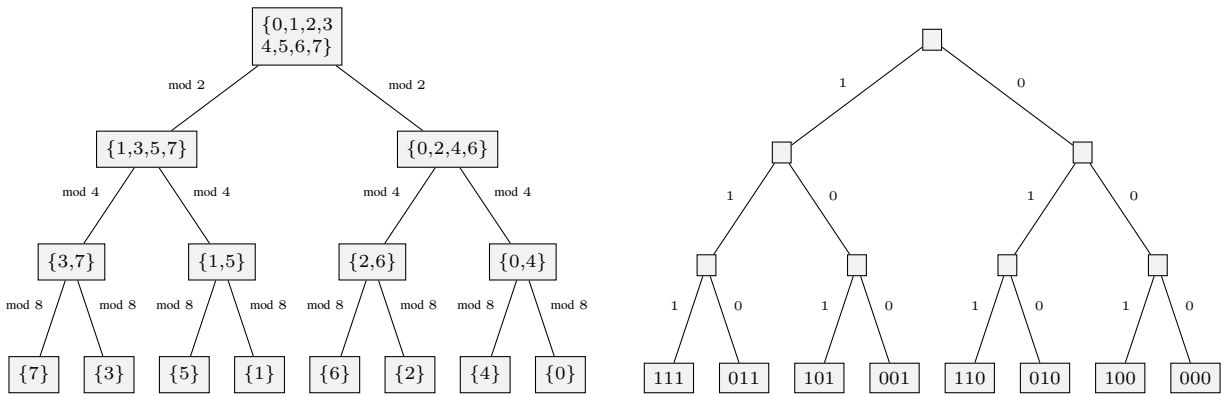
\begin{figure*}[ht]
\centering
\begin{tikzpicture}[scale=0.5]
\node (0) at (0,8) [ minimum size = 0.25cm, fill=gray!10, draw] {$\scriptscriptstyle{\substack{\{0,1,2,3 \\ 4,5,6,7\}}}$};
\node (1.1) at (-4,5) [ minimum size = 0.25cm, fill=gray!10, draw] {$\scriptscriptstyle{\substack{\{1,3,5,7\}}}$};
\node (1.2) at (4,5) [ minimum size = 0.25cm, fill=gray!10, draw] {$\scriptscriptstyle{\substack{\{0,2,4,6\}}}$};
\node (2.1) at (-6,2) [ minimum size = 0.25cm, fill=gray!10, draw] {$\scriptscriptstyle{\substack{\{3,7\}}}$};
\node (2.2) at (-2,2) [ minimum size = 0.25cm, fill=gray!10, draw] {$\scriptscriptstyle{\substack{\{1,5\}}}$};
\node (2.3) at (2,2) [ minimum size = 0.25cm, fill=gray!10, draw] {$\scriptscriptstyle{\substack{\{2,6\}}}$};
\node (2.4) at (6,2) [ minimum size = 0.25cm, fill=gray!10, draw] {$\scriptscriptstyle{\substack{\{0,4\}}}$};
\node (3.1) at (-7,-1) [ minimum size = 0.25cm, fill=gray!10, draw] {$\scriptscriptstyle{\substack{\{7\}}}$};
\node (3.2) at (-5,-1) [ minimum size = 0.25cm, fill=gray!10, draw] {$\scriptscriptstyle{\substack{\{3\}}}$};
\node (3.3) at (-3,-1) [ minimum size = 0.25cm, fill=gray!10, draw] {$\scriptscriptstyle{\substack{\{5\}}}$};
\node (3.4) at (-1,-1) [ minimum size = 0.25cm, fill=gray!10, draw] {$\scriptscriptstyle{\substack{\{1\}}}$};
\node (3.5) at (1,-1) [ minimum size = 0.25cm, fill=gray!10, draw] {$\scriptscriptstyle{\substack{\{6\}}}$};
\node (3.6) at (3,-1) [ minimum size = 0.25cm, fill=gray!10, draw] {$\scriptscriptstyle{\substack{\{2\}}}$};
\node (3.7) at (5,-1) [ minimum size = 0.25cm, fill=gray!10, draw] {$\scriptscriptstyle{\substack{\{4\}}}$};
\node (3.8) at (7,-1) [ minimum size = 0.25cm, fill=gray!10, draw] {$\scriptscriptstyle{\substack{\{0\}}}$};

\draw (node cs:name=0) --node[above left] {$\scriptscriptstyle{\text{mod }2}$} (node cs:name =1.1);
\draw (node cs:name=0) -- node[above right] {$\scriptscriptstyle{\text{mod }2}$}(node cs:name =1.2);

\draw (node cs:name=1.1) --node[above left] {$\scriptscriptstyle{\text{mod }4}$} (node cs:name =2.1);
\draw (node cs:name=1.1) --node[above right] {$\scriptscriptstyle{\text{mod }4}$} (node cs:name =2.2);
\draw (node cs:name=1.2) --node[above left] {$\scriptscriptstyle{\text{mod }4}$} (node cs:name =2.3);
\draw (node cs:name=1.2) --node[above right] {$\scriptscriptstyle{\text{mod }4}$} (node cs:name =2.4);

\draw (node cs:name=2.1) --node[above left] {$\scriptscriptstyle{\text{mod }8}$} (node cs:name =3.1);
\draw (node cs:name=2.1) --node[above right] {$\scriptscriptstyle{\text{mod }8}$} (node cs:name =3.2);
\draw (node cs:name=2.2) --node[above left] {$\scriptscriptstyle{\text{mod }8}$} (node cs:name =3.3);
\draw (node cs:name=2.2) --node[above right] {$\scriptscriptstyle{\text{mod }8}$} (node cs:name =3.4);
\draw (node cs:name=2.3) --node[above left] {$\scriptscriptstyle{\text{mod }8}$} (node cs:name =3.5);
\draw (node cs:name=2.3) --node[above right] {$\scriptscriptstyle{\text{mod }8}$} (node cs:name =3.6);
\draw (node cs:name=2.4) --node[above left] {$\scriptscriptstyle{\text{mod }8}$} (node cs:name =3.7);
\draw (node cs:name=2.4) --node[above right] {$\scriptscriptstyle{\text{mod }8}$} (node cs:name =3.8);


\end{tikzpicture}\hspace{0.5cm}
\begin{tikzpicture}[scale=0.5]
\node (0) at (0,8) [ minimum size = 0.25cm, fill=gray!10, draw] {$\scriptscriptstyle{\substack{}}$};
\node (1.1) at (-4,5) [ minimum size = 0.25cm, fill=gray!10, draw] {$\scriptscriptstyle{\substack{}}$};
\node (1.2) at (4,5) [ minimum size = 0.25cm, fill=gray!10, draw] {$\scriptscriptstyle{\substack{}}$};
\node (2.1) at (-6,2) [ minimum size = 0.25cm, fill=gray!10, draw] {$\scriptscriptstyle{\substack{}}$};
\node (2.2) at (-2,2) [ minimum size = 0.25cm, fill=gray!10, draw] {$\scriptscriptstyle{\substack{}}$};
\node (2.3) at (2,2) [ minimum size = 0.25cm, fill=gray!10, draw] {$\scriptscriptstyle{\substack{}}$};
\node (2.4) at (6,2) [ minimum size = 0.25cm, fill=gray!10, draw] {$\scriptscriptstyle{\substack{}}$};
\node (3.1) at (-7,-1) [ minimum size = 0.25cm, fill=gray!10, draw] {$\scriptscriptstyle{\substack{111}}$};
\node (3.2) at (-5,-1) [ minimum size = 0.25cm, fill=gray!10, draw] {$\scriptscriptstyle{\substack{011}}$};
\node (3.3) at (-3,-1) [ minimum size = 0.25cm, fill=gray!10, draw] {$\scriptscriptstyle{\substack{101}}$};
\node (3.4) at (-1,-1) [ minimum size = 0.25cm, fill=gray!10, draw] {$\scriptscriptstyle{\substack{001}}$};
\node (3.5) at (1,-1) [ minimum size = 0.25cm, fill=gray!10, draw] {$\scriptscriptstyle{\substack{110}}$};
\node (3.6) at (3,-1) [ minimum size = 0.25cm, fill=gray!10, draw] {$\scriptscriptstyle{\substack{010}}$};
\node (3.7) at (5,-1) [ minimum size = 0.25cm, fill=gray!10, draw] {$\scriptscriptstyle{\substack{100}}$};
\node (3.8) at (7,-1) [ minimum size = 0.25cm, fill=gray!10, draw] {$\scriptscriptstyle{\substack{000}}$};

\draw (node cs:name=0) --node[above left] {$\scriptscriptstyle{1}$} (node cs:name =1.1);
\draw (node cs:name=0) -- node[above right] {$\scriptscriptstyle{0}$}(node cs:name =1.2);

\draw (node cs:name=1.1) --node[above left] {$\scriptscriptstyle{1}$} (node cs:name =2.1);
\draw (node cs:name=1.1) --node[above right] {$\scriptscriptstyle{0}$} (node cs:name =2.2);
\draw (node cs:name=1.2) --node[above left] {$\scriptscriptstyle{1}$} (node cs:name =2.3);
\draw (node cs:name=1.2) --node[above right] {$\scriptscriptstyle{0}$} (node cs:name =2.4);

\draw (node cs:name=2.1) --node[above left] {$\scriptscriptstyle{1}$} (node cs:name =3.1);
\draw (node cs:name=2.1) --node[above right] {$\scriptscriptstyle{0}$} (node cs:name =3.2);
\draw (node cs:name=2.2) --node[above left] {$\scriptscriptstyle{1}$} (node cs:name =3.3);
\draw (node cs:name=2.2) --node[above right] {$\scriptscriptstyle{0}$} (node cs:name =3.4);
\draw (node cs:name=2.3) --node[above left] {$\scriptscriptstyle{1}$} (node cs:name =3.5);
\draw (node cs:name=2.3) --node[above right] {$\scriptscriptstyle{0}$} (node cs:name =3.6);
\draw (node cs:name=2.4) --node[above left] {$\scriptscriptstyle{1}$} (node cs:name =3.7);
\draw (node cs:name=2.4) --node[above right] {$\scriptscriptstyle{0}$} (node cs:name =3.8);

\end{tikzpicture}
\caption{Connection to binary representation: $\TT_8$ (on the left) and binary representation (on the right).}
\label{fig:binary_tree_Z_8}
\end{figure*}

Recall that a binary tree is a graph in which each node has at most two children nodes. See Fig \ref{fig:binary_tree_Z_8} for an example. A leaf node of a binary tree has no children. The root of a binary tree is the node with no parent. Each node in a binary tree has a label, and we will work with binary trees whose node labels are subsets of $\mathbb{Z}_N$. For a binary tree $\TT$, we let  $2\TT$ represent the binary tree of $\TT$ with all node label elements multiplied by $2$. Likewise, by $\TT+1$, we mean the binary tree of $\TT$ with all node label elements incremented by $1$.


First, we note that $\mathbb{Z}_N$ naturally defines a complete binary tree by successively splitting modulo $2$. The root of such a binary tree has the label $\mathbb{Z}_N$, and its children are labeled by the even and odd indices of $\mathbb{Z}_N$. More formally, the tree $\TT_N$ representing the set $\mathbb{Z}_N$ is defined recursively as  $(\text{left subtree}, \text{right subtree}, \text{root label})$:
\[
\TT_N = \left(2\TT_{N/2}+1 , 2\TT_{N/2}, \mathbb{Z}_N \right).
\]


We often refer to $\TT_N$ as $\TT$ when $N$ is apparent from the context. 
Note that all node labels are subsets of $\mathbb{Z}_N$ corresponding to congruence classes (we will return to this later). Each node in $\TT_N$, other than the leaves, have exactly two children. 
 When it is clear from the context, we often use the node label to refer to the node itself (i.e., we make no distinction between the node and its label). For instance, `index $i$ is in node $v$' means that `index $i$ is an element in the label of $v$'.

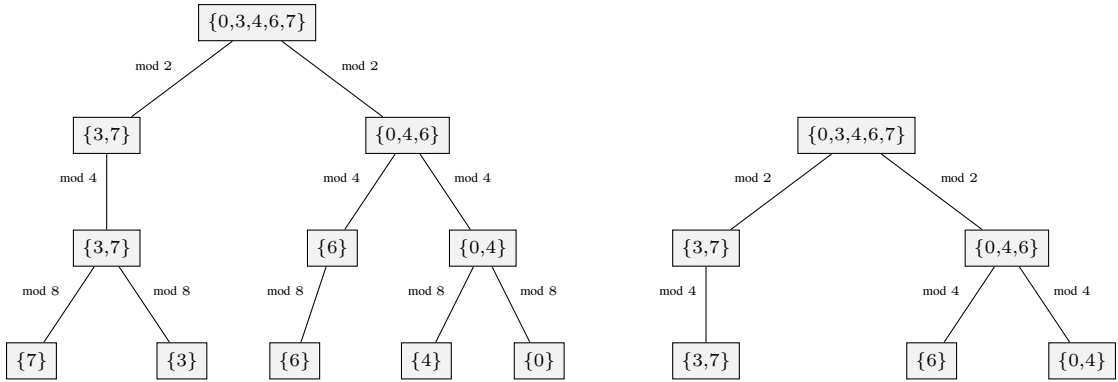
\begin{figure*}[ht]
\centering
\begin{tikzpicture}[scale=0.5]
\node (0) at (0,8) [ minimum size = 0.25cm, fill=gray!10, draw] {$\scriptscriptstyle{\substack{\{0,3,4,6,7\}}}$};
\node (1.1) at (-4,5) [ minimum size = 0.25cm, fill=gray!10, draw] {$\scriptscriptstyle{\substack{\{3,7\}}}$};
\node (1.2) at (4,5) [ minimum size = 0.25cm, fill=gray!10, draw] {$\scriptscriptstyle{\substack{\{0,4,6\}}}$};
\node (2.1) at (-4,2) [ minimum size = 0.25cm, fill=gray!10, draw] {$\scriptscriptstyle{\substack{\{3,7\}}}$};
\node (2.3) at (2,2) [ minimum size = 0.25cm, fill=gray!10, draw] {$\scriptscriptstyle{\substack{\{6\}}}$};
\node (2.4) at (6,2) [ minimum size = 0.25cm, fill=gray!10, draw] {$\scriptscriptstyle{\substack{\{0,4\}}}$};
\node (3.1) at (-6,-1) [ minimum size = 0.25cm, fill=gray!10, draw] {$\scriptscriptstyle{\substack{\{7\}}}$};
\node (3.2) at (-2,-1) [ minimum size = 0.25cm, fill=gray!10, draw] {$\scriptscriptstyle{\substack{\{3\}}}$};
\node (3.3) at (1,-1) [ minimum size = 0.25cm, fill=gray!10, draw] {$\scriptscriptstyle{\substack{\{6\}}}$};
\node (3.7) at (4.5,-1) [ minimum size = 0.25cm, fill=gray!10, draw] {$\scriptscriptstyle{\substack{\{4\}}}$};
\node (3.8) at (7.5,-1) [ minimum size = 0.25cm, fill=gray!10, draw] {$\scriptscriptstyle{\substack{\{0\}}}$};

\draw (node cs:name=0) --node[above left] {$\scriptscriptstyle{\text{mod }2}$} (node cs:name =1.1);
\draw (node cs:name=0) -- node[above right] {$\scriptscriptstyle{\text{mod }2}$}(node cs:name =1.2);

\draw (node cs:name=1.1) --node[above left] {$\scriptscriptstyle{\text{mod }4}$} (node cs:name =2.1);
\draw (node cs:name=1.2) --node[above left] {$\scriptscriptstyle{\text{mod }4}$} (node cs:name =2.3);
\draw (node cs:name=1.2) --node[above right] {$\scriptscriptstyle{\text{mod }4}$} (node cs:name =2.4);

\draw (node cs:name=2.1) --node[above left] {$\scriptscriptstyle{\text{mod }8}$} (node cs:name =3.1);
\draw (node cs:name=2.1) --node[above right] {$\scriptscriptstyle{\text{mod }8}$} (node cs:name =3.2);
\draw (node cs:name=2.3) --node[above left] {$\scriptscriptstyle{\text{mod }8}$} (node cs:name =3.3);
\draw (node cs:name=2.4) --node[above left] {$\scriptscriptstyle{\text{mod }8}$} (node cs:name =3.7);
\draw (node cs:name=2.4) --node[above right] {$\scriptscriptstyle{\text{mod }8}$} (node cs:name =3.8);
\end{tikzpicture}\hspace{1cm}
\begin{tikzpicture}[scale=0.5]
\node (0) at (0,8) [ minimum size = 0.25cm, fill=gray!10, draw] {$\scriptscriptstyle{\substack{\{0,3,4,6,7\}}}$};
\node (1.1) at (-4,5) [ minimum size = 0.25cm, fill=gray!10, draw] {$\scriptscriptstyle{\substack{\{3,7\}}}$};
\node (1.2) at (4,5) [ minimum size = 0.25cm, fill=gray!10, draw] {$\scriptscriptstyle{\substack{\{0,4,6\}}}$};
\node (2.1) at (-4,2) [ minimum size = 0.25cm, fill=gray!10, draw] {$\scriptscriptstyle{\substack{\{3,7\}}}$};
\node (2.3) at (2,2) [ minimum size = 0.25cm, fill=gray!10, draw] {$\scriptscriptstyle{\substack{\{6\}}}$};
\node (2.4) at (6,2) [ minimum size = 0.25cm, fill=gray!10, draw] {$\scriptscriptstyle{\substack{\{0,4\}}}$};

\draw (node cs:name=0) --node[above left] {$\scriptscriptstyle{\text{mod }2}$} (node cs:name =1.1);
\draw (node cs:name=0) -- node[above right] {$\scriptscriptstyle{\text{mod }2}$}(node cs:name =1.2);

\draw (node cs:name=1.1) --node[above left] {$\scriptscriptstyle{\text{mod }4}$} (node cs:name =2.1);
\draw (node cs:name=1.2) --node[above left] {$\scriptscriptstyle{\text{mod }4}$} (node cs:name =2.3);
\draw (node cs:name=1.2) --node[above right] {$\scriptscriptstyle{\text{mod }4}$} (node cs:name =2.4);

\end{tikzpicture}
\caption{Example tree: $\TT_8(\{0,3,6,7\})$ (on the left) and truncated tree $\TT_8^2(\{0,3,6,7\})$ (on the right).}
\label{fig:truncated_tree}
\end{figure*}

We will often work with subtrees of $\TT$ defined as follows. Given $\JJ$, we construct a subtree by taking the intersection of all node labels in $\TT$ with $\JJ$, and discarding any resulting empty labels. We denote such a binary tree by $\TT(\JJ)$ and call it the \emph{congruence tree} for $\JJ$. See Fig \ref{fig:truncated_tree} for example. 

The leaves of this tree have singleton labels: thus, each leaf corresponds to an element of the set $\JJ$. The \emph{level} of any node in $\TT(\JJ)$ is its distance from the root: the root itself is at level $0$, its children are at level $1$, and so on. All the leafs of $\TT_N(\JJ)$ are at level $M$. 

We note that at each level of $\TT(\JJ)$, the indices in $\JJ$ are essentially split into congruence classes; at $l$ levels from the root, the extant $2^l$ nodes correspond to the elements of $\JJ$ that leaves a reminder of $0$, $1, 2, \ldots, 2^{l}-1$ when divided by $2^l$, respectively. This binary tree representation of the set is equivalent to the binary expansion of every element in $\JJ$: tracing the path from the root to the leaf gives us the binary expansion of the leaf. For example, Fig. \ref{fig:binary_tree_Z_8} shows the equivalence between $\TT_8$ and binary expansion of elements in $\ZZ_8$. 


\begin{remark}
\label{rem:tree-2-adic}
We make the following useful observations:
\begin{enumerate}
    \item Leaves $j_a, j_b$ are descendants of a node at level $l$ in $\TT(\JJ)$ if and only if $2^l$ divides $j_a-j_b$; and the first (starting from leaf) common ancestor of $j_a$ and $j_b$ is $l$ if and only if $j_a-j_b$ is an odd multiple of $2^l$.
    \item  If a node $v$ is at level $l$; then $\omega_l^{j_1}=\omega_l^{j_2}$ for any $j_1,j_2$ in $v$. Thus, we can define $\omega_l^v \coloneqq \omega_l^j$ for any $j$ in $v$.
\end{enumerate}
\end{remark}


Given a binary tree $\TT(\JJ)$, we define the truncated tree $\TT^r(\JJ)$ obtained by removing all nodes below level $r$ from $\TT(\JJ)$. Unlike the trees $\TT(\JJ)$, the leaves of $\TT^r(\JJ)$ are at level $r$, and may not be labeled by singletons. For example, see Fig \ref{fig:truncated_tree}. The algorithm we propose later in Section \ref{sec:Hi-DFT-to-DFT} is executed on truncated trees $\TT^r(\JJ)$ with a suitably chosen $r$. 

    
    \emph{Note:} Given the entries of $\JJ$ in binary, we note that computing the congruence structure $\TT^r(\JJ)$ requires accessing the first $r$ bits for each entry in $\JJ$, and so costs $O(kr)$ bit operations. For example, when $r=M=\log N$ (i.e, we need to compute the tree up to the last level), we need $O(k \log N)$ bit operations. 



Given a weight vector $\vec{w}:\mathbb{Z}_N \rightarrow \mathbb{C}$ defined on the indices of $\mathbb{Z}_N$, we consider the $w-$\emph{induced weight} of a node $v$ in $\TT^r(\JJ)$ as
\[
\mu_\JJ(v, \vec{w}) = \sum_{i \in \text{ label of }v} w_i.
\]
Note that
\[
\mu_\JJ(v) := \mu_\JJ(v, \vec{1}) 
\]
is the number of indices in the label of $v$. We often refer to this as the weight of the node $v$ in $\TT(\JJ)$. If a node $v$ has an empty label in $\TT(\JJ)$ (or the node $v$ is not in $\TT(\JJ)$), we set the weight $\mu(v, \vec{w})=0$.

If node $v$ is located at level $l$, we see that $\mu_\JJ(v)$ is simply the number of elements of $\JJ$ that are congruent to $v$ modulo $2^l$. Also, 
\begin{align}
\sum_{v \text{ at level }l}\mu_\JJ(v) = |\JJ| = k, \text{ for any level }r,\text{ and} \quad \nonumber \\  \mu_\JJ(v) = \mu_\JJ(v_1) + \mu_\JJ(v_2), \text{ where }v_1,v_2\text{ are children of }v \label{eq:mu-recursion-level}.
\end{align}

For example, consider the congruence trees in Fig \ref{fig:tree_elementary}: these trees are \emph{perfectly balanced}, i.e., all the nodes at the same level have equal weight. Such tree representations of sets are common: they were used in \cite{kapralov2019dimension} for dimension independent sparse-FFT and in \cite{DS-1} to characterize universal sets. 
We also need the following definitions
\begin{defn} (\emph{Pivots})
We say a level $l$ is a pivot in $\TT(\JJ)$ if there exists a node at level $l$ in $\TT(\JJ)$ with two children. Thus a pivot is a level at which there is a split in the congruence tree. 
\end{defn}
From Remark \ref{rem:tree-2-adic}(1) we conclude the following
\begin{remark}
\label{rem:pivot-as-gcd}
A level $e$ is a pivot for $\TT(\JJ)$ iff $2^e$ is the largest power of $2$ dividing $j_1-j_2$ for some $j_1\neq j_2 \in \JJ$. 
\end{remark}
Clearly, the number of pivots is upper bounded by the number of levels ($\log N$) in the tree. Also, note that in a congruence tree $\TT(\JJ)$, the root node is $\JJ$, and all the leaf nodes are singletons; so there is a lower bound on the number of pivots. 

\begin{proposition}
\label{prop:num-pivots-cong-tree}
The congruence tree $\TT(\JJ)$ has at least $\log |\JJ|$ pivots.
\end{proposition}
See Appendix \ref{proof:prop:num-pivots-cong-tree} for the proof.

\begin{defn}
A congruence tree $\TT(\JJ)$ (and equivalently the set $\JJ$) is called \emph{homogeneous} if it has exactly $\log |\JJ|$ pivots.
\end{defn}
The homogeneous trees in this work are the same as $2-$homogenous trees from \cite{fan2016compact} and equivalent to conforming digit tables from \cite{siripuram2020convolution, 9518104}. We note the following properties of homogeneous congruence trees (again, the proof is deferred to the appendix).
\begin{proposition}
\label{prop:hom-tree-prop}
Suppose $N=2^M$ and $\JJ \subseteq \Z_N$. The following statements are equivalent. 
\begin{enumerate}
    \item The congruence tree $\TT(\JJ)$ is homogeneous.
    \item The children have equal weight at every split in the congruence tree $\TT(\JJ)$. If a node at level $l$ splits, then so do all the nodes at level $l$.

 \item The congruence tree $\TT(\JJ)$ has $s$ pivots (say $r_1, r_2, \ldots, r_s$) with $\max_{v \text{ at level }r} \mu_\JJ(v) = 2^{\sum_s I(r < = r_s)}$.
    \item If $\JJ=\{j_1, j_2, \ldots, j_{2^s}\}$, for any $j_a \neq j_b$ let $2^{e_{ab}}$ be the largest power of $2$ that divides $j_a-j_b$. Then there are $s$ distinct $e_{ab}$. 
   
\end{enumerate}
\end{proposition}

Simple examples of homogeneous sets include consecutive sets of the form $\{0,1,2,\ldots, 2^s\}$ (here the pivots are $\{0,1,2,\ldots, s-1\}$) and periodic sets of the form $\{0, N/2^s, 2N/2^s, \ldots,(2^s-1)N/2^s\} $ (the pivots are $\{M-s,M-s+1,\ldots, M-1\}$): this can be verified from Proposition \ref{prop:hom-tree-prop} (4). Another example is $\JJ=\{3,17,25,27 \}$, which has pivots $\{1,3\}$.

It is known that $\TT(\JJ)$ is homogeneous iff $\JJ$ is spectral (see Definition \ref{def:spectral}). This equivalence has also been proved in more general frameworks; for $\mathbb{Q}_p$ in \cite{fan2016compact} and as the structure of convolution idempotents via digit-tables \cite{siripuram2020convolution,9518104}. To keep the manuscript self-contained, we reproduce a simpler (tailored to our setting) version of the proof that homogeneous trees are equivalent to spectral sets (Proposition \ref{prop:homo-spectral}).

 We often work with sets $\JJ$ for which the tree $\TT(\JJ)$ is partly homogeneous, defined below.
\begin{defn} \label{def:part-H-tree}($\vec{r}-$\emph{homogeneous} trees, $\vec{r}-$part-homogeneous trees, (sub) vector of pivots, and height)
\begin{enumerate}
    \item Supports $\JJ$ that describe a homogeneous tree with pivots $\vec{r}$ are called $\vec{r}-$homogeneous. 
    \item For a vector $\vec{r}$ of pivots with entries arranged in ascending order, we denote by $\vec{r}^{n-}$ the vector obtained by removing the last $n$ entries of $\vec{r}$. Specifically, we denote by $\vec{r}^-$ the vector obtained by removing the last entry from $\vec{r}$. For e.g., with $\vec{r} = (0,1, 4, 5, 9)$, we have $\vec{r}^-=  (0,1,4,5)$ and $\vec{r}^{3-} = (0,1)$. We denote by $r_{\text{max}}$ the largest (last) entry in $\vec{r}$ and by $\Sizer$ the number of pivots in $\vec{r}$. Note that for an $\vec{r}-$homogeneous $\JJ$ we must have $\Sizer = \log |\JJ|$; and that $\vec{r}^{(\Sizer - n)-}$ is a vector obtained by retaining only the first $n$ pivots from $\vec{r}$. Finally, we say $r \in \vec{r}$ to mean $r$ is one of the entries of $\vec{r}$.
    \item An index set $\JJ$ is said to be $\vec{r}-$part-homogeneous if there are no pivots in $\TT(\JJ)$ within the first $r_{\text{max}}$ levels other than those in $\vec{r}$. In particular, $\JJ$ can have (additional) pivots larger than $r_{\text{max}}$; and it can be that some $r \in \vec{r}$ is not a pivot for $\TT(\JJ)$. Clearly, any $\vec{r}-$homogeneous set is $\vec{r}^{i-}-$ part homogeneous for any $i$; and every index set is $\{0,1,2,\ldots\}-$part-homogeneous.
    \item We also find it useful to define the $\vec{r}-$height (or simply height) of a node $v$ in a $\vec{r}-$part-homogeneous tree as the number pivots between $v$ and $r_{\text{max}}$; with the endpoints $v$ and $r_{\text{max}}$ included.
    \end{enumerate}

\end{defn}
The key idea relevant to the current discussion is that (part) homogeneity of $\JJ$ can be exploited to compute the DFT coefficients efficiently. We will discuss this in the coming sections.

\begin{figure*}[ht]
\centering

\begin{tikzpicture}[scale=0.5]
\node (0) at (0,8) [ minimum size = 0.25cm, fill=gray!10, draw] {$\scriptscriptstyle{\substack{\{3,17, \\ 25,27, 35\}}}$};
\node (1) at (0,5) [minimum size = 0.25cm, fill=gray!10, draw] {$\scriptscriptstyle{\substack{\{3,17, \\ 25,27, 35\}}}$};
\node (2.1) at (-4,2) [ minimum size = 0.5cm, fill=gray!10,
draw] {$\scriptscriptstyle{\{17,25\}}$};
\node (2.2) at (4,2) [minimum size = 0.5cm, fill=gray!10, draw]
{$\scriptscriptstyle{\{3,27, 35\}}$};
\node (3.1) at (4,-1) [minimum size = 0.5cm, fill=gray!10, draw]
{$\scriptscriptstyle\{3,27, 35\}$};
\node (3.2) at (-4,-1) [minimum size = 0.5cm, fill=gray!10, draw]
{$\scriptscriptstyle\{17,25\}$};
\node (4.2) at (6,-4) [minimum size = 0.5cm, fill=gray!10, draw]
{$\scriptscriptstyle\{3, 35\}$};
\node (4.3) at (2,-4) [minimum size = 0.5cm, fill=gray!10, draw]
{$\scriptscriptstyle\{27\}$};\
\node (4.0) at (-6,-4) [minimum size = 0.5cm, fill=gray!10, draw]
{$\scriptscriptstyle\{17\}$};
\node (4.1) at (-2,-4) [minimum size = 0.5cm, fill=gray!10, draw]
{$\scriptscriptstyle\{25\}$};
\node (5.2) at (8,-7) [minimum size = 0.5cm, fill=gray!10, draw]
{$\scriptscriptstyle\{35\}$};
\node (5.4) at (4,-7) [minimum size = 0.5cm, fill=gray!10, draw]
{$\scriptscriptstyle\{3\}$};
\node (5.3) at (2,-7) [minimum size = 0.5cm, fill=gray!10, draw]
{$\scriptscriptstyle\{27\}$};\
\node (5.0) at (-6,-7) [minimum size = 0.5cm, fill=gray!10, draw]
{$\scriptscriptstyle\{17\}$};
\node (5.1) at (-2,-7) [minimum size = 0.5cm, fill=gray!10, draw]
{$\scriptscriptstyle\{25\}$};
\draw (0) -- node[left] {$\scriptscriptstyle{\text{mod }2}$}(1);
\draw (node cs:name=1) --node[above left] {$\scriptscriptstyle{\text{mod }4}$} (node cs:name =2.1);
\draw (node cs:name=1) -- node[above right] {$\scriptscriptstyle{\text{mod }4}$}(node cs:name =2.2);
\draw (node cs:name=2.1) -- node[above left] {$\scriptscriptstyle{\text{mod }8}$}(node cs:name =3.2);
\draw (node cs:name=2.2) -- node[above left] {$\scriptscriptstyle{\text{mod }8}$}(node cs:name =3.1);
\draw (node cs:name=3.2) -- node[above left] {$\scriptscriptstyle{\text{mod }16}$}(node cs:name =4.0);
\draw (node cs:name=3.2) -- node[above right] {$\scriptscriptstyle{\text{mod }16}$}(node cs:name =4.1);
\draw (node cs:name=3.1) -- node[above right] {$\scriptscriptstyle{\text{mod }16}$}(node cs:name =4.2);
\draw (node cs:name=3.1) -- node[above left] {$\scriptscriptstyle{\text{mod }16}$}(node cs:name =4.3);
\draw (node cs:name=4.0) -- node[above left] {$\scriptscriptstyle{\text{mod }32}$}(node cs:name =5.0);
\draw (node cs:name=4.1) -- node[above right] {$\scriptscriptstyle{\text{mod }32}$}(node cs:name =5.1);
\draw (node cs:name=4.2) -- node[above right] {$\scriptscriptstyle{\text{mod }32}$}(node cs:name =5.2);
\draw (node cs:name=4.3) -- node[above left] {$\scriptscriptstyle{\text{mod }32}$}(node cs:name =5.3);
\draw (node cs:name=4.2) -- node[above left] {$\scriptscriptstyle{\text{mod }32}$}(node cs:name =5.4);


\draw[->] (1.5,8) --node[right = 11mm] {$\mathbf{level = 0, height = 2}$} (4.5,8);
\draw[->] (1.5,5) --node[right = 11mm] {$\mathbf{level = 1, height = 2}$} (4.5,5);
\draw[->] (5.5,2) --node[right = 11mm] {$\mathbf{level = 2, height = 1}$} (8.5,2);
\draw[->] (5.5,-1) --node[right = 11mm] {$\mathbf{level = 3, height = 1}$} (8.5,-1);
\draw[->] (7.5,-4) --node[right = 11mm] {$\mathbf{level = 4, height = 0}$} (10.5,-4);
\draw[->] (9.5,-7) --node[right = 11mm] {$\mathbf{level = 5, height = 0}$} (10.5,-7);

\end{tikzpicture}

\caption{An example $\{1,3\}-$part-homogeneous congruence tree for $\JJ= \{3,17,25,27,35\}$, with the levels and heights of each node indicated.}
\label{fig:homologus_tree_height}
\end{figure*}
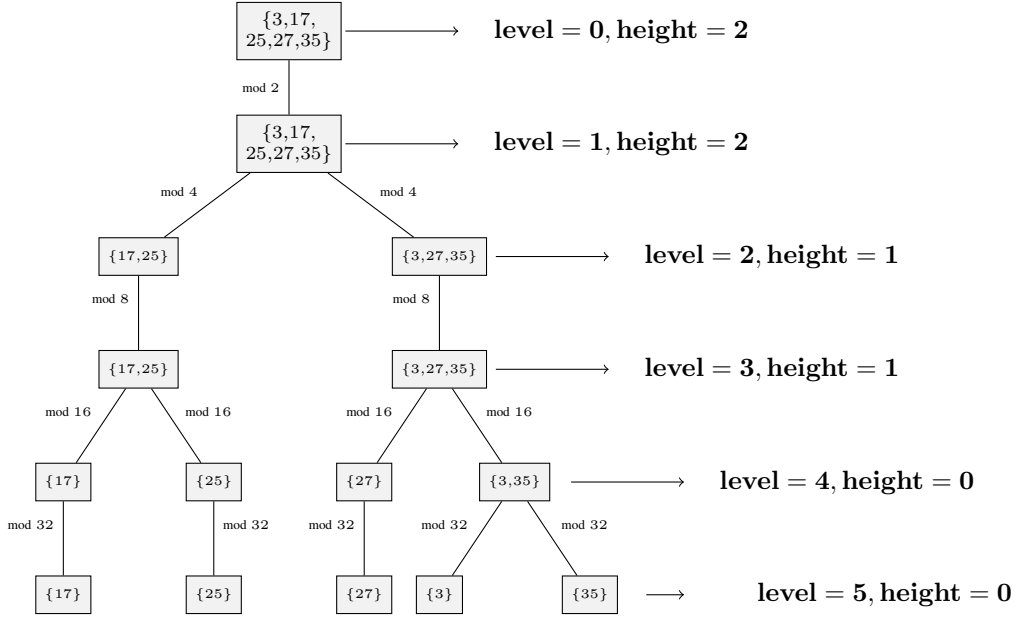

For a homogeneous tree, the height of any node $v$ is the number of splits at and below the level of $v$ in the congruence tree $\TT(\JJ)$; and in general, for a $\vec{r}-$part-homogeneous tree, the height of any node $v$ is the number of splits at and below the level of $v$ in the truncated congruence tree $\TT^{r_{\text{max}}+1}(\JJ)$. If the homogeneous tree is $\TT_N$, then the height and level of any node sum to $M$. See Fig \ref{fig:homologus_tree_height} for an illustration.
By part homogeneity, the number of pivots from root till height $n$ is at most $\Sizer-n$, and so we have the following observation.
\begin{remark}
\label{rem:part-homogenous-number-of-nodes}
If $\JJ$ is $\vec{r}-$part-homogeneous, then the number of nodes in $\TT(\JJ)$ at height $n$ is at most $2^{\Sizer-n}$. If $\JJ$ is $\vec{r}-$homogeneous, then the number of nodes at height $n$ is exactly $2^{\Sizer-n}$.
\end{remark}

\section{Downsampling and congruence trees}
\label{sec:down-samp-cong-tree}
This section explores the connection between downsampling and congruence trees. It is well known that uniform downsampling of the signal in the time domain leads to aliasing in frequency; we observe (details below) that the structure of the congruence trees captures this aliasing pattern. We also generalize these observations to specific non-uniform downsampling patterns. We find it helpful to introduce the notion of Homogeneity-induced-DFT, which we set up later in the section.

Consider $f \in \mathbb{C}^N$; here,  $f$ represents an $N-$length time-domain discrete signal. Let $\F f$ denote the DFT of $f$, and let us assume that the DFT $\F f$ is supported on $\JJ$, as in the introduction. Now, sampling the signal $f$ uniformly in the time domain results in the sampled signal (denoted $f_{\downarrow m}$)
\[
f_{\downarrow m}[k] = f[2^mk], \text{ for }k=0,1,2\ldots
\]
The downsampling factor $2^m$ also corresponds to the resolution in the time domain. A lower value of $m$ implies a higher resolution, i.e., we collect more samples in the time domain. For example, if $m=0$: we take all the time domain samples, and if $m=1$: we take half of the samples.  

It is well known that the down-sampling process leads to aliasing \cite{osgood2018lectures} of the frequency domain signal $\F f$ :
\begin{equation}
\F f_{\downarrow m}[n] = 2^{M-m}\sum_{r=0}^{2^m}\F f[n + r2^{M-m}],
\end{equation}

The right-hand side above is the sum of all DFT coefficients with indices congruent to $n$ modulo $2^{M-m}$. In the notation being considered here, this is simply the sum of all weights at the corresponding node in $\TT(\JJ)$; so that 
\begin{equation}
\F f_{\downarrow m} [n] = \mu_\JJ(v, \F f), 
\label{eq:aliasing-full-tree}
\end{equation}

\begin{figure*}[ht]
\centering

\begin{tikzpicture}[scale = 0.35]
\node[scale=0.25]  (0) at (0,8) [ minimum size = 0.25cm, fill=gray!10, draw] {};
\node[above right,font=\tiny] at (0,8) {$\F f_{\downarrow 8}[0] = \sum_{i=0}^7\F f[i]$};
\node[scale=0.25] (1.1) at (-12,4) [ minimum size = 0.25cm, fill=gray!10, draw] {};
\node[above left,font=\tiny] at (-12,4) { \(\F f_{\downarrow 4}[1] = \sum_{i \text{ odd }} \F f[i]\)}; 

\node[scale=0.25] (1.2) at (12,4) [ minimum size = 0.25cm, fill=gray!10, draw] {};
\node[above right,font=\tiny] at (12,4) { \(\F f_{\downarrow 4}[1] = \sum_{i \text{ even }} \F f[i]\)};
\node (2.1)[scale=0.25]  at (-18,1) [ minimum size = 0.25cm, fill=gray!10, draw] {};
\node [above left,font=\tiny]  at (-18,1) {$\F f_{\downarrow 2}[3] = \F f[3]+\F f[7]$};

\node (2.2)[scale=0.25]  at (-6,1) [ minimum size = 0.25cm, fill=gray!10, draw] {};
\node [above right,font=\tiny]  at (-6,1) {$\F f_{\downarrow 2}[1] = \F f[1]+\F f[5]$};

\node (2.3)[scale=0.25]  at (6,1) [ minimum size = 0.25cm, fill=gray!10, draw] {};
\node [below left,font=\tiny]  at (6,1) {\(\F f_{\downarrow 2}[2] = \F f[2]+\F f[6]\)};

\node (2.4)[scale=0.25]  at (18,1) [ minimum size = 0.25cm, fill=gray!10, draw] {};
\node [above right,font=\tiny]  at (18,1) {\(\F f_{\downarrow 2}[0] = \F f[0]+\F f[4]\)};

\node (3.1)[scale=0.25]  at (-20,-2) [ minimum size = 0.25cm, fill=gray!10, draw] {};
\node [left,font=\tiny]  at (-20,-2)  {$\F f[7]$};
\node (3.2)[scale=0.25]  at (-16,-2) [ minimum size = 0.25cm, fill=gray!10, draw] {};
\node [right,font=\tiny]   at (-16,-2)  {$\F f[3]$};
\node (3.3)[scale=0.25]  at (-8,-2) [ minimum size = 0.25cm, fill=gray!10, draw] {};
\node [left,font=\tiny] at (-8,-2)  {$\F f[5]$};
\node (3.4)[scale=0.25]  at (-4,-2) [ minimum size = 0.25cm, fill=gray!10, draw] {};
\node [right,font=\tiny]  at (-4,-2)  {$\F f[1]$};
\node (3.5)[scale=0.25]  at (4,-2) [ minimum size = 0.25cm, fill=gray!10, draw] {};
\node [left,font=\tiny]  at (4,-2) {$\F f[6]$};
\node (3.6)[scale=0.25]  at (8,-2) [ minimum size = 0.25cm, fill=gray!10, draw] {};
\node [right,font=\tiny] at (8,-2)  {$\F f[2]$};
\node (3.7)[scale=0.25]  at (16,-2) [ minimum size = 0.25cm, fill=gray!10, draw] {};
\node [left,font=\tiny] at (16,-2)  {$\F f[4]$};
\node (3.8)[scale=0.25]  at (20,-2) [ minimum size = 0.25cm, fill=gray!10, draw] {};
\node [right,font=\tiny]  at (20,-2) {$\F f[0]$};

\draw (node cs:name=0) --node[above left] {} (node cs:name =1.1);
\draw (node cs:name=0) -- node[above right] {}(node cs:name =1.2);

\draw (node cs:name=1.1) --node[above left] {} (node cs:name =2.1);
\draw (node cs:name=1.1) --node[above right] {} (node cs:name =2.2);
\draw (node cs:name=1.2) --node[above left] {} (node cs:name =2.3);
\draw (node cs:name=1.2) --node[above right] {} (node cs:name =2.4);

\draw (node cs:name=2.1) --node[above left] {} (node cs:name =3.1);
\draw (node cs:name=2.1) --node[above right] {} (node cs:name =3.2);
\draw (node cs:name=2.2) --node[above left] {} (node cs:name =3.3);
\draw (node cs:name=2.2) --node[above right] {} (node cs:name =3.4);
\draw (node cs:name=2.3) --node[above left] {} (node cs:name =3.5);
\draw (node cs:name=2.3) --node[above right] {} (node cs:name =3.6);
\draw (node cs:name=2.4) --node[above left] {} (node cs:name =3.7);
\draw (node cs:name=2.4) --node[above right] {} (node cs:name =3.8);

\end{tikzpicture}
\caption{Connection to aliasing: $\TT^8$.}
\label{fig:aliasing_tree_Z_8}
\end{figure*}
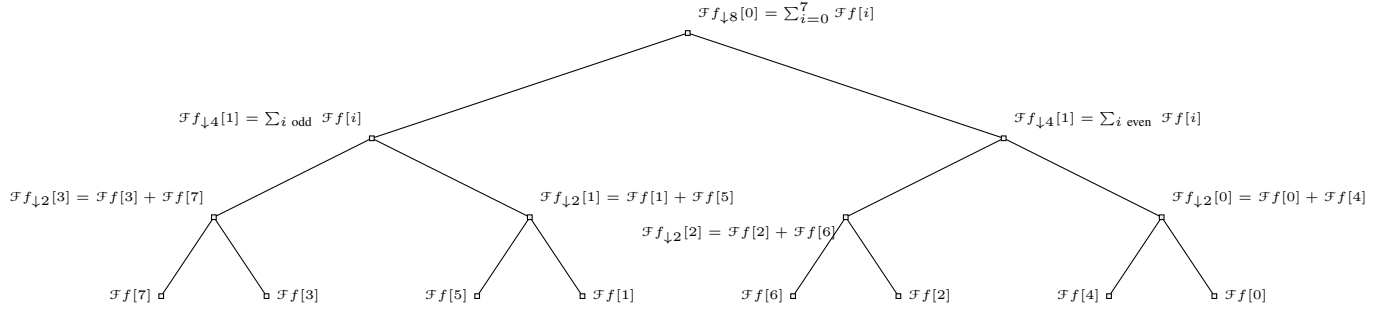
Where $v$ is the node congruent to $n$ at level $M-m$. In other words, if we consider the DFT coefficients in $\F f$ as the weights on the indices, then the weights of the nodes at level $M-m$ correspond to the DFT coefficients of the signal downsampled by $2^m$ in the time domain. Thus, the induced weights at level $l$ can be found by computing the DFT of the downsampled signal $f_{\downarrow (M-l)}$. Since this is a $2^l$ length signal, the computational complexity of finding the induced weights is $c_1l2^{l}$ by a radix-$2$ FFT (where $c_1 = 3/2$).Note that in this case, we uniformly sample the signal in the time domain to obtain $2^l$ samples.


By sampling to $2^l$ samples in the time domain, we incur aliasing in the frequency domain, and the pattern of this aliasing is captured by the structure of $\TT(\JJ)$, as discussed previously. For some structures, the aliasing may not be too \emph{heavy}, i.e., only a few DFT coefficients get aliased. We discuss the simplest example of such a structure next.

\subsection{Elementary sets}
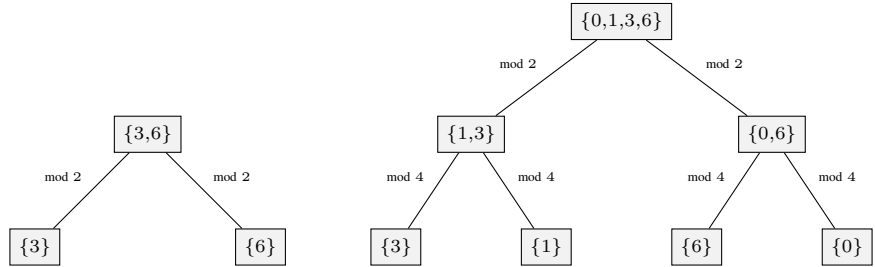
\begin{figure*}[ht]
\centering
\begin{tikzpicture}[scale=0.5]
\node (0) at (0,8) [ minimum size = 0.25cm, fill=gray!10, draw] {$\scriptscriptstyle{\substack{\{3, 6\}}}$};
\node (1.1) at (-3,5) [ minimum size = 0.25cm, fill=gray!10, draw] {$\scriptscriptstyle{\substack{\{3\}}}$};
\node (1.2) at (3,5) [ minimum size = 0.25cm, fill=gray!10, draw] {$\scriptscriptstyle{\substack{\{6\}}}$};

\draw (node cs:name=0) --node[above left] {$\scriptscriptstyle{\text{mod }2}$} (node cs:name =1.1);
\draw (node cs:name=0) -- node[above right] {$\scriptscriptstyle{\text{mod }2}$}(node cs:name =1.2);


\end{tikzpicture}\hspace{1cm}
\begin{tikzpicture}[scale=0.5]
\node (0) at (0,8) [ minimum size = 0.25cm, fill=gray!10, draw] {$\scriptscriptstyle{\substack{\{0,1,3,6\}}}$};
\node (1.1) at (-4,5) [ minimum size = 0.25cm, fill=gray!10, draw] {$\scriptscriptstyle{\substack{\{1,3\}}}$};
\node (1.2) at (4,5) [ minimum size = 0.25cm, fill=gray!10, draw] {$\scriptscriptstyle{\substack{\{0,6\}}}$};
\node (2.1) at (-6,2) [ minimum size = 0.25cm, fill=gray!10, draw] {$\scriptscriptstyle{\substack{\{3\}}}$};
\node (2.2) at (-2,2) [ minimum size = 0.25cm, fill=gray!10, draw] {$\scriptscriptstyle{\substack{\{1\}}}$};
\node (2.3) at (2,2) [ minimum size = 0.25cm, fill=gray!10, draw] {$\scriptscriptstyle{\substack{\{6\}}}$};
\node (2.4) at (6,2) [ minimum size = 0.25cm, fill=gray!10, draw] {$\scriptscriptstyle{\substack{\{0\}}}$};

\draw (node cs:name=0) --node[above left] {$\scriptscriptstyle{\text{mod }2}$} (node cs:name =1.1);
\draw (node cs:name=0) -- node[above right] {$\scriptscriptstyle{\text{mod }2}$}(node cs:name =1.2);

\draw (node cs:name=1.1) --node[above left] {$\scriptscriptstyle{\text{mod }4}$} (node cs:name =2.1);
\draw (node cs:name=1.1) --node[above right] {$\scriptscriptstyle{\text{mod }4}$} (node cs:name =2.2);
\draw (node cs:name=1.2) --node[above left] {$\scriptscriptstyle{\text{mod }4}$} (node cs:name =2.3);
\draw (node cs:name=1.2) --node[above right] {$\scriptscriptstyle{\text{mod }4}$} (node cs:name =2.4);


\end{tikzpicture}

\caption{Example trees of Elementary sets:}
\label{fig:tree_elementary}
\end{figure*}

We say $\JJ \subseteq \Z_N$, for $N=2^M$ is an elementary set if 
\begin{enumerate}
    \item $|\JJ| = 2^r$ for some $r$, and 
    \item The set $\JJ \mod 2^r$ = $\{0,1,2,\ldots,2^r-1 \}$. 
\end{enumerate}
In other words, if $\JJ$ is an elementary set, it contains exactly one element from each congruence class modulo $2^r$. Since the nodes in $\TT(\JJ)$ at level $r$ correspond to the congruence classes mod $2^r$, if $\JJ$ is elementary, all the nodes at level $r$ are singletons. See Fig \ref{fig:tree_elementary} for examples of elementary sets of sizes $2^1$ and $2^2$, respectively.

Since all the nodes at level $r$ are singletons, the DFT-induced weights at level $r$ are simply the DFT coefficients (i.e., there is no aliasing). So if $\JJ$ is an elementary set, the DFT of signals in $\BB^\JJ$ can be computed with $c_1 r2^r = c_1 |\JJ| \log |\JJ|=c_1 k \log k$ operations. We can construct a family of elementary sets $\famJ$ by collecting all elementary sets of a given size $k$. For this family, we need to compute the binary tree up to level $r$, which can be done in $O(kr) = O(k \log k)$ bit operations, and  then compute a $2^r-$ length DFT, which can be done in $c_1 k \log k$ operations. Thus such a family will be Fourier computable. In some sense, this family can be interpreted as being ideal w.r.t. Fourier computability. We discuss this and generalizations in Section \ref{sec:Hi-DFT-to-DFT}. 

But first, we note that elementary sets of size $2^r$ describe homogeneous trees with pivots $(0,1,2,\ldots, r-1)$. We can ask if the Fourier computability of elementary sets extends to arbitrary homogeneous trees (i.e., those with pivots not necessarily in the first few positions). This is the topic for the rest of this section.

Let us start with an example: consider a homogeneous congruence tree with $s<M/2$ pivots at the bottom-most levels $\{M-s, M-s+1, \ldots, M-1\}$. The corresponding set $\JJ \subseteq \Z_N$ must be of size $|\JJ| = 2^s$, so ideally, we desire an algorithm to compute the DFT of signals $f \in \BB^\JJ$ that uses $\text{(const)} s 2^s$ operations. Now suppose we downsample $f$ uniformly (to $2^s$ samples) and obtain the DFT coefficients. Since there is only one node at level $s$ in $\TT(\JJ)$ (the splits start at $M-s>s$), the DFT of the downsampled signal has, at most, one non-zero coefficient. In other words, the downsampled signal $f_{\downarrow 2^s}$ is a (scalar) multiple of $\vec{1}$. This is true of shifts of $f$ as well, and so $f$ must be periodic with period $2^s$. In particular, sampling $f$ uniformly requires us to take a lot of redundant samples (i.e., sampling the same entry in each period). A better approach would be to take the first $2^s$ samples in $f$, which would involve all the values in $f$ in a given period. This suggests that the sampling patterns for homogeneous trees, in general, need to be non-uniform. We investigate these sampling patterns next. We define the sampling pattern $\II_{\vec{r}}$ associated with a $\vec{r}-$part-homogeneous $\JJ$:

\usetikzlibrary{positioning,fit,calc} 
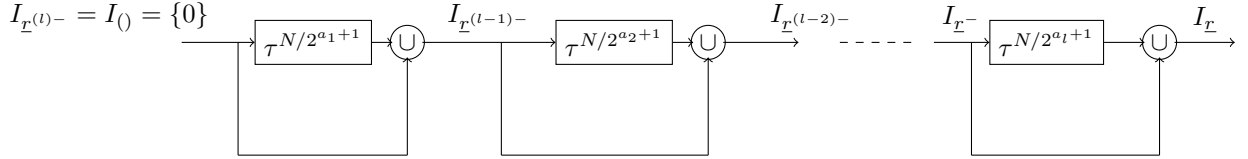
\begin{figure*}[htb]
\centering
\begin{tikzpicture}[scale=0.5]

\node (s0) at (3.5,20) [ minimum size = 0.25cm, draw] {$\tau^{N/2^{a_1+1}}$};
\draw[->] (0,20) --node[above left] {$I_{\vec{r}^{(l)-}} = I_{()} = \{0\}$} (node cs:name =s0); \draw (6,20) node[circle,minimum size=0.5, inner sep=1.5pt,fill=white,draw]{$\cup$} ;
\draw[->] (node cs:name =s0) -- (5.6,20);
\draw (1.5,20) --  (1.5, 17);
\draw (1.5,17) --  (6, 17);
\draw [->] (6,17) -- (6, 19.6);

\node (s1) at (11.5,20) [ minimum size = 0.25cm, draw] {$\tau^{N/2^{a_2+1}}$};
\draw[->] (6.4,20) --node[above] {$I_{\vec{r}^{(l-1)-}}$} (node cs:name =s1);
\draw (14,20) node[circle,minimum size=0.5, inner sep=1.5pt,fill=white,draw]{$\cup$} ;
\draw[->] (node cs:name =s1) -- (13.6,20);
\draw (8.5,20) --  (8.5, 17);
\draw (8.5,17) --  (14, 17);
\draw [->] (14,17) -- (14, 19.6);
\draw[->] (14.4,20) --node[above right] {$I_{\vec{r}^{(l-2)-}}$}  (16.4, 20);

 \draw[dashed] (17.5,20) --  (19.5, 20);

\node (sr) at (23,20) [ minimum size = 0.25cm, draw] {$\tau^{N/2^{a_l+1}}$};
 \draw[->] (20,20) --node[above] {$I_{\vec{r}^-}$} (node cs:name =sr); \draw (26,20) node[circle,minimum size=0.5, inner sep=1.5pt,fill=white,draw]{$\cup$} ;
 \draw[->] (node cs:name =sr) -- (25.6,20);
 \draw (21,20) --  (21, 17);
 \draw (21,17) --  (26, 17);
 \draw [->] (26,17) --  (26, 19.6);

 \draw[->] (26.4,20) --node[above] {$I_{\vec{r}}$}  (28, 20);

\end{tikzpicture}

\caption{General structure of the downsampling patterns. The values of $\vec{r} = \{a_1, a_2, \ldots, a_l\}$ are picked based on the pivot locations in the family $\mathpzc{J}$.}
\label{fig:indicators_non_uniform}
\end{figure*}

\usetikzlibrary{positioning,fit,calc} 
\begin{figure*}[htb]
\centering
\begin{tikzpicture}[scale=0.5]

\node (s0) at (3.5,20) [ minimum size = 0.25cm, draw] {$\tau^{N/2}$};
\draw[->] (0,20) --node[above left] {$I_{\vec{r}^{(l)-}} = I_{()} = \{0\}$} (node cs:name =s0); \draw (6,20) node[circle,minimum size=0.5, inner sep=1.5pt,fill=white,draw]{$\cup$} ;
\draw[->] (node cs:name =s0) -- (5.6,20);
\draw (1.5,20) --  (1.5, 17);
\draw (1.5,17) --  (6, 17);
\draw [->] (6,17) -- (6, 19.6);

\node (s1) at (11.5,20) [ minimum size = 0.25cm, draw] {$\tau^{N/4}$};
\draw[->] (6.4,20) --node[above] {$I_{\vec{r}^{(l-1)-}}$} (node cs:name =s1);
\draw (14,20) node[circle,minimum size=0.5, inner sep=1.5pt,fill=white,draw]{$\cup$} ;
\draw[->] (node cs:name =s1) -- (13.6,20);
\draw (8.5,20) --  (8.5, 17);
\draw (8.5,17) --  (14, 17);
\draw [->] (14,17) -- (14, 19.6);
\draw[->] (14.4,20) --node[above right] {$I_{\vec{r}^{(l-2)-}}$}  (16.4, 20);

 \draw[dashed] (17.5,20) --  (19.5, 20);

\node (sr) at (23,20) [ minimum size = 0.25cm, draw] {$\tau^{N/2^{l}}$};
 \draw[->] (20,20) --node[above] {$I_{\vec{r}^-}$} (node cs:name =sr); \draw (26,20) node[circle,minimum size=0.5, inner sep=1.5pt,fill=white,draw]{$\cup$} ;
 \draw[->] (node cs:name =sr) -- (25.6,20);
 \draw (21,20) --  (21, 17);
 \draw (21,17) --  (26, 17);
 \draw [->] (26,17) --  (26, 19.6);

 \draw[->] (26.4,20) --node[above] {$I_{\vec{r}}$}  (28, 20);

\end{tikzpicture}

\caption{The structure of the downsampling patterns when $a_1=0, a_2=1, \ldots$ in Fig \ref{fig:indicators_non_uniform}. In this case the downsampling patterns are uniform.}
\label{fig:indicators_uniform}
\end{figure*}
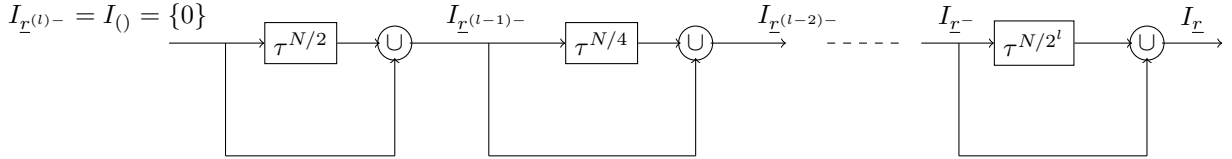

\begin{defn}
\label{def:r_sampling_pattern}
  (pivoted sampling patterns)  The $\vec{r}-$pivoted sampling pattern is defined recursively as
\(
\II_{\vec{r}} =  \II_{\vec{r}^-} \cup \tau^{2^{M-1-r_{\text{max}}}}\II_{\vec{r}^-},\) \(\II_{()}= \{0\}.
\)
\end{defn}

 See Fig \ref{fig:indicators_non_uniform}: we obtain $\II_{\vec{r}}$ as a union of $\II_{\vec{r}^-}$ and translate by a power of $2$. When this process is repeated on  $\II_{\vec{r}^-}$, the power of $2$ used is different. Thus the elements in $\II$ are sums of distinct powers of $2$, and hence distinct. As a consequence, the number of samples $|\II_{\vec{r}}|$ is $2^\Sizer$, and the number of samples in $\II_{\vec{r}^{n-}}$ is $2^{\Sizer-n}$.  

An alternate (non-recursive) way to define the $\vec{r}-$pivoted sampling pattern is
\[
\II_{\vec{r}} = \bigcup_{b_1,\ldots, b_s \in \{0,1\}} \left\{\sum_{k=1}^s b_k 2^{M-1-r_k} \right\}.
\]

\begin{figure*}[ht]  
\centering 

\begin{tikzpicture}[scale=1]
\tikzstyle{every node}=[font=\small]
\draw  (0,0) -- (7.5,0) node [below] {$n$};
\draw[->]  (0,0) node [below] {$0$} -- (0,1.5);
\draw[->]  (1,0) node [below] {$2$} -- (1,1.5);
\draw[->]  (4,0) node [below ] {$8$} -- (4,1.5);
\draw[->]  (5,0) node [below] {$10$} -- (5,1.5);
\end{tikzpicture} 
\begin{tikzpicture}[scale=1]
\tikzstyle{every node}=[font=\small]
\draw  (0,0) -- (7.5,0) node [below] {$n$};
\draw[->]  (0,0) node [below] {$0$} -- (0,1.5);
\draw[->]  (0.5,0) node [below] {$1$} -- (0.5,1.5);
\draw[->]  (2,0) node [below ] {$4$} -- (2,1.5);
\draw[->]  (2.5,0) node [below] {$5$} -- (2.5,1.5);
\draw[->]  (4,0) node [below ] {$8$} -- (4,1.5);
\draw[->]  (4.5,0) node [below] {$9$} -- (4.5,1.5);
\draw[->]  (6,0) node [below ] {$12$} -- (6,1.5);
\draw[->]  (6.5,0) node [below] {$13$} -- (6.5,1.5);
\end{tikzpicture} 
\caption{Example sampling patterns $1_{\II_{\vec{r}}}$ for $N =16$: on the left pivots $\vec{r} = \{1,3\}$, and on the right pivots $\vec{r} = \{0,2,3\}$.}
\label{fig:multicoset-sampling}  
\end{figure*}
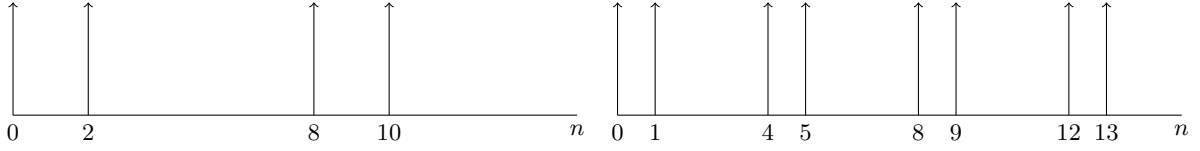

The Fig. \ref{fig:multicoset-sampling} shows the example sampling patterns $1_{\II_{\vec{r}}}$ for $\vec{r} = \{1,3\}$ (on the left)
and $\vec{r} = \{0,2,3\}$ on the right (here, $N = 16$). We can write $1_{\II_{\vec{r}}}$ using the convolution of indicators w.r.t each pivot. Let us consider the example on the left: $1_{\II_{\{1,3\}}} = 1_{\II_{\{1\}}} * 1_{\II_{\{3\}}}$. In general, we can write

\[1_{\II_{\vec{r}}} = 1_{\II_{\vec{r}^{-}}} * 1_{\II_{\left(r_{\text{max}}\right)}}.\]

Such sampling patterns are known as multi-coset sampling patterns, and have been investigated in the signal processing literature in the context of sub-Nyquist sampling \cite{Venk-Bresler}, with applications, including power-spectrum-blind sensing \cite{Aria-multi-coset}, array processing and DoA estimation \cite{kochman2011finite}, for e.g.


Note that $\II_{\vec{r}}$ is itself homogeneous: this follows by applying Proposition \ref{prop:hom-tree-prop}, the largest powers of $2$ that divide differences $m-n$ for $m\neq n \in \II$ are $2^{M-1-r}$ for some $r \in \vec{r}$. Thus $\II_{\vec{r}}$ is $(M-1-\vec{r})-$homogeneous.
 

Now consider an $f \in \BB^\JJ$, which we sample at locations $\II_{\vec{r}}$ in the time domain. The original (unsampled) signal $f$ has nonzero entries only at locations $\JJ$ in the frequency domain; we would like to investigate what happens to these coefficients after sampling at $\II_{\vec{r}}$ in the time domain. The following definition helps us formalize the answer to this question.

\begin{defn} \label{def:Hi-DFT}(Homogeneity-induced DFT) For any $\vec{r}-$part-homogeneous $\JJ$, we define the Homogeneity-induced DFT (Hi-DFT) of any $f\in \BB^\JJ$ at height $n$, denoted $\F_\JJ^n f$; as
\begin{equation}
\label{eq:r-down-sampling}    
  \F_\JJ^n f \coloneqq \left(\F\left(f1_{\II_{\vec{r}^{n-}}}\right)\right)_{\JJ} = \F(\JJ, \II_{\vec{r}^{n-}}) f_{\II_{\vec{r}^{n-}}}, 
\end{equation}
for $0\leq n \leq \Sizer$. Part-homogeneous sets and pivoted sampling patterns are defined in Def \ref{def:part-H-tree} and Def \ref{def:r_sampling_pattern}.

\end{defn}
Note that $\F_\JJ^n f$ has $|\JJ|$ entries, one corresponding to each element of $\JJ$. As such, it is convenient to index the array $\F_\JJ^n f$ by the elements of $\JJ$ (as opposed to linear indexing from $0$ to $|\JJ|-1$). Note that to compute the Hi-DFT, we need the samples of $f$ on $\II_{\vec{r}}^{n-}$. Computing the Hi-DFT at height $n$ (naively) using Definition \ref{def:Hi-DFT} requires multiplying the $|\JJ|\times 2^{\Sizer-n}$ matrix $\F(\JJ, \II_{\vec{r}^{n-}})$ with a vector, and hence costs $O(|\JJ|2^{\Sizer-n})$. Also note that at the extreme, when $n=\Sizer$, the downsampling pattern is $\II_{()}=\{0\}$, and so all the values in $\F_\JJ^n f$ are just $f(0)$.
\subsection{Some Examples and properties of the Hi-DFT}
The Hi-DFT is a transform corresponding to certain structured Fourier submatrices. In particular, the we note that the Hi-DFT is not the same as a smaller sized DFT matrix. We consider the following examples and observations. 
\begin{enumerate}
    \item When $\JJ$ is an elementary set of size $2^s$; this is homogeneous with pivots $\vec{s} =(0,1,2,\ldots,s-1)$ so that the sampling pattern above is \(\II_{\vec{s}} = \cup_{b_s \in \{0,1\}}\left\{\sum_{k=1}^s b_k 2^{M-k} \right\} = \cup_{b \in [0:2^s-1]} \{b2^{M-s}\},\)
which is a uniform downsampling pattern. In this case, $\II=\II_{\vec{s}}$ is a subgroup of $\Z_N$. Note also that the submatrix $\F(\JJ,\II)$ involved in the definition of the Hi-DFT above is a smaller DFT matrix.
\item Following up on the above if $\JJ=\Z_N$ (an extreme example of an elementary set), then the corresponding sampling pattern $\II=\II_{(0,1,2,\ldots, M-1)} = \Z_N$; and the sub-pivoted sampling pattern $\II^{n-}$ is a subgroup (uniform-downsampling pattern) of $\Z_N$ with $N/2^n$ samples.  when $\JJ = \Z_N$, the Hi-DFT (at height $n=0$) reduces to the standard DFT.  For other heights $n$, we have
\(
\F_{\Z_N}^n f (m)= \F_{N/2^n} f_{\downarrow 2^n}(m), \text{ for any }m\in \Z_N;
\)
thus the Hi-DFT at height $n$ is the same as the $N/2^n$ point DFT of downsampled $f$. The value of $n$, in this case, has a downsampling-based interpretation: we will elaborate on this later in the section. 
\item If the set of pivots is at the bottom-most levels $\vec{s}' = (M-s, M-s+1, \ldots, M-1)$, then \[\II_{\vec{s}'} = \cup_{b_1,\ldots, b_s \in \{0,1\}} \left\{\sum_{k=1}^s b_k 2^{k-1} \right\} = \cup_{b \in [0 : 2^{s-1}]} \{b\},\]
thus the samples $\II = \II_{\vec{s}'}$ are the first $2^s$ locations, as expected. While $\II$ is not a subgroup, it can be verified that the submatrix $\F(\JJ,\II)$ is a smaller DFT matrix (in this case, the frequency support $\JJ$ is a (shifted) subgroup).
\item However, for arbitrary pivots, neither of these observations holds in general. For e.g. if $N=10$, $\JJ=\{84,305,725,992\}$ we have the pivots $(0,2)$. In this case $\II_{(0,2)} = \{0,128,512,640\}$; neither $\II$ or $\JJ$ are subgroups of $\Z_N$. Also the submatrix $\F^{-1}(\II, \JJ)$ is 

\[
2^{-8}\begin{pmatrix}
0.25 & 	0.25 &	0.25 &	0.25 \\
0.25 & 	-0.25 &	0.25 &	-0.25 \\
0.25 & 	0.1768 +  0.1768  & -0.25 &	-0.1768 - i 0.1768 \\
0.25 & 	- 0.1768 - i 0.1768 &	- 0.25 &	0.1768 + i 0.1768 \\
\end{pmatrix},
\]

which is not a smaller DFT matrix. The key observation we make later (Theorem \ref{thm:computation_Hi-DFT}, Corollary \ref{cor:block-structure}) is that submatrices with rows and columns given by $\JJ$ and $\II_{\vec{r}}$, though not DFT matrices themselves, have computational properties similar to DFT matrices.
\end{enumerate}
 
We also note the following related property for the Hi-DFT at height $n=0$, following the discussion in the Subsection \ref{sec:preliminary_observations_system}, \eqref{eq:submatrix-method}:
\begin{equation}
\label{eq:Hi-DFT-and-DFT}
\F_\JJ^0 f = \F(\JJ, \II_{\vec{r}}) f_{\II_{\vec{r}}} = \F(\JJ, \II_{\vec{r}}) \F^{-1}(\II_{\vec{r}}, \JJ) (\F f)_\JJ.
\end{equation}

Before we proceed, we address the following questions (which were already addressed in the context of DFT):

\begin{enumerate}
    \item How are the computed values of the Hi-DFT $\F_\JJ^n f$ (for arbitrary $n$) related to the DFT $\F f$ ?
    \item Is there a way to compute the Hi-DFT $\F_\JJ^n f$ efficiently ?
\end{enumerate}
The answers to these questions form the basis for the design of our algorithm. We answer both of these questions in turn. The rest of this section is devoted to answering 1) above, and the next section answers 2).
\subsection{The aliasing pattern}
We define
 \(   h_{\vec{r}^{n-}} = \F1_{\II_{\vec{r}^{n-}}} \)
as the aliasing pattern corresponding to $\II_{\vec{r}^{n-}}$. We have, by definition of $\II_{\vec{r}^{n-}}$,

\begin{equation}\label{eq:ind_dft}
\begin{split}
    h_{\vec{r}^{n-}}(m) 
    = \left(1+ e^{-2\pi i m/2^{r_{s-n}+1}} \right) \F1_{\II_{\vec{r}^{(n+1)-}}}(m).
\end{split}
\end{equation}
 From definition \ref{def:r_sampling_pattern}, we know that $\F 1_{\II_{()}} = 1$. By applying \eqref{eq:ind_dft} recursively we get that the aliasing pattern is given by
  \begin{equation}\label{eq:pivoted-aliasing-pattern}
 h_{\vec{r}^{n-}}(m) = \prod_{i = 1}^{s-n} \left(1+ e^{-2\pi i m/2^{r_{i}+1}} \right) 
 \end{equation}
Note that this can be seen as a consequence of the sampling pattern $1_{\II_{\vec{r}}}$ being a convolution of smaller sampling patterns, as observed previously.

The zeroes of the aliasing pattern $h_{\vec{r}^{n-}}$ have a consequence for the aliasing in the DFT of the sampled signal \eqref{eq:r-down-sampling}.
The following lemma states the relevant property.
\begin{lemma}
\label{lem:aliasing-pattern}
For an $\vec{r}-$part-homogeneous $\JJ$, the subpivoted sampling pattern $\II_{\vec{r}^{n-}}$ defined in \ref{def:r_sampling_pattern}; the aliasing pattern  $h_{\vec{r}^{n-}} = \F 1_{\II_{\vec{r}^{n-}}}$ satisfies, for any $j_1, j_2 \in \JJ$
\[
h_{\vec{r}^{n-}}(j_1-j_2)=\begin{cases} 0 \text{ if }j_1-j_2 = \text{(odd) }2^{r}\text{ for some }r \in \vec{r}^{n-},\\
2^{\Sizer-n}\text{ otherwise. }
\end{cases}
\]

\end{lemma}
\begin{IEEEproof}
From Remark \ref{rem:tree-2-adic}, we know that for any $j_1, j_2 \in \JJ$, the difference $j_1-j_2$ is an odd multiple of $2^r$ for some pivot $r$ of $\JJ$: this $r$ is the largest power of $2$ dividing $j_1-j_2$. Also, recall that this largest power $r$ is the level of the first (starting from leaves) common ancestor of $j_1$ and $j_2$ in the congruence tree $\TT(\JJ)$. We have the following two cases
\begin{enumerate}
    \item The first common ancestor of $j_1$ and $j_2$ is within the first $r_{\Sizer-n}$ levels (this ancestor has to be at one of the levels in $\vec{r}^{n-}$ due to part-homogeneity). Thus $j_1-j_2 = \text{(odd) }2^{r}$ for some $r \in \vec{r}^{n-}$. In this case, since
    \[
    1 + e^{2\pi i (j_1-j_2)/2^{r+1}} = 1+ e^{i\pi} = 0,
    \]
    we have $h_{\vec{r}^{n-}}(j_1-j_2) =0$ from \eqref{eq:pivoted-aliasing-pattern}.
    \item The first common ancestor of $j_1$ and $j_2$ is after level $r_{\Sizer-n}$,  i.e. $j_1-j_2 = \text{(odd) }2^{r}$ for some $r > r_{\Sizer-n}$. Here
    \(
    1+ e^{2\pi i (j_1-j_2)/2^{r_i+1}} = 2 \text{ for any }r_i \in \vec{r}^{n-},
    \)
    and so 
    \(
    h_{\vec{r}^{n-}}(j_1-j_2) = 2^{\Sizer - n}.
    \)
\end{enumerate}
\end{IEEEproof}

Now, for $f \in \BB^\JJ$, the Hi-DFT is given by
\begin{align*}
\F_\JJ^n f (m) &= (\F f *  h_{\vec{r}^{n-}}) (m)\\
& = \sum_{j \in \KK} \F f(j)  h_{\vec{r}^{n-}}(m-j), \text{ for }m \in \JJ;
\end{align*}
which is the convolution of the DFT $\F f$ with the aliasing pattern $h_{\vec{r}^{n-}}$. We now have from Lemma \ref{lem:aliasing-pattern},
\[
\F_\JJ^n f (m) = 2^{\Sizer}\sum_{ j\in  \JJ'(m)} \F f(j) /2^n,
\]

where $\JJ'(m) \subseteq \JJ$ is the set of all indices $j\in \JJ$ that satisfy $j = \text{(odd) }2^r + m$ for some $r \notin \vec{r}^{n-}$. As observed in the proof of Lemma \ref{lem:aliasing-pattern}, this is the same as all the indices $j \in \JJ$, whose first common ancestor with $m$ is after the top $\Sizer-n$ pivots in $\TT(\JJ)$: this is the set of all coefficients which contribute to the aliasing.

Thus we conclude
\begin{equation}
\F_\JJ^n f (m) = 2^{\Sizer}\mu_\JJ(v, \F f)/2^n, 
\label{eq:aliasing-homogeneous-tree}
\end{equation}
Where $v$ is the ancestor of $m$ at height $n$. This is similar to the observation for $\TT^N$ in \eqref{eq:aliasing-full-tree}, where the DFT of the uniformly downsampled coefficients were related to the DFT-induced weights in the tree $\TT^N$. The above equation \eqref{eq:aliasing-homogeneous-tree} generalizes this observation to part-homogeneous index sets as well: if $\JJ$ is part-homogeneous, then the Hi-DFT coefficients are the DFT-induced weights in the tree $\TT(\JJ)$. This answers the question how the computed values  $\F_\JJ^n f$ are related to the DFT $\F f$.

\begin{figure*}[ht]
\centering
\resizebox{\textwidth}{!}{%
\begin{tikzpicture}[scale=0.5]
\node (0) at (0,8) [ minimum size = 0.25cm, fill=gray!10, draw] {$\scriptscriptstyle{\substack{\{3,17, \\ 25,27, 35\}}}$};
\node (1) at (0,5) [minimum size = 0.25cm, fill=gray!10, draw] {$\scriptscriptstyle{\substack{\{3,17, \\ 25,27, 35\}}}$};
\node (2.1) at (-4,2) [ minimum size = 0.5cm, fill=gray!10,
draw] {$\scriptscriptstyle{\{17,25\}}$};
\node (2.2) at (4,2) [minimum size = 0.5cm, fill=gray!10, draw]
{$\scriptscriptstyle{\{3,27, 35\}}$};
\node (3.1) at (4,-1) [minimum size = 0.5cm, fill=gray!10, draw]
{$\scriptscriptstyle\{3,27, 35\}$};
\node (3.2) at (-4,-1) [minimum size = 0.5cm, fill=gray!10, draw]
{$\scriptscriptstyle\{17,25\}$};
\node (4.2) at (6,-4) [minimum size = 0.5cm, fill=gray!10, draw]
{$\scriptscriptstyle\{3, 35\}$};
\node (4.3) at (2,-4) [minimum size = 0.5cm, fill=gray!10, draw]
{$\scriptscriptstyle\{27\}$};\
\node (4.0) at (-6,-4) [minimum size = 0.5cm, fill=gray!10, draw]
{$\scriptscriptstyle\{17\}$};
\node (4.1) at (-2,-4) [minimum size = 0.5cm, fill=gray!10, draw]
{$\scriptscriptstyle\{25\}$};
\node (5.2) at (8,-7) [minimum size = 0.5cm, fill=gray!10, draw]
{$\scriptscriptstyle\{35\}$};
\node (5.4) at (4,-7) [minimum size = 0.5cm, fill=gray!10, draw]
{$\scriptscriptstyle\{3\}$};
\node (5.3) at (2,-7) [minimum size = 0.5cm, fill=gray!10, draw]
{$\scriptscriptstyle\{27\}$};\
\node (5.0) at (-6,-7) [minimum size = 0.5cm, fill=gray!10, draw]
{$\scriptscriptstyle\{17\}$};
\node (5.1) at (-2,-7) [minimum size = 0.5cm, fill=gray!10, draw]
{$\scriptscriptstyle\{25\}$};
\draw (0) -- node[left] {$\scriptscriptstyle{\text{mod }2}$}(1);
\draw (node cs:name=1) --node[above left] {$\scriptscriptstyle{\text{mod }4}$} (node cs:name =2.1);
\draw (node cs:name=1) -- node[above right] {$\scriptscriptstyle{\text{mod }4}$}(node cs:name =2.2);
\draw (node cs:name=2.1) -- node[above left] {$\scriptscriptstyle{\text{mod }8}$}(node cs:name =3.2);
\draw (node cs:name=2.2) -- node[above left] {$\scriptscriptstyle{\text{mod }8}$}(node cs:name =3.1);
\draw (node cs:name=3.2) -- node[above left] {$\scriptscriptstyle{\text{mod }16}$}(node cs:name =4.0);
\draw (node cs:name=3.2) -- node[above right] {$\scriptscriptstyle{\text{mod }16}$}(node cs:name =4.1);
\draw (node cs:name=3.1) -- node[above right] {$\scriptscriptstyle{\text{mod }16}$}(node cs:name =4.2);
\draw (node cs:name=3.1) -- node[above left] {$\scriptscriptstyle{\text{mod }16}$}(node cs:name =4.3);
\draw (node cs:name=4.0) -- node[above left] {$\scriptscriptstyle{\text{mod }32}$}(node cs:name =5.0);
\draw (node cs:name=4.1) -- node[above right] {$\scriptscriptstyle{\text{mod }32}$}(node cs:name =5.1);
\draw (node cs:name=4.2) -- node[above right] {$\scriptscriptstyle{\text{mod }32}$}(node cs:name =5.2);
\draw (node cs:name=4.3) -- node[above left] {$\scriptscriptstyle{\text{mod }32}$}(node cs:name =5.3);
\draw (node cs:name=4.2) -- node[above left] {$\scriptscriptstyle{\text{mod }32}$}(node cs:name =5.4);
\end{tikzpicture}
\begin{tikzpicture}[scale=0.5]
\node[scale=0.25] (0) at (0,8) [ minimum size = 0.25cm, fill=gray!10, draw] {};
\node[above right,font=\tiny] at (0,8) {};

\node[scale = 0.25] (1) at (0,5) [minimum size = 0.25cm, fill=gray!10, draw] {};
\node[above right,font=\tiny] at (0,5) {$\F_\JJ^2 f[] = \sum_{i \in \JJ}\F f[i]$};

\node[scale = 0.25] (2.1) at (-4,2) [ minimum size = 0.5cm, fill=gray!10,draw] {};
\node[above left,font=\tiny] at (-4,2) {};

\node[scale = 0.25] (2.2) at (4,2) [minimum size = 0.5cm, fill=gray!10, draw] {};
\node[above right,font=\tiny] at (4,2) {};

\node[scale = 0.25] (3.1) at (4,-1) [minimum size = 0.5cm, fill=gray!10, draw]{};
\node[above right,font=\tiny] at (4,-1) {$\substack{\F_\JJ^1 f[3] =\F_\JJ^1 f[27] =\F_\JJ^1 f[35] \\= 2\F f[3] + 2\F f[27] + 2\F f[35]}$};

\node[scale = 0.25] (3.2) at (-4,-1) [minimum size = 0.5cm, fill=gray!10, draw]{};
\node[above left,font=\tiny] at (-4,-1) {$\substack{\F_\JJ^1 f[17] =\F_\JJ^1 f[25]  \\= 2\F f[17] + 2\F f[25]}$};

\node[scale = 0.25] (4.2) at (6,-4) [minimum size = 0.5cm, fill=gray!10, draw]{};
\node[right,font=\tiny] at (6,-4) {$\substack{\F_\JJ^0 f[3] =\F_\JJ^0 f[35] \\= \F f[3] + \F f[35]}$};

\node[scale = 0.25] (4.3) at (2,-4) [minimum size = 0.5cm, fill=gray!10, draw]
{};
\node[right,font=\tiny] at (2,-4) {$\F^0 f[27]$};

\node[scale = 0.25] (4.0) at (-6,-4) [minimum size = 0.5cm, fill=gray!10, draw]
{};
\node[left,font=\tiny] at (-6,-4) {$\F^0 f[17]$};

\node[scale = 0.25] (4.1) at (-2,-4) [minimum size = 0.5cm, fill=gray!10, draw]{};
\node[left,font=\tiny] at (-2,-4) {$\F^0 f[25]$};

\node[scale = 0.25] (5.4) at (7.5,-7) [minimum size = 0.5cm, fill=gray!10, draw]{};
\node[right,font=\tiny] at (7.5,-7) {$\F f[35]$};

\node[scale = 0.25] (5.2) at (5,-7) [minimum size = 0.5cm, fill=gray!10, draw]{};
\node[right,font=\tiny] at (5,-7) {$\F f[3]$};

\node[scale = 0.25] (5.3) at (2,-7) [minimum size = 0.5cm, fill=gray!10, draw]
{};
\node[right,font=\tiny] at (2,-7) {$\F f[27]$};

\node[scale = 0.25] (5.0) at (-6,-7) [minimum size = 0.5cm, fill=gray!10, draw]
{};
\node[left,font=\tiny] at (-6,-7) {$\F f[17]$};

\node[scale = 0.25] (5.1) at (-2,-7) [minimum size = 0.5cm, fill=gray!10, draw]{};
\node[left,font=\tiny] at (-2,-7) {$\F f[25]$};

\draw (0) -- node[left] {}(1);
\draw (node cs:name=1) --node[above left] {} (node cs:name =2.1);
\draw (node cs:name=1) -- node[above right] {}(node cs:name =2.2);
\draw (node cs:name=2.1) -- node[above left] {}(node cs:name =3.2);
\draw (node cs:name=2.2) -- node[above left] {}(node cs:name =3.1);
\draw (node cs:name=3.2) -- node[above left] {}(node cs:name =4.0);
\draw (node cs:name=3.2) -- node[above right] {}(node cs:name =4.1);
\draw (node cs:name=3.1) -- node[above right] {}(node cs:name =4.2);
\draw (node cs:name=3.1) -- node[above left] {}(node cs:name =4.3);

\draw (node cs:name=4.0) -- node[above left] {}(node cs:name =5.0);
\draw (node cs:name=4.1) -- node[above right] {}(node cs:name =5.1);
\draw (node cs:name=4.2) -- node[above right] {}(node cs:name =5.2);
\draw (node cs:name=4.2) -- node[above right] {}(node cs:name =5.4);
\draw (node cs:name=4.3) -- node[above left] {}(node cs:name =5.3);

\end{tikzpicture}
}
\caption{ An example $\{1,3\}-$part-homogeneous tree. Here $\JJ= \{3,17,25,27, 35\}$ (on the left) and $N=64$. The aliasing pattern on downsampling with $\II_{\vec{r}^{n-}}$ for $f\in\BB^\JJ$ is on the right; this pattern is similar to that for trees $\TT^N$. The downsampling patterns used are $\II_{\vec{r}} = \{0,2,8,10\}$, $\II_{\vec{r}^-} = \{0,8\}$ and $\II_{\vec{r}^{2-}} = \{0\}$.}
\label{fig:homologus_tree}
\end{figure*}
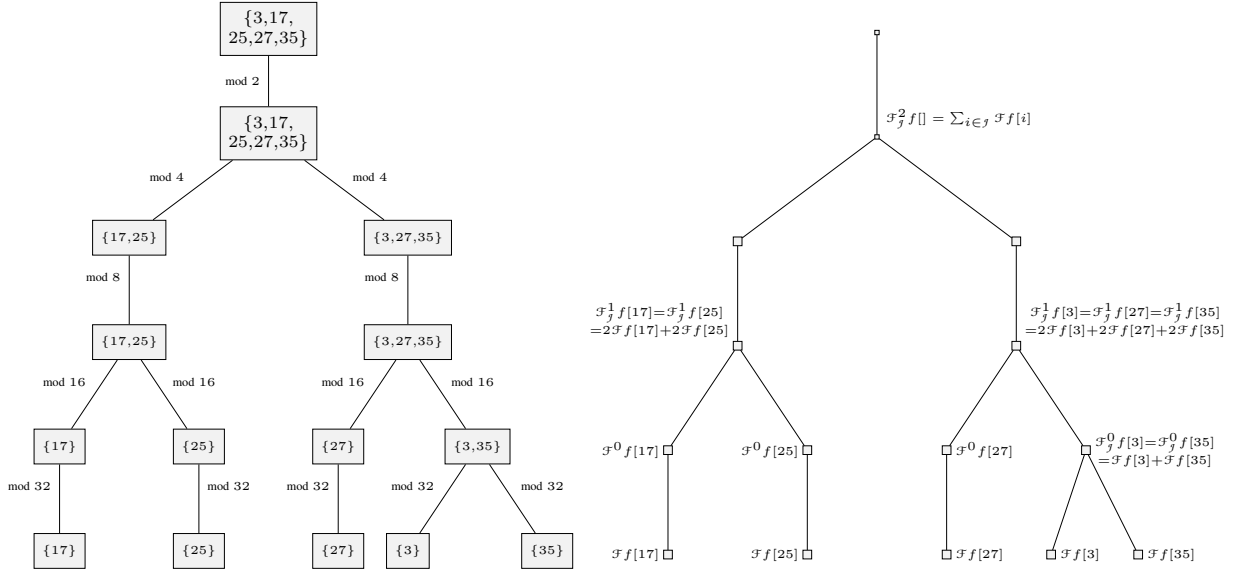

Take the following example: set $N=64$ and $\JJ = \{3,17,25,27, 35\}$. The corresponding congruence tree is shown in Fig \ref{fig:homologus_tree}, and clearly, the set $\JJ$ is $\{1,3\}-$part-homogeneous. We construct the sampling patterns $\II_{\vec{r}^{2-}} = \{0\}$, $\II_{\vec{r}^{-}} = \{0,8\}$ and $\II_{\vec{r}} = \{0,2,8,10\}$ as in Definition \ref{def:r_sampling_pattern}. With these sampling patterns, the corresponding Hi-DFT coefficients (denoted $\F_\JJ^2 f, \F_\JJ^1 f$ and $\F^0_\JJ f$ respectively, computed using \eqref{eq:aliasing-homogeneous-tree}) are given in Fig \ref{fig:homologus_tree}. As we can see, the downsampling pattern $\II_{\vec{r}^{n-}}$ can be interpreted as operating at height $n$ of the tree: the DFT coefficients on downsampling with such an $\II_{\vec{r}^{n-}}$ are the induced weights at height $n$.

We note the following consequences of the observation made from \eqref{eq:aliasing-homogeneous-tree}. First, start with an $\vec{r}-$ homogeneous $\JJ$, and apply \eqref{eq:aliasing-homogeneous-tree} to $n=0$. By homogenity, all nodes at height $0$ are singletons, so we get that $\F_\JJ^0 f$ (the Hi-DFT at height $0$)  is essentially the (scaled) DFT coefficients $\F f$ evaluated on $\JJ$:
\[
\F_\JJ^0 f = |\JJ| (\F f)_\JJ.
\]
This can be interpreted as saying that there is no aliasing on the coefficients $\JJ$: the structured downsampling pattern $\II_{\vec{r}}$ we picked ensures that aliasing, though present in coefficients outside $\JJ$, does not effect the coefficients within $\JJ$. This is also the key motivation behind earlier applications of multi-coset sampling \cite{Venk-Bresler,Aria-multi-coset,kochman2011finite}. 

Now, combining this observation with \eqref{eq:Hi-DFT-and-DFT}, we get that 
\begin{equation}
\label{eq:hom-implies-spectral}
     \F(\JJ, \II_{\vec{r}}) \F^{-1}(\II_{\vec{r}}, \JJ)  = |\JJ| \text{Id},
\end{equation}
and since the two matrices above are square and related by a conjugate transpose, we have that $ \F(\JJ, \II_{\vec{r}})$ is unitary; and so $\JJ$ is spectral. Thus homogeneity of $\JJ$ implies the spectrality of $\JJ$. The following proposition (already known in the literature) establishes the equivalence between homogeneity and spectrality.
\begin{proposition}\label{prop:homo-spectral}
An index set $\JJ\subseteq \Z_N $ is homogenous if and only if it is spectral.
\end{proposition}




Our observation in \eqref{eq:aliasing-homogeneous-tree} generalizes the known result above to arbitrary $n$; and provides a tree-based interpretation of the Hi-DFT coefficients.

As another observation from \eqref{eq:aliasing-homogeneous-tree}, note that $\F_\JJ^n f (m)$ depends only on the ancestor of $m$ at height $n$. Thus the vector $\F_\JJ^n f$ is more naturally indexed by nodes at height $n$ in $\TT(\JJ)$. For any node $v$ at height $n$, we unambiguously define $\F_\JJ^n f(v)\coloneqq  \F_\JJ^n f(m)$ for any descendant $m$ of $v$. Note that there are at most $2^{\Sizer-n}$ nodes at height $n$, the Hi-DFT $\F_\JJ^n f$ has at most $2^{\Sizer-n}$ values, indexed by the nodes at height $n$. We can make this precise: let $\JJ_1 \subseteq \JJ$ be obtained by taking the (labels of the) left branch of each split at height $1$ in $\TT(\JJ)$ (and discarding the right branch), and likewise, $\JJ_2  \subseteq \JJ$ be obtained by taking the right branch of each split at height $1$. The observation above can be written as \(\F_\JJ^1 f (\JJ_1) = \F_\JJ^1f(\JJ_2)\). Also, note that by construction, $\JJ_1$ and $\JJ_2$ are both $\vec{r}^--$part-homogeneous, and so by both $\F_\JJ^1 f(\JJ_1)$ and $\F_{\JJ_1}^0 f$ are both sampling $f$ in $\II_{\vec{r}^-}$ and evaluating the resulting DFT on $\JJ_1$. So by  Definition, $\F_\JJ^1 f (\JJ_1) = \F_{\JJ_1}^0 f$; and likewise $\F_\JJ^1f(\JJ_2)=\F_{\JJ_2}^0 f $. Thus we have the following for any $f \in \BB^\JJ$:
\begin{equation}
\label{eq:Hi-DFT-dynamic-programming}
\F_\JJ^1f(\JJ_1) =  \F_\JJ^1f(\JJ_2) = \F_{\JJ_1}^0 f = \F_{\JJ_2}^0 f. 
\end{equation}

We close this subsection with the generalization of \eqref{eq:aliasing-homogeneous-tree}.

\begin{lemma} \label{lem:shift-mod-Hi-DFT}(Shift-modulation-aliasing property for Hi-DFT)
For $\vec{r}-$part-homogeneous $\JJ$, for height $0\leq n \leq \Sizer$ in $\TT(\JJ)$ , and $a\in \Z_N$; consider the Hi-DFT of the shifted signal 
\[
\F_\JJ^{n,a}f = \F_\JJ^{n} (\tau^af)
\]
This is the same as the (modulated DFT)-induced weights in $\TT(\JJ)$: 
\begin{equation}
\F_\JJ^{n,a} f (m) = \frac{2^\Sizer}{2^n}\mu_\JJ(v, \underline{\omega}^{-a}\F f), 
\label{eq:aliasing-homogeneous-tree-shift}
\end{equation}
\end{lemma}
where $v$ is the ancestor of $m$ at height $n$. We know that the above property holds for $\TT^N$, the above lemma generalizes this to arbitrary part-homogeneous trees.
\begin{figure*}[ht]
\centering
\resizebox{1\textwidth}{!}{%
\begin{tikzpicture}[scale=0.5]
\node (0) at (0,8) [ minimum size = 0.25cm, fill=gray!10, draw] {$\scriptscriptstyle{\substack{\{3,17, \\ 25,27\}}}$};
\node (1) at (0,5) [minimum size = 0.25cm, fill=gray!10, draw] {$\scriptscriptstyle{\substack{\{3,17, \\ 25,27\}}}$};
\node (2.1) at (-4,2) [ minimum size = 0.5cm, fill=gray!10,
draw] {$\scriptscriptstyle{\{17,25\}}$};
\node (2.2) at (4,2) [minimum size = 0.5cm, fill=gray!10, draw]
{$\scriptscriptstyle{\{3,27\}}$};
\node (3.1) at (4,-1) [minimum size = 0.5cm, fill=gray!10, draw]
{$\scriptscriptstyle\{3,27\}$};
\node (3.2) at (-4,-1) [minimum size = 0.5cm, fill=gray!10, draw]
{$\scriptscriptstyle\{17,25\}$};
\node (4.2) at (6,-4) [minimum size = 0.5cm, fill=gray!10, draw]
{$\scriptscriptstyle\{3\}$};
\node (4.3) at (2,-4) [minimum size = 0.5cm, fill=gray!10, draw]
{$\scriptscriptstyle\{27\}$};\
\node (4.0) at (-6,-4) [minimum size = 0.5cm, fill=gray!10, draw]
{$\scriptscriptstyle\{17\}$};
\node (4.1) at (-2,-4) [minimum size = 0.5cm, fill=gray!10, draw]
{$\scriptscriptstyle\{25\}$};

\draw (0) -- node[left] {}(1);
\draw (node cs:name=1) --node[above left] {} (node cs:name =2.1);
\draw (node cs:name=1) -- node[above right] {}(node cs:name =2.2);
\draw (node cs:name=2.1) -- node[above left] {}(node cs:name =3.2);
\draw (node cs:name=2.2) -- node[above left] {}(node cs:name =3.1);
\draw (node cs:name=3.2) -- node[above left] {}(node cs:name =4.0);
\draw (node cs:name=3.2) -- node[above right] {}(node cs:name =4.1);
\draw (node cs:name=3.1) -- node[above right] {}(node cs:name =4.2);
\draw (node cs:name=3.1) -- node[above left] {}(node cs:name =4.3);

\end{tikzpicture}\hspace{1cm}
\begin{tikzpicture}[scale=0.5]
\node (0) at (0,8) [ minimum size = 0.25cm, fill=gray!10, draw] {$\scriptscriptstyle{\substack{\{3,17\}}}$};
\node (1) at (0,5) [minimum size = 0.25cm, fill=gray!10, draw] {$\scriptscriptstyle{\substack{\{3,17\}}}$};
\node (2.1) at (-2,2) [ minimum size = 0.5cm, fill=gray!10,
draw] {$\scriptscriptstyle{\{17\}}$};
\node (2.2) at (2,2) [minimum size = 0.5cm, fill=gray!10, draw]
{$\scriptscriptstyle{\{3\}}$};

\draw (0) -- node[left] {}(1);
\draw (node cs:name=1) --node[above left] {} (node cs:name =2.1);
\draw (node cs:name=1) -- node[above right] {}(node cs:name =2.2);
\end{tikzpicture}\hspace{1cm}
\begin{tikzpicture}[scale=0.5]
\node (0) at (0,8) [ minimum size = 0.25cm, fill=gray!10, draw] {$\scriptscriptstyle{\substack{\{25,27\}}}$};
\node (1) at (0,5) [minimum size = 0.25cm, fill=gray!10, draw] {$\scriptscriptstyle{\substack{\{25,27\}}}$};
\node (2.1) at (-2,2) [ minimum size = 0.5cm, fill=gray!10,
draw] {$\scriptscriptstyle{\{25\}}$};
\node (2.2) at (2,2) [minimum size = 0.5cm, fill=gray!10, draw]
{$\scriptscriptstyle{\{27\}}$};

\draw (0) -- node[left] {}(1);
\draw (node cs:name=1) --node[above left] {} (node cs:name =2.1);
\draw (node cs:name=1) -- node[above right] {}(node cs:name =2.2);
\end{tikzpicture}

}
\caption{An example $\vec{r}-$homogeneous tree (on the left) for $\JJ= \{3,17,25,27\}$, here the set of pivots $\Vec{r} = \{1,3\}$  and $N=32$. The $\vec{r}^{-}-$homologous trees with pivot $\Vec{r}^{-} = \{1\}$: $\JJ_1 = \{3,17\}$ and $\JJ_2 = \{25,27\}$ (on the centre and right respectively).}
\label{fig:homologus_tree_split}
\end{figure*}
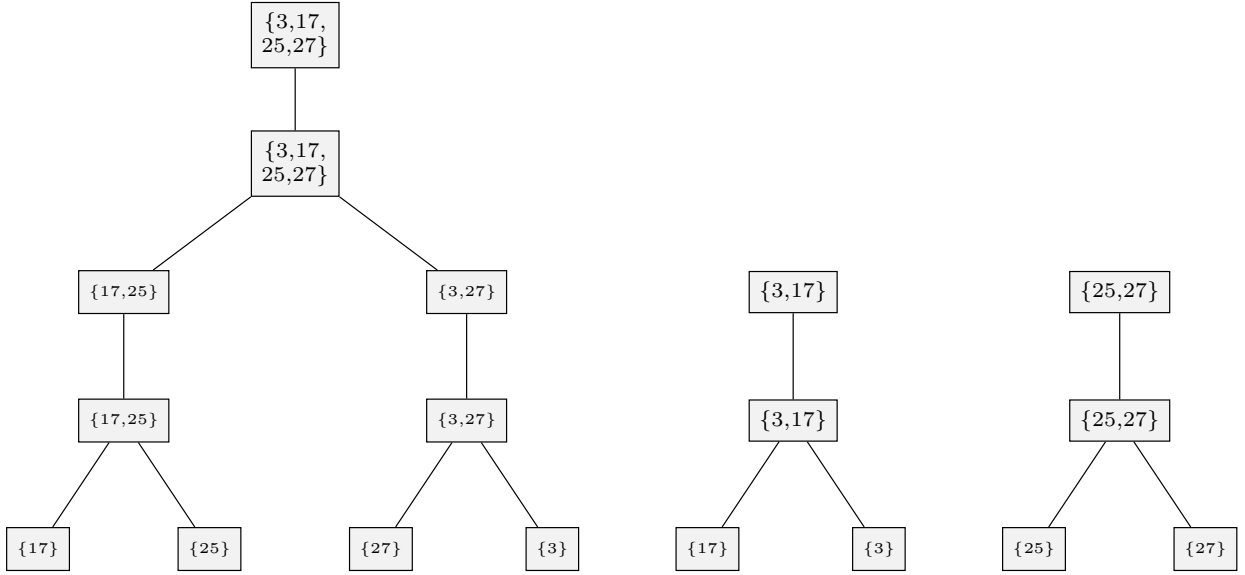
\section{Computing the Homogenity-induced DFT}
\label{Sec:compute-Hi-DFT}
 In this section, we show how to compute the Hi-DFT $\F_\JJ^n f$ efficiently. The main conclusion from this section is that the complexity of Hi-DFT scales only with the number of pivots $\Sizer$ and not the locations of these pivots. 

\begin{theorem}\label{thm:computation_Hi-DFT}
If $\JJ$ is $\vec{r}-$part-homogeneous, then for any $f \in \BB^\JJ$,  (with sample access to $f$) the DFT-induced weights $\mu_\JJ(v, \F f)$ for nodes $v$ at height $n$ (or equivalently, the Hi-DFT $\F_\JJ^n f$) can be computed with $1.5\ A\log A +A $ arithmetic operations, where $A=|\II_{\vec{r}^{n-}}| = 2^{\Sizer-n}$ is the number of time-domain samples.
\end{theorem}

The key idea in the proof is to use the recursive structure of the homogeneous tree to split the problem of computing the Hi-DFT induced weights into smaller problems in such a way that the results of the smaller problems can be combined. To make this precise, we state the following lemma.

\begin{lemma}
\label{lem:rec-hom-tree-split}
Consider an $\vec{r}-$part-homogeneous $\JJ$. Then, for any height $ 0\leq n<\Sizer$, and any node $j$ at height $n$,
\begin{gather*}
    \F_\JJ^{n}f (j) = 2\left(\F_\JJ^{n+1} f (j) + \omega_{r+1}^{j}\F_\JJ^{n+1, a_n}f (j)\right) \text{ if }j\text{ is a left descendant},\\ 
    \F_\JJ^{n}f (j) = 2\left(\F_\JJ^{n+1} f ( j) - \omega_{r+1}^{j}\F_\JJ^{n+1, a_n}f (j)\right) \text{ if }j \text{ is a right descendant;}
\end{gather*}
where $a_n= 2^{M-r_{\Sizer - n} - 1}$ and $r = r_{\Sizer-n}$ be the $n^{th}$ pivot from the end. Here we say that node $j$ is a left (right) descendant if $j$ is in the left (right) branch of the preceding split in $\TT(\JJ)$.

\end{lemma}

Now start with a $\vec{r}-$part-homogeneous $\JJ$, and consider the problem of computing the induced weights at height $n=0$ for $f$. Note that there are $2^\Sizer$ values to compute, one for each node at height $n=0$. We do this by computing the induced weights at height $n=1$ for $f$ and $\tau^{a_0}f$ i.e., $\F_\JJ^1 f$ and $\F_\JJ^{1,a_0} f$ (since there are $2^{\Sizer-1}$ nodes at height $1$; each of these two vectors has $2^{\Sizer-1}$ values). We combine the resulting computed values $\F_\JJ^1 f$ and $\F_\JJ^{1,a_0} f$ as per Lemma \ref{lem:rec-hom-tree-split} to obtain the induced weights at height $0$. Recall that computing $\F_\JJ^1 f$ uses samples of $f$ from $\II_{\vec{r}}^{-}$; and likewise computing $\F_\JJ^{1,a_0} f$ uses samples of $f$ from $\tau^{a_0}\II_{\vec{r}}^{-}$. Thus, $\F_\JJ^0 f$ is being computed with samples from $\II_{\vec{r}} = \II_{\vec{r}}^{-}\cup\tau^{a_0}\II_{\vec{r}}^{-}$; which is consistent with our expectation.

This can proceed recursively; to find the induced weights at height $n=1$, we use the induced weights at height $n=2$ for $f$ and $\tau^{a_1}f$, and so on. The algorithm terminates when we reach $n=\Sizer$: as observed previously, the weights $\F_\JJ^{\Sizer} (\tau^a f)$ at height $n=\Sizer$  reduces to the sample values of $f$. This algorithm is summarized in Alg \ref{alg:Hi-DFT}.

When $\JJ$ is homogeneous, the Hi-DFT $\F f_\JJ^0$ at height $n=0$ is just the DFT coefficients, and the number of time-domain samples is $|\II_{\vec{r}}|=|\JJ|$, so we have the following. 
\begin{corollary}
\label{cor:dft_homogenious}
If $\JJ$ is homogeneous, then given $\TT(\JJ)$ and sample access to $f \in \BB^\JJ$, the DFT coefficients $\F f$ can be computed with $O(|\JJ|\log|\JJ|)$ arithmetic operations.
\end{corollary}

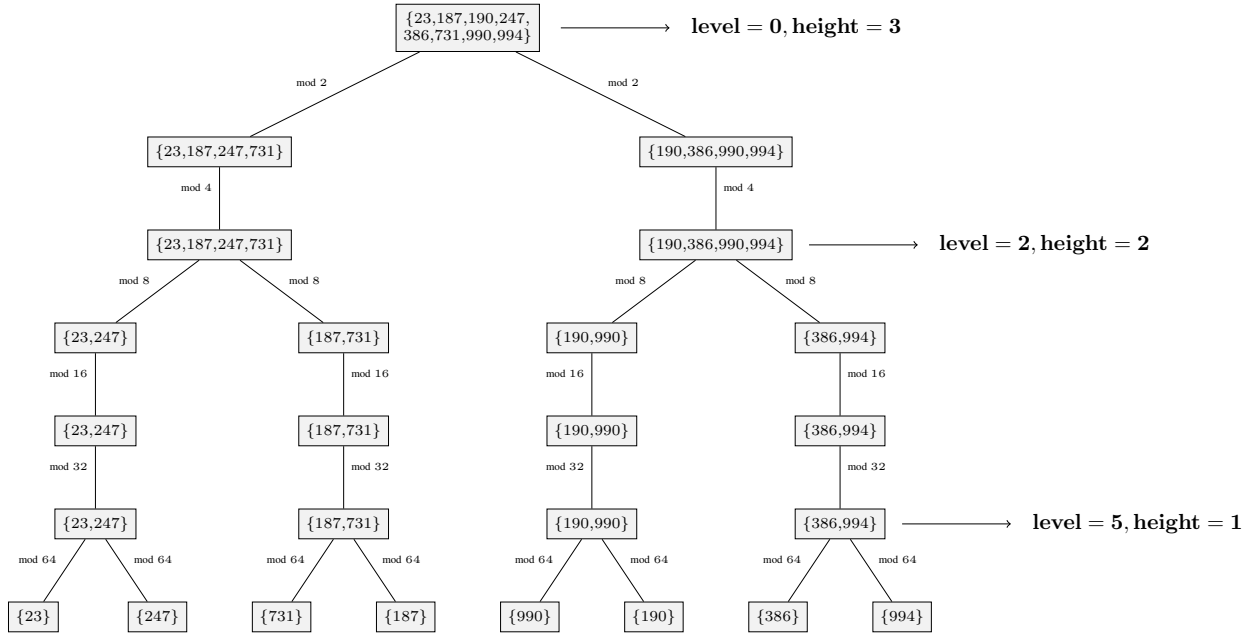
\begin{figure*}[ht]
\centering
\resizebox{1\textwidth}{!}{%
\begin{tikzpicture}[scale=0.5]
\node (0) at (0,12) [ minimum size = 0.25cm, fill=gray!10, draw] {$\scriptscriptstyle{\substack{\{23, 187, 190, 247,\\ 386, 731, 990, 994\}}}$};

\node (1.1) at (-8,8) [ minimum size = 0.25cm, fill=gray!10, draw] {$\scriptscriptstyle{\substack{\{23, 187, 247, 731\}}}$};
\node (1.2) at (8,8) [ minimum size = 0.25cm, fill=gray!10, draw] {$\scriptscriptstyle{\substack{\{190, 386, 990, 994\}}}$};

\draw (node cs:name=0) --node[above left] {$\scriptscriptstyle{\text{mod }2}$} (node cs:name =1.1);
\draw (node cs:name=0) -- node[above right] {$\scriptscriptstyle{\text{mod }2}$}(node cs:name =1.2);

\node (2.1) at (-8,5) [ minimum size = 0.25cm, fill=gray!10, draw] {$\scriptscriptstyle{\substack{\{23, 187, 247, 731\}}}$};
\node (2.2) at (8,5) [ minimum size = 0.25cm, fill=gray!10, draw] {$\scriptscriptstyle{\substack{\{190, 386, 990, 994\}}}$};

\draw (node cs:name=1.1) --node[above left] {$\scriptscriptstyle{\text{mod }4}$} (node cs:name =2.1);
\draw (node cs:name=1.2) -- node[above right] {$\scriptscriptstyle{\text{mod }4}$}(node cs:name =2.2);

\node (3.1) at (-12,2) [ minimum size = 0.25cm, fill=gray!10, draw] {$\scriptscriptstyle{\substack{\{23,247\}}}$};
\node (3.2) at (-4,2) [ minimum size = 0.25cm, fill=gray!10, draw] {$\scriptscriptstyle{\substack{\{187,731\}}}$};
\node (3.3) at (4,2) [ minimum size = 0.25cm, fill=gray!10, draw] {$\scriptscriptstyle{\substack{\{190,990\}}}$};
\node (3.4) at (12,2) [ minimum size = 0.25cm, fill=gray!10, draw] {$\scriptscriptstyle{\substack{\{386,994\}}}$};

\draw (node cs:name=2.1) --node[above left] {$\scriptscriptstyle{\text{mod }8}$} (node cs:name =3.1);
\draw (node cs:name=2.1) --node[above right] {$\scriptscriptstyle{\text{mod }8}$} (node cs:name =3.2);
\draw (node cs:name=2.2) --node[above left] {$\scriptscriptstyle{\text{mod }8}$} (node cs:name =3.3);
\draw (node cs:name=2.2) --node[above right] {$\scriptscriptstyle{\text{mod }8}$} (node cs:name =3.4);

\node (4.1) at (-12,-1) [ minimum size = 0.25cm, fill=gray!10, draw] {$\scriptscriptstyle{\substack{\{23,247\}}}$};
\node (4.2) at (-4,-1) [ minimum size = 0.25cm, fill=gray!10, draw] {$\scriptscriptstyle{\substack{\{187,731\}}}$};
\node (4.3) at (4,-1) [ minimum size = 0.25cm, fill=gray!10, draw] {$\scriptscriptstyle{\substack{\{190,990\}}}$};
\node (4.4) at (12,-1) [ minimum size = 0.25cm, fill=gray!10, draw] {$\scriptscriptstyle{\substack{\{386,994\}}}$};

\draw (node cs:name=3.1) --node[above left] {$\scriptscriptstyle{\text{mod }16}$} (node cs:name =4.1);
\draw (node cs:name=3.2) --node[above right] {$\scriptscriptstyle{\text{mod }16}$} (node cs:name =4.2);
\draw (node cs:name=3.3) --node[above left] {$\scriptscriptstyle{\text{mod }16}$} (node cs:name =4.3);
\draw (node cs:name=3.4) --node[above right] {$\scriptscriptstyle{\text{mod }16}$} (node cs:name =4.4);

\node (5.1) at (-12,-4) [ minimum size = 0.25cm, fill=gray!10, draw] {$\scriptscriptstyle{\substack{\{23,247\}}}$};
\node (5.2) at (-4,-4) [ minimum size = 0.25cm, fill=gray!10, draw] {$\scriptscriptstyle{\substack{\{187,731\}}}$};
\node (5.3) at (4,-4) [ minimum size = 0.25cm, fill=gray!10, draw] {$\scriptscriptstyle{\substack{\{190,990\}}}$};
\node (5.4) at (12,-4) [ minimum size = 0.25cm, fill=gray!10, draw] {$\scriptscriptstyle{\substack{\{386,994\}}}$};

\draw (node cs:name=4.1) --node[above left] {$\scriptscriptstyle{\text{mod }32}$} (node cs:name =5.1);
\draw (node cs:name=4.2) --node[above right] {$\scriptscriptstyle{\text{mod }32}$} (node cs:name =5.2);
\draw (node cs:name=4.3) --node[above left] {$\scriptscriptstyle{\text{mod }32}$} (node cs:name =5.3);
\draw (node cs:name=4.4) --node[above right] {$\scriptscriptstyle{\text{mod }32}$} (node cs:name =5.4);

\node (6.1) at (-14,-7) [ minimum size = 0.25cm, fill=gray!10, draw] {$\scriptscriptstyle{\substack{\{23\}}}$};
\node (6.2) at (-10,-7) [ minimum size = 0.25cm, fill=gray!10, draw] {$\scriptscriptstyle{\substack{\{247\}}}$};
\node (6.3) at (-6,-7) [ minimum size = 0.25cm, fill=gray!10, draw] {$\scriptscriptstyle{\substack{\{731\}}}$};
\node (6.4) at (-2,-7) [ minimum size = 0.25cm, fill=gray!10, draw] {$\scriptscriptstyle{\substack{\{187\}}}$};
\node (6.5) at (2,-7) [ minimum size = 0.25cm, fill=gray!10, draw] {$\scriptscriptstyle{\substack{\{990\}}}$};
\node (6.6) at (6,-7) [ minimum size = 0.25cm, fill=gray!10, draw] {$\scriptscriptstyle{\substack{\{190\}}}$};
\node (6.7) at (10,-7) [ minimum size = 0.25cm, fill=gray!10, draw] {$\scriptscriptstyle{\substack{\{386\}}}$};
\node (6.8) at (14,-7) [ minimum size = 0.25cm, fill=gray!10, draw] {$\scriptscriptstyle{\substack{\{994\}}}$};


\draw (node cs:name=5.1) --node[above left] {$\scriptscriptstyle{\text{mod }64}$} (node cs:name =6.1);
\draw (node cs:name=5.1) --node[above right] {$\scriptscriptstyle{\text{mod }64}$} (node cs:name =6.2);
\draw (node cs:name=5.2) --node[above left] {$\scriptscriptstyle{\text{mod }64}$} (node cs:name =6.3);
\draw (node cs:name=5.2) --node[above right] {$\scriptscriptstyle{\text{mod }64}$} (node cs:name =6.4);
\draw (node cs:name=5.3) --node[above left] {$\scriptscriptstyle{\text{mod }64}$} (node cs:name =6.5);
\draw (node cs:name=5.3) --node[above right] {$\scriptscriptstyle{\text{mod }64}$} (node cs:name =6.6);
\draw (node cs:name=5.4) --node[above left] {$\scriptscriptstyle{\text{mod }64}$} (node cs:name =6.7);
\draw (node cs:name=5.4) --node[above right] {$\scriptscriptstyle{\text{mod }64}$} (node cs:name =6.8);

\draw[->] (3,12) --node[right = 11mm] {$\mathbf{level = 0, height = 3}$} (6.5,12);

\draw[->] (11,5) --node[right = 11mm] {$\mathbf{level = 2, height = 2}$} (14.5,5);

\draw[->] (14,-4) --node[right = 11mm] {$\mathbf{level = 5, height = 1}$} (17.5,-4);

\end{tikzpicture}}
\caption{Homogeneous tree for $\JJ= \{23, 187, 190, 247, 386, 731, 990, 994\}$, here the set of pivots $\Vec{r} = \{0,2,5\}$  and $N=1024$.}
\label{fig:ex_hom_butterfly}
\end{figure*}
As a nontrivial example, consider the set $\JJ = \{23, 187, 190, 247, 386, 731, 990, 994\}$, which is $(0,2,5)-$ homogenous. The computational structure of Algorithm \ref{alg:Hi-DFT} (the butterfly diagram) is shown in Fig \ref{fig:ex_hom_butterfly}. This is very similar to radix-2, albeit with different connections and twiddle factors. This algorithm is what we refer to as `Modified radix-2' in Fig \ref{fig:framework_alg1}.

In fact, when $\JJ=\Z_N$ and $n=0$, Lemma \ref{lem:rec-hom-tree-split} reduces to 
\begin{gather*}
    \F_{N}f (j) = 2\left(\F_{N/2} f_{\downarrow 2} (j) + \omega_M^{j_1}\F_{N/2} (\tau f)_{\downarrow 2} (j)\right) \text{ if }j\in [0:N/2-1],\\ 
    \F_{N}f (j) = 2\left(\F_{N/2} f_{\downarrow 2} (j) - \omega_M^{j_1}\F_{N/2} (\tau f)_{\downarrow 2} (j)\right) \text{ if }j\in [N/2:N-1]. 
\end{gather*}

This is the same as the first step in the decimation in time radix-2 FFT. Lemma \ref{lem:rec-hom-tree-split} is a generalization of this well-known relationship. Thus, Algorithm \ref{alg:Hi-DFT} can be interpreted as a generalization of the decimation in time radix-2 FFT for the DFT to the Hi-DFT. 

Conceptually, Algorithm \ref{alg:Hi-DFT} operates by splitting the problem of finding the induced weights in $\TT(\JJ)$ into smaller subproblems. To make this precise, let $\JJ_1$ and $\JJ_2$ be as in \eqref{eq:Hi-DFT-dynamic-programming}; then Lemma \ref{lem:rec-hom-tree-split} for $n=0$, along with \eqref{eq:Hi-DFT-dynamic-programming}, gives us 
\begin{gather*}
\F_{\JJ}^0 f (j) = 2 \left(\F_{\JJ_1}^0f (j) + \omega_{r+1}^j\F_{\JJ_1}^0(\tau^af)  (j) \right)\text{ if }j\in \JJ_1 \\
\F_{\JJ}^0 f (j) = 2 \left(\F_{\JJ_1}^0f (j) - \omega_{r+1}^j\F_{\JJ_1}^0(\tau^af)  (j) \right)\text{ if }j\in \JJ_2.
\end{gather*}
Recall that the matrix corresponding to $\F_\JJ^0$ is $\F(\JJ, \II_{\vec{r}})$. We order the rows $\JJ$ by listing the elements of $\JJ_1$ first, followed by elements of $\JJ_2$; likewise, the columns $\II_{\vec{r}}$ by listing the elements of $\II_{\vec{r}^-}$ followed by $\tau^a \II_{\vec{r}^-}$. Then we see the following structure for the submatrix $\F(\JJ, \II_{\vec{r}})$.

\begin{corollary}
\label{cor:block-structure}
Suppose that $N$ is a power of $2$ and that $\JJ\subseteq \mathbb{Z}_N$ is  $\vec{r}-$part-homogenous. Then with a suitable ordering of the rows and columns, submatrix $\F(\JJ, \II_{\vec{r}})$ has the form 
\[
\F(\JJ, \II_{\vec{r}}) = \begin{pmatrix}
I & D \\ I & -D
\end{pmatrix}\begin{pmatrix}
\F(\JJ_1, \II_{\vec{r}^-}) & 0 \\ 0 & \F(\JJ_1, \II_{\vec{r}^-})
\end{pmatrix},
\]
where $D$ is a diagonal matrix; and $\JJ_1 \subseteq \JJ$ is  $\vec{r}^--$part-homogeneous.
\end{corollary}
This particular structure is also discussed in our earlier work \cite{9518104}, albeit with proof using digit tables. Theorem \ref{thm:computation_Hi-DFT} is a generalization of our result from \cite{9518104}. Note that a submatrix decomposition similar to the one in Corollary \ref{cor:block-structure} holds for the DFT matrix \cite{osgood2018lectures,cooley1965algorithm}. Thus the Hi-DFT, even though not the same as a DFT, has a recursive structure
similar to the DFT. The ordering of the rows $\JJ$ and columns $\II_{\vec{r}}$ required for the structure from Corollary \ref{cor:block-structure} is also similar to bit-reversal (see \cite{9518104} for details).

In summary, if there is available information that the pivots in $\TT(\JJ)$ are within $\vec{r}$, some spectral information (induced DFT weights) can be computed efficiently. The number of arithmetic operations required to obtain this spectral information depends on the number of pivots ($\Sizer$) but not on the locations $\vec{r}$. Homogenous sets are extreme examples when the number of pivots is the smallest possible leading to efficient DFT computation. Thus we note that homogenous sets are Fourier-computable.



\begin{corollary}\label{cor:spectral-Fourier-computable}
Let $\mathcal{S}_{\vec{r}}$ be the family of $\vec{r}-$homogeneous spectral sets. Then $\mathcal{S}_{\vec{r}}$ is Fourier computable.
\end{corollary}





\begin{algorithm}[ht]
\caption{Homogeneity induced DFT Algorithm (\texttt{Hi-DFT})}
\label{alg:Hi-DFT}
  \begin{algorithmic}
  \STATE \textbf{function} $(\F f)_\JJ$ = Hi-DFT($f_\II,\vec{r}, \JJ, N$)
  \STATE \hspace{0.25cm} $n$ = length($f_\II$);
  \STATE \hspace{0.25cm} \textbf{If} $n==1$
  \STATE \hspace{0.5cm} $(\F f)_\JJ = f_\II$;
  \STATE \hspace{0.25cm} \textbf{else}
  \STATE \hspace{0.5cm} $f1 = f_\II (0:n/2-1)$, and $f2 = f_\II (n/2-1:n-1)$;
  \STATE \hspace{0.5cm} $f1 = $ Hi-DFT($f1,\vec{r}^-, \JJ(0:n/2-1), N$);
  \STATE \hspace{0.5cm} $f2 = $Hi-DFT($f2,\vec{r}^-, \JJ(0:n/2-1), N$);
  \STATE \hspace{0.5cm} $a =N/2^{r_{max}+1}$, where $r_{max} = max(\vec{r})$;
  \STATE \hspace{0.5cm} $\omega = \exp(2*pi*1i*a/N)$;
  \STATE \hspace{0.5cm} $\omega = \omega^{\JJ(0:n/2-1)}$;
  \STATE \hspace{0.5cm} $(\F f)_\JJ = [f1+\omega .* f2 ; f1-\omega .* f2]$
  \STATE \hspace{0.5cm} $\vec{r} = \vec{r}^-$
  \STATE \hspace{0.25cm} \textbf{end}
  \STATE \textbf{end}
    \end{algorithmic}
\end{algorithm}









\section{From Hi-DFT to DFT}
\label{sec:Hi-DFT-to-DFT}
\begin{figure}[ht]
\centering
\begin{tikzpicture}[scale=0.5]
\node (0) at (0,8) [ minimum size = 0.25cm, fill=gray!10, draw] {$\scriptscriptstyle{\substack{\{0,1,6,7,512\}}}$};
\node (1.1) at (-4,5) [ minimum size = 0.25cm, fill=gray!10, draw] {$\scriptscriptstyle{\substack{\{1,7\}}}$};
\node (1.2) at (4,5) [ minimum size = 0.25cm, fill=gray!10, draw] {$\scriptscriptstyle{\substack{\{0,6,512\}}}$};
\node (2.1) at (-6,2) [ minimum size = 0.25cm, fill=gray!10, draw] {$\scriptscriptstyle{\substack{\{7\}}}$};
\node (2.2) at (-2,2) [ minimum size = 0.25cm, fill=gray!10, draw] {$\scriptscriptstyle{\substack{\{1\}}}$};
\node (2.3) at (2,2) [ minimum size = 0.25cm, fill=gray!10, draw] {$\scriptscriptstyle{\substack{\{6\}}}$};
\node (2.4) at (6,2) [ minimum size = 0.25cm, fill=gray!10, draw] {$\scriptscriptstyle{\substack{\{0,512\}}}$};

\draw (node cs:name=0) --node[above left] {$\scriptscriptstyle{\text{mod }2}$} (node cs:name =1.1);
\draw (node cs:name=0) -- node[above right] {$\scriptscriptstyle{\text{mod }2}$}(node cs:name =1.2);

\draw (node cs:name=1.1) --node[above left] {$\scriptscriptstyle{\text{mod }4}$} (node cs:name =2.1);
\draw (node cs:name=1.1) --node[above right] {$\scriptscriptstyle{\text{mod }4}$} (node cs:name =2.2);
\draw (node cs:name=1.2) --node[above left] {$\scriptscriptstyle{\text{mod }4}$} (node cs:name =2.3);
\draw (node cs:name=1.2) --node[above right] {$\scriptscriptstyle{\text{mod }4}$} (node cs:name =2.4);


\end{tikzpicture}
\caption{Example tree: $\TT_2(\{0,1,6,7,512\})$ (on the right), here $N=1024$.}
\label{fig:tree_uniform_sampling}
\end{figure}
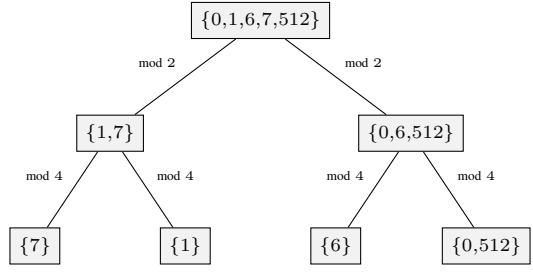

In the previous section, we saw that if the number of pivots is small, then the DFT-induced weights (or the Hi-DFT) can be computed efficiently. When $\JJ$ is spectral/homogeneous, the Hi-DFT coefficients are the same as the DFT coefficients (and hence can be computed fast). In this section, we attempt to generalize this to approximately homogeneous sets. The key idea is to contain $\JJ$ in a union of shifted homogeneous sets. We start with an example and discuss a simple algorithm, which we then analyze.
First, the key point for consideration is the choice of the pivots $\vec{r}$. Recall, for example, that any set is $\{0,1,2,\ldots,r\}-$part-homogeneous, for any $r$. The choice of $r$ is relevant to what the corresponding Hi-DFT values are. For example, take $N=1024$, and the set $\JJ=\{0,512,1,6, 7\}$ (see Fig \ref{fig:tree_uniform_sampling} for its congruence tree). We can, for instance, pick $r = 10=\log 1024$: the advantage of this choice of $r$ is that the DFT coefficients are completely isolated at level $r$, and so the DFT-induced weights at level $r$ will coincide with the DFT. However, to compute the weights, we have to use $2^{10}=1024$, i.e., all the samples in the time domain. On the other hand, suppose we take $r = 2$ and compute the DFT-induced weights at level $2$. In this case, the 3 frequency coefficients  at locations $\{1,6,7\}$ are isolated and the remaining two at $\{0,512\}$ are aliased. We can identify the coefficients $\F f(1), \F f (6), \F f(7)$, but we only know the sum $\F f(0) + \F f(512)$ of the remaining two DFT coefficients. To resolve the values of $\F f(0)$ and $\F f(512)$, we can shift the time domain signal and repeat the above process to get new induced weights. This gives us the value of $\F f(0) + \F f (512) e^{512i/N}$. Thus we have a $2 \times 2$ Vandermonde system of equations in $\F f(0)$ and $\F f(512)$, which can be solved easily by $5$ operations. 

The total number of computations, in this case, is $29$: the two (Hi-)DFTs (one for downsampled signal and one for the downsampled shifted signal) of size $4$ need $12$ computations each and the $2\times2$ system of equations can be solved with $5$ computations. Note that we need access to $8$ time domain samples. Now compare this to the full DFT, which takes $15\times1024$ computations, and needs access to 1024 time domain samples.

To generalize this, we note that for node $v$ at level $r$, the number of DFT coefficients aliased into node $v$ is $\mu_\JJ(v)$. Let $\mu^\star_r = \max_{\substack{v \text{ at level }r}}\mu_\JJ(v)$ be the largest number of aliased coefficients.
As discussed in the example above, we then repeat the downsampling process on $\tau f, \tau^2 f, \ldots, \tau^{\mu^\star_r-1}f$, to find the induced weights $\mu_\JJ(v, \F \tau^k f)$ for each node $v$ at level $r$.  From the basic properties of the DFT \eqref{eq:DFT_down_sampling}, this process results in a system of equations at each node in level $r$. If $v$ has label $\{\mathpzc{l}_1, \mathpzc{l}_2, \ldots \}$, the system of equations obtained at node $v$ is
\begin{equation}
\label{eq:alg1_system}
\underbrace{\mu_\JJ(v, \F \tau^k f)}_{\substack{\text{computed from the} \\ \text{shifted-downsampled signal}}} = \sum_m\F f(\mathpzc{l}_m)e^{2\pi i k \mathpzc{l}_m/N}, 
\end{equation}
for $k=0,1,2,\ldots \mu_\JJ(v)-1$, with unknowns  $\F f(\mathpzc{l}_1),\F f(\mathpzc{l}_2), \ldots $
Since this system is Vandermonde, the unknowns $\F f(\mathpzc{l}_m)$ can be recovered by solving this system.
We refer to the algorithm as SHIFT-AND-SAMPLE \texttt{(SAS)}, outlined below and summarized in Algorithm \ref{alg:shift-sample}.

This idea of repeating the sampling process with shifts has been used extensively in the sparse-FFT literature \cite{pawar-ramachandran,ghazi2013sample,nearlyoptimalsparse} (to name a few). The key difference is that algorithms for the sparse-FFT need to isolate the frequency coefficients (i.e., completely avoid or exploit aliasing to obtain singleton \emph{frequency bins}) whereas here, since we know the locations already, we can allow for aliasing and solve the resulting systems of equations.

\usetikzlibrary{positioning,fit,calc} 
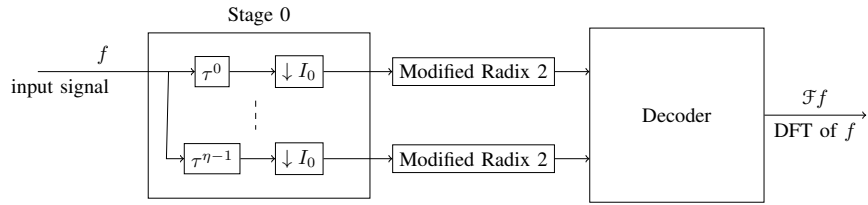
\begin{figure*}[ht]
\centering
\resizebox{0.7\textwidth}{!}{%
\begin{tikzpicture}[scale=0.5]

\node (s0_11) at (0,20) [ minimum size = 0.25cm, draw] {$\tau^0$};
\node (s0_12) at (3,20) [ minimum size = 0.25cm, draw] {$\downarrow I_0$};
\node (s0_13) at (9,20) [ minimum size = 0.25cm, draw] {Modified Radix 2};
\draw[->] (node cs:name=s0_11) --node[above left] {} (node cs:name =s0_12); 
\draw[->] (node cs:name=s0_12) --node[above left] {} (node cs:name =s0_13);

\node (s0_21) at (0,17) [ minimum size = 0.25cm, draw] {$\tau^{\eta-1}$};
\node (s0_22) at (3,17) [ minimum size = 0.25cm, draw] {$\downarrow I_0$};
\node (s0_23) at (9,17) [ minimum size = 0.25cm, draw] {Modified Radix 2};
\draw[->] (node cs:name=s0_21) --node[above left] {} (node cs:name =s0_22); 
\draw[->] (node cs:name=s0_22) --node[above left] {} (node cs:name =s0_23);

\draw [dashed](1.5,19) --node[above right] {} (1.5,18);
\node (s_0) [draw,inner xsep=8mm,inner ysep=4mm,fit=(s0_11) (s0_22),label={above:Stage 0}]{};

\draw[->] (-6,20) --node[above left] {$f$} node[below left] {input signal} (s0_11);
\draw[->] ($(s0_11)+(-1.5,0)$) -- ($(s0_21.west)+(-0.6,0)$) |- (s0_21.west);

\node (s_dec) at (16,18.5) [ minimum width=3cm,minimum height=3cm, draw] {Decoder};

\draw[->] (s0_13) -- ($(s_dec.west)+(0,1.5)$);
\draw[->] (s0_23) -- ($(s_dec.west)+(0,-1.5)$);

\draw[->] (s_dec.east) --node[above] {$\F f$} node[below] {DFT of $f$} ($(s_dec.east)+(3.5,0)$);
\end{tikzpicture}
}
\caption{Framework for our algorithm \ref{alg:shift-sample}. The signal is repeatedly shifted and downsampled (we shift $\mu_r^{\star}$ times, where $\mu_r^{\star}$ is maximum aliasing at level $r$). The decoder involves solving  Vandermonde system of equations involving Fourier submatrix.}
\label{fig:framework_alg1}
\end{figure*}

The choice of the level/resolution $r$ can be seen as striking a compromise between trimming of radix-2 computation graphs (Section \ref{sec:preliminary_observations_trim}) and the solving the system of equations (Section \ref{sec:preliminary_observations_system}): if $r=0$, we are devolving to the approach in Section \ref{sec:preliminary_observations_system}, whereas for $r$ such that there is complete isolation we are just trimming the computation graph, as in Section \ref{sec:preliminary_observations_trim}.  In general, we can also apply this to a general choice of $\vec{r}$ (in the preceding discussion, we set $\vec{r}$ to be consecutive) so long as $\JJ$ is $\vec{r}-$part-homogeneous: in this case, the induced weights will be found by applying Algorithm \ref{alg:Hi-DFT}.

\begin{algorithm}[ht]
\caption{Shift and Sample Algorithm (\texttt{SAS}) for balanced partly homogeneous families $(\famJ_N, k_N)$/ UoE}
\label{alg:shift-sample}
  \begin{algorithmic}
  \STATE \textbf{} Pick pivots $\vec{r}\coloneqq \vec{r}_N$ such that all $\JJ\in \famJ_N$ are $\vec{r}-$part-homogeneous. Set $r = r_{\text{max}}+1$.
  \STATE \textbf{Input: } Support $\JJ \in \famJ_N$, sample access to $f \in \BB^\JJ$.
  \STATE \textbf{Output: } $\F f(\JJ)$: The DFT coefficients of the signal $f$ on $\JJ$
  \STATE 0: Compute the truncated binary tree $\TT_{r}(\JJ)$
  \STATE 1: Set $\mu^{*}_r = \max_{\substack{i=0 \text{ to }2^{r}-1}}\mu_\JJ(v_i)$ at level $r$;
  \STATE 2: \textbf{for} $j = 0$ to $\mu_r^\star-1$
  \STATE 3: \hspace{0.25cm} Compute the Hi-DFT of the shifted signal: $\F_\JJ^{0,j}f$; using Algorithm \ref{alg:Hi-DFT}
  \STATE 4: \textbf{end for}
  \STATE 5: \textbf{for} $i = 0$ to $2^{r}-1$
  \STATE 6: \hspace{0.25cm} solve the $\mu(v_i)\times\mu(v_i)$ system \eqref{eq:alg1_system} at node $v_i$, level $r$;
  \STATE 7: \textbf{end for}
    \end{algorithmic}
\end{algorithm}


\begin{lemma}\label{lemma:computation_shift_sample}
If $\vec{r}$ is picked such that $\JJ$ is $\vec{r}-$part-homogeneous, Algorithm \ref{alg:shift-sample} outputs the DFT coefficients of any $f\in \BB^\JJ$; and 
steps 2-7 in Algorithm \ref{alg:shift-sample} (\texttt{SAS}) cost \begin{equation}\label{eq:alg1bnd}
\begin{split}
  c_1\Sizer2^\Sizer\mu_r^\star + c_2\sum_{\substack{v \text{ at level }r}} \mu_\JJ^2(v) \\ \leq 
   2^\Sizer\mu^\star_r \left(c_1\Sizer + c_2\mu^\star_r\right)\end{split}
\end{equation} computations (where $c_1 = 3/2$ and $c_2 = 6$). Here \(\mu^\star_r = \max_{\substack{v \text{ at level }r_{\text{max}}}}\mu_\JJ(v)\) is the largest number of aliased coefficients at level $r_{\text{max}}$. Also, note that the sample complexity is $\mu_r^*k$.
\end{lemma}



 The overall complexity above has two components: one involving the DFT computation, and one involving solving the system of equations. These depend on the value of $\mu_r^\star$ and $\Sizer$: lower values are desirable. For the value of $\mu_r^\star$ to be small, all the nodes at level $r$ must have roughly equal weight; in other words, the congruence tree must be balanced; and we need to pick a small number of pivots $\vec{r}$ such that the complexity above is the lowest while ensuring that every $\JJ \in \famJ_N$ is $\vec{r}-$part-homogeneous. As an example, for the complexity bound computed in Lemma \ref{lemma:computation_shift_sample} to be $O(k \log k)$ (here $k=|\JJ|$), we may pick the number of pivots $\Sizer \approx \log k$. We see an example of families for which this choice establishes Fourier computability. 
\subsection{Families with small number of pivots}
The following theorem formalizes the preceding discussion.
\begin{theorem}
\label{thm:part-hom-Fourier-computability}
Consider a (deterministic) family of sets $\{(\famJ_N, k_N) \}$ for which $\exists$  constants $\alpha_1, \alpha_2$ and a set of pivots $\vec{r}_N$ such that 
\begin{enumerate}
    \item $\Sizern=\alpha_1+ \log k_N$,
    \item for each $\JJ \in \famJ_N$, $\mu_r^\star (\JJ) \leq \alpha_2$ for $r=\text{max}(\vec{r}_N) +1 $ and
    \item each $\JJ \in \famJ_N$ is $\vec{r}_N-$part homogeneous.
\end{enumerate} Here \(\mu^\star_r = \max_{\substack{v \text{ at level }r}}\mu_\JJ(v)\) is the largest number of aliased coefficients at level $r$. Then $(\famJ_N, k_N)$ is Fourier-computable. We refer to such families as balanced partly-homogeneous families. In these families, there are at most $\log|\JJ| + \text{(const)}$ pivots needed to achieve reasonable isolation of the indices.
\end{theorem}
For example, any family with $\log k_N + \alpha_1$ pivots is clearly balanced partly homogenous (in this case, we have complete isolation with $\alpha_2=1$, and $\JJ$ is contained in a homogeneous set of size $k_N2^{\alpha_1}$). One such example is the family of \emph{universal} sets \cite{DS-1}, defined below.

\begin{defn}
\label{def:universal_sets}
\emph{Universal sets \cite{DS-1}} We say that $\JJ \subseteq \Z_N$ is universal if any square submatrix of $\F$ with columns indexed by $\JJ$ is invertible.
\end{defn}
 Universal sets $\JJ$ have the property \cite{DS-1} 
\[
|\mu_\JJ(v)-\mu_\JJ(w)| \leq 1 \text{ for any nodes }v,w\text{ at the same level.}
\]
For example, sets of consecutive integers and arithmetic progressions (with an odd common difference) are universal. Universal sets of size $k$, from the property above (combined with \eqref{eq:mu-recursion-level}), have pivots $\{0,1,2\ldots, \lceil \log k\rceil-1\}$, and hence Fourier computable.

Another example of such families involves arithmetic progressions of arbitrary common difference.
\begin{corollary}\label{cor:AP}
The family of arithmetic progressions is defined as
\begin{equation}
\JJ = \{a, a+s, a+2s, \ldots, a+(k_N-1)s \}\subseteq \Z_N, \famJ_N =\{\JJ\}.
\label{eq:arithmetic-prog-def}
\end{equation}
such that $|\JJ|=k_N$. This family is Fourier computable.
\end{corollary}


In this context, it is also appropriate to consider generalized arithmetic progressions, which have been studied extensively in the context of number theory \cite{tao2006additive}.

\begin{defn}\label{def:gap}
A generalized arithmetic progression (GAP) is a set of the form 
\begin{equation}
    \JJ = \{a+n_1s_1+n_2s_2+\ldots+n_ds_d: 0\leq n_j \leq N_j-1 \} \nonumber
\end{equation}
\end{defn}
We say that this is a generalized arithmetic progression of dimension $d$, with common differences $s_1, s_2, \ldots, s_d$. The generalized arithmetic progression is \emph{proper} if $|\JJ|=N_1 N_2\ldots N_{d}$ (i.e., there, all the sums above are distinct). Note that the sampling patterns $\II_{\vec{r}}$ defined in \ref{def:r_sampling_pattern} are generalized arithmetic progressions with $s_i$ as powers of $2$, and $N_i=2$ for all $i$. In general, however, there is no assumption on $s_i$ or $N_i$. 

Generalized arithmetic progressions are studied in the context of additive combinatorics. Of particular note are inverse sum set problems \cite{sanders2013structure}, which ask if $\JJ$ has a small sumset, i.e. 
\[
|\JJ+\JJ| = |\{j+j': j\in \JJ, j'\in \JJ \}|\leq C|\JJ|,
\]
for some constant $C$, then what can be said about the structure of $\JJ$ ? The constant $C$ is often refereed to as the doubling constant and sets $\JJ$ satisfying the above property are said to have additive structure. Indeed, if $\JJ$ is a GAP of the form in Defn \ref{def:gap}, then we see that
\[
\JJ+\JJ \subseteq \{a+n_1s_1+n_2s_2+\ldots+n_ds_d: 0\leq n_j \leq 2N_j-2 \},
\]
and so $|\JJ+\JJ| \leq 2^d |\JJ|$; and thus generalized arithmetic progressions (of small dimension) have additive structure.

The celebrated Freiman's theorem \cite{gordon1975ga, ruzsa1999analog, green2007freiman} states that any set (of integers) with additive structure is contained in a generalized arithmetic progression of size  $C'|\JJ|$ and dimension $C''$, where $C', C''$ depend only on the doubling constant $C$ of $\JJ$ (and in particular, they do not depend on the size of $\JJ$). In the setting of our problem, this motivates us to investigate Fourier computability for generalized arithmetic progressions; which can be connected to the Fourier computability of sets with additive structure.

We provide a preliminary result in this direction. 
\begin{lemma}
\label{lem:gap-complexity}
For the family of singleton generalized arithmetic progressions defined as
\[
\famJ_N=\{ \JJ\}, \JJ\subseteq \Z_N \text{ is a proper GAP of dimension }d;
\]
there exists an $O(k_N\log^2 k_N)$ structured-DFT algorithm.  
\end{lemma}

The proof bounds the number of pivots in a generalized arithmetic progression, followed by applying Algorithm \ref{alg:Hi-DFT}. Note that the constant in the above algorithm depends on $d$ (and scales as $2^{O(d \log d)}$: see the proof in Appendix \ref{proof:gap-complexity}). Lemma \ref{lem:gap-complexity} assumes that $d$ does not grow with $n$, though the proof goes through even when $k_N$ grows at least as fast as $d^d$. We would also like to note that for an arbitrary GAP, the pivots are not essentially consecutive, so the generalization of radix-2 (Algorithm \ref{alg:Hi-DFT}) is necessary for this result to hold. We can then conclude the following by an application of Freiman's theorem to Lemma \ref{lem:gap-complexity}.

\begin{theorem}
\label{thm:additive-structure-complexity}
Consider the singleton family of sets with doubling constant $C$
\[
\famJ_N=\{\JJ\}, \JJ\subseteq \Z_N \text{ satisfies }|\JJ+\JJ|\leq C|\JJ|.
\]
There exists a $O(k_N\log^2k_N)$ structured DFT algorithm for this family.
\end{theorem}

Lemma \ref{lem:gap-complexity} and Theorem \ref{thm:additive-structure-complexity} assert that the number of pivots is small enough for Algorithm \ref{alg:Hi-DFT} to work in $O(k\log^2k)$. However, note that we still need to address \emph{how} to find the pivots efficiently. This is the main reason for the families being defined as singleton families: this allows us to push the complexity of computing the pivots to pre-processing. Generalizing Lemma \ref{lem:gap-complexity} to families that contain all GAPs of a given dimension would involve accounting for the complexity of pivot computation as well. This is still a challenge, which we hope to explore later. Other possible directions include generalizations when the dimension $d$ (or the doubling constant $C$ in Theorem \ref{thm:additive-structure-complexity}) is allowed to grow (slowly) with $N$.



\subsection{Union of homogeneous sets}
We next discuss another family for which Algorithm \ref{alg:shift-sample} applies. Recall that elementary sets (discussed in Section \ref{sec:down-samp-cong-tree}) give the simplest examples of Fourier computable families. We can ask if a union of elementary sets has Fourier computability.

\begin{lemma} (Fourier computability of union of elementary sets) \label{lem:UoE1}
Consider a (deterministic) family $(\famJ_N,k_N)$ satisfying such that each $\JJ \in \famJ_N$ is a union of the form
\begin{equation}
\label{eq:UoE1}
\JJ = \bigcup_{i = 0}^{a_N}\bigcup_{j = 0}^{\eta_i-1}\JJ_{ij},
\end{equation}
with each $\JJ_{ij}$ is an elementary set of size $2^i$; and $a_N\leq \log k_N + \alpha$ for some constant $\alpha$. If for all $i$, we have $\eta_i \leq C$ for some constant $C$, then this family is Fourier computable. We refer to such families as UoE (union-of-elementary).
\end{lemma}

Thus each set $\JJ$ in a UoE family is a union of many elementary sets, $\eta_i$ of which have size $2^i$. What makes Lemma \ref{lem:UoE1} appealing is that 
\emph{every} index has a decomposition into elementary sets. If $\JJ=\{j_1, j_2,\ldots,j_k\}$ then (in the absence of any structural information on $\JJ$) we may write
\(
\JJ = \{j_1\} \cup \{j_2\} \cup \ldots \cup \{j_k\},
\)
which is a union of $k$ elementary sets of size $2^0$: here $\eta_0=k$ and $\eta_1=\eta_2=\ldots=0$. If some structural information on $\JJ$ is available that allows us to decompose $\JJ$ into a union of elementary sets as in \eqref{eq:UoE1} while keeping the values of $\eta_i$ small, then Lemma \ref{lem:UoE1} guarantees Fourier computability.

Note that each set $\JJ_{ij}$ in the union is elementary, i.e., it contains exactly one element from each congruence class modulo $2^{i}$ and has pivots $\{0,1,2,\ldots,i-1\}$. As an example, we have the following (here $\eta_i=1$ for $i=0,1,2,3$)
\begin{align*}
\JJ &= \{1,2,3,4,8,9,10,13,22,23,31 \} \\
&= \{0 \} \cup \{31,22\} \cup \{3,4,9,10\} \cup \{1,2,3,4,8,13,22,23\}.
\end{align*}
Note that there is no requirement for the union to be disjoint (as in the example above). We have
\[
k_N= |\JJ|\leq \sum_{i = 0}^{a_N}\sum_{j=0}^{\eta_i-1} {|\JJ_{ij}|} = \sum_{i=0}^{a_N} \eta_i 2^i \leq C \left(2^{a_N+1}-1\right).
\]
So that $\log(k_N)$ is $O(a_N)$. The condition $a_N \leq \log k_N + \alpha$ is to ensure that there is `not too much overlap' in the individual sets $\JJ_i$ (i.e., the inequality above is not satisfied too comfortably). If the sets $\JJ_i$ are disjoint, this condition is automatically satisfied. 

Note that any split in $\TT(\JJ_{ij})$ is retained in $\TT(\JJ)$, and therefore $\JJ$ has pivots at locations $\{0,1,2,\ldots,a_N-1\}$ (it could, of course, have additional pivots as well). 

Also, note that the family from Lemma \ref{lem:UoE1} is not necessarily balanced partly-homogeneous, so Theorem \ref{thm:part-hom-Fourier-computability} does not directly apply.
The following example is for a UoE family which is not balanced partly homogeneous. This family is constructed by including as many large powers of $2$ as possible in the index set.
\begin{remark}
\label{rem:UOE-not-balanced}
Consider, for example, the family $\famJ$ defined by
\[
\JJ = \bigcup_{i=0}^{a_N} \bigcup_{j=0}^{2^i}\{2^{a_N+i} + j\}, \quad \famJ_N = \{\JJ\}.
\]
 Then $\famJ$ is UoE but not balanced partly homogeneous.

\end{remark}



However, applying Algorithm \ref{alg:shift-sample} with a carefully chosen $\vec{r}$ establishes the Fourier computability. The key idea is that choice of pivots $\vec{r}$ with small size (i.e., smaller $\Sizer$) lead to the fast computation of the Hi-DFT but might increase the amount of aliasing (i.e., $\mu^\star)$, which increases the complexity of solving the system of equations. We need to pick the number of pivots $\Sizer$ that strikes the best possible compromise between these two competing requirements from \eqref{eq:alg1bnd}. We elaborate on this next.

First, for UoE families, we next bound the amount of aliasing. 

\begin{proposition}
\label{prop:UoE-mu-r-star-bound}
For an index set $\JJ$ in a UoE as in Lemma \ref{lem:UoE1}, consider a node $v$ at level $s\leq a_N$ in $\TT(\JJ)$. Then \(\mu_\JJ(v) \leq C (s + 2^{a_N-s+1}-1)\).
\end{proposition}

 At level $s=a_N$, Proposition \ref{prop:UoE-mu-r-star-bound} implies that there is at most $O(a_N) = O(\log k_N)$ aliasing. So if we pick $\vec{r}=\{0,1,2\ldots, a_N-1\}$ to Apply algorithm \ref{alg:shift-sample}, then the complexity bound from Lemma \ref{lemma:computation_shift_sample} becomes $O(k_N \log ^2 k_N)$ (The example family from Remark \ref{rem:UOE-not-balanced} achieves this complexity, so the bound is also tight). To reduce the complexity, we can attempt to  take even less samples, i.e. we pick $\vec{r} = \{0,1,2,\ldots,s-1 \}$ for some $s \leq a_N$. We effectively move up the tree to operate at level $s$ (instead of $a_N$). This may reduce the complexity of the DFT computation (as we evaluate the DFT of a smaller length signal now), but certainly increases the complexity of solving the system of equations, owing to an increase in the weights $\mu_\JJ$ as we move up the tree. We can see what happens at the extreme choices: for $s=1$, the complexity of DFT computation is constant, whereas the complexity of solving the system of equations dominates (and is in fact $O(k^2)$ as observed before), and at $s=a_N$ the DFT complexity dominates. We may ask if there is a possible compromise on the resolution $s$ that can enable Algorithm \ref{alg:shift-sample} to achieve an $O(k \log k)$ complexity. 
 
 Proposition \ref{prop:UoE-mu-r-star-bound} helps us quantify the aliasing as we move up the tree. In particular, if we move up the tree (starting at $a_N$) by $\log \log k_N$ levels, i.e., we pick $a_N-s\approx \log \log k_N$, then the amount of aliasing increases (additively) by at most $O(\log k_N)$ (when compared to $s=a_N$), while the complexity of computing the DFT ($s2^{s}$) decreases (multiplicatively) by a factor of $\log k_N$. We see that this is sufficient to establish Fourier computability. 

We close this section will a generalization of Lemma \ref{lem:UoE1} to union of homogeneous sets.
\begin{theorem}
\label{thm:UoE2}
(Fourier computability of union of homogeneous sets)
Given $\vec{l}_N-$homogenous $\KK_N$, consider a (deterministic) family $(\famJ_N,k_N)$ such that  \begin{enumerate}
    \item each $\JJ \in \famJ_N$ is a union of the form
\begin{equation}
\label{eq:UoE2}
\JJ = \bigcup_{i = 0}^{a_N}\bigcup_{j = 0}^{\eta_i-1}\JJ_{ij},
\end{equation}
where each $\JJ_{ij}$ is $\vec{l}^{(\Sizeln-i)-}-$homogeneous,
    \item $a_N\leq \log k_N + \alpha$ for some constant $\alpha$, 
    \item $\JJ \subseteq \KK_N$, and
    \item $\eta_i \leq C$ for some constant $C$,
\end{enumerate} then the family $(\famJ_N,k_N)$ is Fourier computable. We refer to such families as UoH (union-of-homogeneous).
\end{theorem}
When $\vec{l}_N = (0,1,2,\ldots,M-1)$ (so that $\KK_N = \Z_N$), Theorem \ref{thm:UoE2} reduces to Lemma \ref{lem:UoE1}. In this case, each of the $\J_{ij}$ has consecutive pivots and hence is elementary. The only condition in Theorem \ref{thm:UoE2} that does have a direct analogy to those in Lemma \ref{lem:UoE1} is 3); which requires that the constituent homogeneous sets $\JJ_{ij}$ be subsets of the base homogeneous set $\KK_N$. Recall that a union of sets retains the pivots of the constituent sets but could introduce additional pivots. Condition 3) is imposed to ensure that the number of pivots in the union does not drastically increase. The condition $\JJ \subseteq \KK_N$ ensures that the set $\JJ$ is $\vec{l}_N^{i-}-$part-homogeneous for any $i$. Note that this condition is trivially satisfied when $\KK_N=\Z_N$, which is the case for Lemma \ref{lem:UoE1}.

\subsection{Random subsets of homogeneous sets}
\label{sec:stoch-analysis}





Suppose we are given homogeneous sets $\KK_N \subseteq \Z_N$.In this section, we will analyze the performance of Algorithm \ref{alg:shift-sample} on the stochastic families $\KK_N^{\downarrow k}$. Recall that any index set in the family $\KK_N^{\downarrow k}$ is obtained by independently selecting elements from $\KK_N$ with probability $k/|\KK|$. Since $\Z_N$ itself is homogeneous, therefore the analysis in this section applies to $\Z_N^{\downarrow k}$ as well. Recall (from the discussion in Section \ref{sec:problem_setup}) that the sizes of sets in $\KK_N^{\downarrow k}$  are $\Theta(k)$, and we hope for Algorithm \ref{alg:shift-sample} to have a (worst case) complexity of $O(k \log k)$.




As in the case of Section \ref{sec:Hi-DFT-to-DFT}, the key question is the choice of pivots $\vec{r}$ to apply Algorithm \ref{alg:shift-sample}. Suppose $\KK_N$ has pivots $\vec{l}$; since any $\JJ \in \KK^{\downarrow k}_N$ is $\vec{l}^{n-}-$part-homogeneous for any $n$,  we can attempt to apply Algorithm \ref{alg:shift-sample} with $\vec{r}=\vec{l}^{(\Sizel-n)-}$ for some $n$; as in the discussion for Theorem \ref{thm:UoE2}. The complexity of finding the DFT induced weights (Hi-DFT) is $n2^{n}$ and the weights $\mu_\JJ$ will be random variables. The overall complexity also includes the random variable $\mu_{r}^\star$ (for $r=r_{\text{max}}+1 = l_n + 1$), which we try to bound next.

Consider the congruence trees $\TT(\KK_N)$ and $\TT(\JJ)$; for any node in $\TT(\JJ)$, we consider the $\vec{l}-$height (i.e. nodes at level $l_{\text{max}}+1$ are mapped to height $0$). From the assumed probability model, the leaves of $\TT(\KK_N)$ are present in $\TT(\JJ)$ with probability $p = \frac{k}{|\KK|}$; and we have $2^h$ of such leaves contributing to each node $i$ at height $h$.  Let $2^{n'}=|\KK|/2^n$; here $n'$ is the height of nodes at level $r_{\text{max}}+1$. For any node $v$ at height $n'$, the weight $\mu_\JJ(v) \sim \text{Binomial}(2^{n'},p)$ is thus a Binomial random variable with mean $ \lambda \coloneqq k/2^{n}$.

The probability that $\mu_\JJ(v)$ is large (or small) can be bounded by some well-known concentration inequalities \cite{harvey2022randomized,mukhopadhyay2020probability}. We see that $\mu_r^\star$ is $O(\log k)$. 



\begin{lemma}
\label{lem:bound_mu_r_star}
Given homogeneous set $\KK$ with pivots $\vec{l}$
, and assume $\JJ \in \KK_N^{\downarrow k}$. For $r = l_{n} +1$ as in the preceding discussion, let \(\mu^\star_r = \max_{\substack{v \text{ at level }r}}\mu_\JJ(v)\) be the largest number of aliased coefficients at level $r$ in $\TT(\JJ)$, as before. If $\lambda = k/2^n$ is $\Omega(1)$ and $O(\log k)$, then  
\begin{gather}
    \Prob(\mu_r^\star \geq c \log k) \rightarrow 0, \label{eq:mu-r-star-ub0}
\end{gather}
 where $c$ is a constant.
 
\end{lemma}

 By using Le Cam's theorem for Poisson approximation \cite{le1960approximation} \cite[Section 2.3]{ross2007second}, we may be able to give tighter bounds on $\mu_r^\star$ for some ranges of $r$: for e.g., 
 the maxima of $x$ Poisson random variables is known to oscillate between two consecutive integers close to $\log x/\log \log x $ \cite{anderson1970extreme, kimber1983note}. We do not need the oscillation property for the present discussion, so we only work with the (simpler) tail bounds. 
 
The upper bound on $\mu_r^\star$ from \eqref{eq:mu-r-star-ub0} is sufficient to establish the following:
\begin{theorem}\label{thm:hom_rand_subset}
Given homogeneous sets $\KK_N \subseteq \Z_N$, the (stochastic) family $(\KK_N^{\downarrow k_N}, k_N)$ is Fourier computable. 
\end{theorem}
Note that the average number of samples required by Algorithm \ref{alg:shift-sample} for these families is $E(\mu_r^*)k_N = O(k_N)$. 




\section{Converse results}
In this section, we attempt to understand the fundamental limitations of structured-FFT algorithms. We start with the following questions
\begin{enumerate}
    \item We motivated the target $O(k\log k)$ complexity by comparison with the well-known $O(N \log N)$ complexity for the DFT. Is it possible to establish that a structured DFT algorithm has to have a complexity of $\Omega(k \log k)$ ?
    \item Can we give an example of families which are provably not Fourier computable?
\end{enumerate}
While we are unable to conclusively answer these questions in full, we present some partial results along these lines. To the best of our knowledge, there are not many results in the literature that give lower bounds on computational complexity: for example, it is not even known if $O(N \log N)$ complexity of radix-2 is optimal for computing the (full) DFT. There are some lower bounds on computational complexity for sparse-FFT (see \cite[Corollary~5.2]{nearlyoptimalsparse} for e.g.), which applies a reduction to compute the DFT via a sparse-FFT algorithm; and thereby obtains a lowers bound on the sparse-FFT algorithm by assuming that $O(N\log N)$ is the optimal complexity for DFT. Note that Algorithm \ref{alg:Hi-DFT} gives a $O(k \log k)$ complexity for computing the DFT of signals in $\BB^\JJ$ for homogeneous sets $\JJ$ of size $k$. As remarked earlier, this includes the radix-2 FFT (since $\Z_N$ itself is homogeneous); and thus is a generalization of the standard radix-2. While we do not have proof that any structured-DFT algorithm has a complexity of $\Omega(k \log k)$, we can see that this is the case with the shift and sample framework from Fig \ref{fig:framework_alg1}.

\begin{lemma}
\label{lem:con_SAS}
Assume that $n\times n$ Vandermonde systems from \eqref{eq:alg1_system} cannot be solved faster than $n\log n$. Then, on any family $(\famJ_N,k_N)$, Algorithm \ref{alg:shift-sample} has a complexity of $\Omega(k_N \log k_N)$, irrespective of the choice of pivots.
\end{lemma}
For question 2) above, we can naturally consider supports $\JJ$, which has the largest possible number of pivots; one such example is
\begin{equation}
    \JJ^\star = \{1,2,4,8,\ldots, 2^{M-1}\}. \label{eq:powers-of-2-set}
\end{equation}
Note that $\JJ^\star$ has size $M$, and pivots $\{0,1,2,\ldots,M-1\}$ (i.e. every level in $\TT(\JJ^\star)$ is a pivot). To start, we can check if Algorithm \ref{alg:shift-sample} can be applied to such sets. The following lemma answers this in the negative.
\begin{lemma} \label{lem:con_1}
Assume that the $N-$ point DFT cannot be computed faster than $N\log N$. Then, for the family $\famJ=\{\JJ^\star\}$ with $\JJ^\star$ as defined in \eqref{eq:powers-of-2-set}, Algorithm \ref{alg:shift-sample} has complexity at least as much as that of solving \eqref{eq:submatrix-method}. Thus Algorithm \ref{alg:shift-sample} is at least as slow as the submatrix method from Section \ref{sec:preliminary_observations_system}.
\end{lemma}
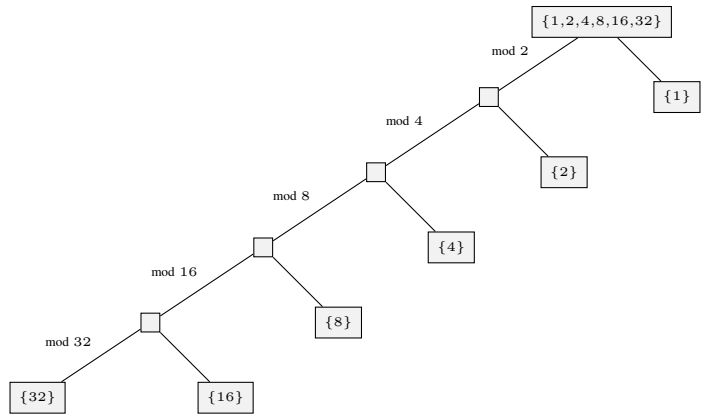
\begin{figure}[ht]
\centering
\begin{tikzpicture}[scale=0.5]
\node (0) at (8,11) [ minimum size = 0.25cm, fill=gray!10, draw] {$\scriptscriptstyle{\{1,2,4,8,16,32\}}$};
\node (1.1) at (5,9) [ minimum size = 0.25cm, fill=gray!10, draw] {};
\node (1.2) at (10,9) [ minimum size = 0.25cm, fill=gray!10, draw] {$\scriptscriptstyle{\{1\}}$};
\node (2.1) at (2,7) [ minimum size = 0.25cm, fill=gray!10, draw] {};
\node (2.2) at (7,7) [ minimum size = 0.25cm, fill=gray!10, draw] {$\scriptscriptstyle{\{2\}}$};
\node (3.1) at (-1,5) [ minimum size = 0.25cm, fill=gray!10, draw] {};
\node (3.2) at (4,5) [ minimum size = 0.25cm, fill=gray!10, draw] {$\scriptscriptstyle{\{4\}}$};
\node (4.1) at (-4,3) [ minimum size = 0.25cm, fill=gray!10, draw] {};
\node (4.2) at (1,3) [ minimum size = 0.25cm, fill=gray!10, draw] {$\scriptscriptstyle{\{8\}}$};
\node (5.1) at (-7,1) [ minimum size = 0.25cm, fill=gray!10, draw] {$\scriptscriptstyle{\{32\}}$};
\node (5.2) at (-2,1) [ minimum size = 0.25cm, fill=gray!10, draw] {$\scriptscriptstyle{\{16\}}$};

\draw (node cs:name=0) --node[above left] {$\scriptscriptstyle{\text{mod }2}$} (node cs:name =1.1);
\draw (node cs:name=0) --node[above left] {} (node cs:name =1.2);

\draw (node cs:name=1.1) --node[above left] {$\scriptscriptstyle{\text{mod }4}$} (node cs:name =2.1);
\draw (node cs:name=1.1) --node[above right] {} (node cs:name =2.2);
\draw (node cs:name=2.1) --node[above left] {$\scriptscriptstyle{\text{mod }8}$} (node cs:name =3.1);
\draw  (node cs:name=2.1) --node[above right] {} (node cs:name =3.2);
\draw (node cs:name=3.1) --node[above left] {$\scriptscriptstyle{\text{mod }16}$} (node cs:name =4.1);
\draw (node cs:name=3.1) --node[above right] {} (node cs:name =4.2);
\draw (node cs:name=4.1) --node[above left] {$\scriptscriptstyle{\text{mod }32}$} (node cs:name =5.1);
\draw (node cs:name=4.1) --node[above right] {} (node cs:name =5.2);

\end{tikzpicture}

\caption{Example set $\JJ^\star = \{1,2,4,8,16,32\}$ for $N = 64$.}
\label{fig:ex-converse}
\end{figure}

Since the best-known complexity (as far as we are aware) for solving the Vandermonde system with Fourier submatrices is $6k^2$, the Fourier computability of $\famJ^\star$ is not settled.

To investigate this further, recall that there were two key aspects (as posed in Section \ref{sec:problem_setup}) that went into the design of Algorithm \ref{alg:Hi-DFT}; which enables fast DFT computation for homogeneous support sets:
\begin{enumerate}
    \item A downsampling pattern $\II$ ( taken as $\II=\II_{\vec{r}}$ for $\vec{r}-$homogeneous sets) had the isolating/no-aliasing property (Lemma \ref{lem:aliasing-pattern}): in particular that 
    \(
    h(j_1-j_2)=(\F 1_{\II})(j_1-j_2) = 0 \text{ for }j_1\neq j_2 \in \JJ,
    \)
    so that computing $\F(f 1_\II)$ on indices $\JJ$ would give the DFT coefficients of $f \in \BB^\JJ$.
    \item There was an efficient algorithm to compute the product $\F(\JJ,\II)f_{\II}$ for the matrix $\F(\JJ,\II)$.
\end{enumerate}
The following lemma asserts that it is impossible to pick a sampling pattern $\II$ that achieves isolation for $\JJ^\star$.
\begin{lemma}\label{lem:conv_isolation}
With $\JJ^\star$ as defined in \eqref{eq:powers-of-2-set}, any $\II\subseteq \Z_N$ that satisfies the isolating property
\[
h(j_1-j_2)=0 \text{ for }j_1\neq j_2 \in \KK \subseteq \JJ^\star, \quad h=\F 1_\II 
\]
has size at least $2^{|\KK|}$.
\end{lemma}
In particular, we need $\II= \Z_N$ to achieve complete isolation, thus making the sample complexity (and hence computational complexity) $\Omega(N)$. So we will be unable to satisfy the requirement 1) above. Thus the proposed framework (even with modifications allowing for arbitrary downsampling patterns $\II$) \emph{cannot} establish the (non) Fourier computability of $\JJ^\star$. We can then consider replacing the downsampling (multiplication with $1_\II$) multiplication by a sparse vector $\vec{x} \in \mathbb{C}^N$, but then for the condition 2) to satisfy, we may have to assume \(\text{supp}(\vec{x}) \subseteq \II_{\vec{r}},\) and \(\JJ^\star \subseteq \KK \) for some $\vec{r}-$homogeneous $\KK$. But any spectral set $\KK$ containing $\JJ^\star$ has to have pivots at every level, and so $\KK = \Z_N$, leading to $\Omega(N)$ sample and computational complexity.






\label{sec:converse_results}

\section{Conclusion and Future work}
\label{sec:conclusion}
We investigate the DFT computation of signals with known frequency support of size $k$. When the frequency support is homogeneous, we show that an $O(k\log k)$ complexity is achievable by a generalization of the radix-2 FFT. We then modify the algorithm to work with approximate homogeneity and establish $O(k\log k)$ (and $O(k\log ^2k)$ in some cases) complexity for a variety of support structures.

Fourier computability seems closely connected to the additive structure. The structure $\JJ^\star$ (from Section \ref{sec:converse_results}) for which we are unable to establish Fourier computability is a set with large doubling (we can check that $|\JJ^\star + \JJ^\star| \geq |\JJ^\star|^2/2$); and on the other end, the sets for which we are able to establish Fourier computability seem to have some additive structure. We hope to explore this connection further.

\appendix
\subsection{Proof of Definition \ref{def:fam_r_sets}}\label{proof:def_fam_r_sets}
To see this, note that for $\JJ \in \KK_N^{\downarrow k} $, by Chebyshev's inequality,
\begin{align*}
\Prob\left(|k_N/2 \leq |\JJ| \leq 3k_N/2\right) &= \Prob\left(\left||\JJ| - k_N\right| \leq k_N/2\right)\\
&\geq 1-\frac{4\Var|\JJ|}{k_N^2} \geq 1 -\frac{4}{k_N} \rightarrow 1.
\end{align*}
Thus this is a family as in Definition \ref{def:fam_idx_sets} (with $\alpha = 1/2, \beta =3/2$).\hfill\IEEEQEDhere
\subsection{Proof of Proposition \ref{prop:num-pivots-cong-tree}}
\label{proof:prop:num-pivots-cong-tree}
For each level $r$ in $\TT(\JJ)$, let 
\begin{equation}
\label{eq:mu-r-star-def}
\mu_r^\star = \max_{v \text{ at level }r}\mu_\JJ(v)
\end{equation}
be the largest weight of any node at level $r$. Since the root is the only node at level $0$, we have $\mu_0^\star = |\JJ|$; and since all the leaves are singletons, we have $\mu_{\log N}^\star = 1$. Also, by the binary structure of the tree, we must have
\begin{equation}
\label{eq:hom-tree-mu-r-star}
\mu_{r+1}^\star \begin{cases} \geq \frac{\mu_{r}^\star}{2}\text{ if }r \text{ is a pivot} \\ =\mu_{r}^\star \text{ otherwise.}  \end{cases} 
\end{equation}
Thus 
\[
1= \mu_{\log N}^\star \geq \frac{\mu_0^\star}{2^{\text{ no. of pivots}}} = \frac{|\JJ|}{2^{\text{ no. of pivots}}},
\]
so that the number of pivots is at least $\log |\JJ|$. \hfill \IEEEQEDhere

\subsection{Proof of Proposition \ref{prop:hom-tree-prop}}

1 $\iff$ 2 follows from \eqref{eq:hom-tree-mu-r-star}. The number of pivots is exactly $\log |\JJ|$ iff $\mu_{r+1}^\star = \mu_r^
\star/2$, at every pivot $r$. and so all the nodes at a pivot level must split into two children of the same size. Similarly 1$\iff$3 follows by noting that homology is equivalent to the recursion $\mu_{r+1}^\star = \mu_r^\star/2^{ I(r \text{ is a pivot})}$. Solving this recursion with $\mu_{\log N}^\star=1$ gives us \(\mu_r^\star = 2^{\text{ no. of pivots below }r }\), which is the desired result. For 1$\iff$4, note that the set of pivots is the set $\{e_{ab}\}$. \hfill \IEEEQEDhere

\subsection{Proof of Proposition \ref{prop:homo-spectral}}
The proof Homogeneous $\implies$ Spectral was done as part of the discussion for \eqref{eq:hom-implies-spectral}. Here we prove that if $\JJ$ is spectral, then $\JJ$ is homogeneous.

If $\JJ$ is spectral, then we have a square submatrix $\F(\II,\JJ)$ which is unitary. Let $\JJ=\{j_1, j_2, \ldots, j_k\}$. Now we have for any $j_s\neq j_t \in \JJ$,
\[
\sum_{m \in \II}e^{-2\pi i m (j_s-j_t)/N}= 0.
\]
Consider the polynomial $p(x) = \sum_{m\in \II}x^m$. Since this polynomial has integer coefficients and has $e^{2\pi i (j_s-j_t)/N}$ as a root, it must be divisible by $\Phi_{N'}(x)$ (where $\Phi_n$ is the $n^{th}$ cyclotomic polynomial\cite{isaacs2009algebra}) with $N'= N/\text{gcd}(j_s-j_t,N)$. Let $S=\{n: \Phi_{2^n}(x) | p(x)\}$. Note that $n\in S$ iff there exist $j_s\neq j_t \in \JJ$ such that $j_s-j_t = \text{(odd)}2^{M-n}$; thus $n \in S$ iff $M-n$ is a pivot. So $|S|$ is the number of pivots in $\JJ$.

Since $\Phi_{2^n}(1) = 2$ for any $n$, we must have $2^{|S|}\leq |\JJ|$ or $|S| \leq \log |\JJ|$. But then there are at least $\log |\JJ|$ pivots; so we must have $|S|= \text{number of pivots} = \log |\JJ|$, and so $\JJ$ is homogeneous. 

As mentioned earlier, this equivalence is known in a more general context. The proof here is adapted from  \cite{coven1999tiling} (see also  \cite{newman1977tesselation} and \cite{laba2002spectral}).
\hfill\IEEEQEDhere

\subsection{Proof of Lemma \ref{lem:shift-mod-Hi-DFT}}
First, note that for any sampling pattern $\II$,
\((\tau^{a} f)1_\II = \tau^{a}(f1_{\tau^{-a} \II}) \). We also have
\(\F1_{\tau^{-a}\II_{\vec{r}^{n-}}}(m) = \F \left(\tau^{-a} 1_{\II_{\vec{r}^{n-}}}\right)(m) = e^{-2\pi i am/N}h_{\vec{r}^{n-}}(m), \) and so
\begin{align*}\F_\JJ^{n,a} f (m) &= e^{2\pi i am/N} \F (f1_{\tau^{-a} \II}), \text{ for }m \in \JJ; \\
&=e^{2\pi i am/N}\sum_{j \in \KK} \F f(j)  e^{-2\pi i a(m-j)/N}h_{\vec{r}^{n-}}(m-j);
\\&=|\JJ|\sum_{ j\in  \JJ'(m)} e^{2\pi i aj/N}\F f(j) /2^n, \text{ from Lemma \ref{lem:aliasing-pattern};}
\end{align*}
where $\JJ'(m)$ is the set of nodes with a common ancestor to $m$ at height $n$, similar to the discussion for \eqref{eq:aliasing-homogeneous-tree}. \hfill \IEEEQEDhere

\subsection{Proof of Lemma \ref{lem:rec-hom-tree-split}}
Set $r = r_{\Sizer-n}$ be the $n^{th}$ pivot from the end, and set $a=a_n$. We have for any node $j$ at height $n+1$ in $\TT(\JJ)$
\[
\mu_\JJ(j, \vec{w}) = \mu_\JJ(j_1, \vec{w}) + \mu_\JJ(j_2, \vec{w}),
\]
where $j_1$ and $j_2$ are the left and right descendants (respectively) of node $j$ at height $n$. Now we apply Lemma \ref{lem:shift-mod-Hi-DFT}  \eqref{eq:aliasing-homogeneous-tree-shift}  with $\vec{w} = \F f$ and $\vec{w} = \underline{\omega}^{-a} \F f$: we get 
\begin{gather*}
\F_\JJ^{n+1} f (j) =  \frac{1}{2}\left(\F^n f(j_1) + \F^n f(j_2)\right), \quad \\ \F_\JJ^{n+1,a} f (j) = \frac{\omega^{-j_1}_{r+1}}{2}\left( \F^n f(j_1) + \omega^{j_1-j_2}_{r+1}\F^n f(j_2)\right).
\end{gather*}
Here we used that $\underline{\omega}^{-a} =\underline{\omega}_{r+1}^{-1}$. 
 Now, we see that $j_1-j_2$ must be an odd multiple of $2^{r}$ by definition; and so $\omega_r^{j_1-j_2} = -1$. Substituting this in the equation above gives us 
 \begin{gather*}
    \F_\JJ^{n}f (j_1) = 2\left(\F_\JJ^{n+1} f (j) + \omega_{r+1}^{j_1}\F_\JJ^{n+1, a_n}f (j)\right),\quad \\
    \F_\JJ^{n}f (j_2) = 2\left(\F_\JJ^{n+1} f ( j) - \omega_{r+1}^{j_2}\F_\JJ^{n+1, a_n}f (j)\right).
\end{gather*}
 Now note that $\F_\JJ^{n+1} f (j) = \F_\JJ^{n+1} f (j_1) = \F_\JJ^{n+1} f (j_2)$, giving us the desired result. \hfill \IEEEQEDhere

\subsection{Proof of Theorem \ref{thm:computation_Hi-DFT}}
Let $T(n)$ be the number of arithmetic operations required by Algorithm \ref{alg:Hi-DFT} to compute the Hi-DFT $\F_\JJ^n f$ for $f \in \BB^\JJ$. Then from Lemma \ref{lem:rec-hom-tree-split}, to compute $\F_\JJ^nf$ we need to compute $\F_\JJ^{n+1}f, \F_\JJ^{n+1, a_n}f$ (each of which has $2^{\Sizer-n-1}$ values), followed by $2\times 2^{\Sizer-n-1}$ multiplications and $2^{\Sizer-n-1}$ additions. So 
\(
T(n) = 2T(n+1) + 3\times 2^{\Sizer-n-1}.
\)
Using $T(\Sizer) = 1$, and solving the recurrence above, we get that \(T(n)=2^{-n}\left(2^\Sizer - 3\times 2^{\Sizer - 1}(n-\Sizer) \right) = 2^{\Sizer -n}\left(1+1.5(\Sizer-n) \right) = 1.5 A \log A + A.  \)
\hfill \IEEEQEDhere

\subsection{Proof of Corollary \ref{cor:spectral-Fourier-computable}}
Given $\vec{r}$, and access to the elements of an $\vec{r}-$homogeneous $\JJ$; computing the tree $\TT(\JJ)$ needs access to the bits in positions $\vec{r}$ from each element of $\JJ$. Thus the tree $\TT(\JJ)$ can be computed in $O(|\JJ| \Sizer)=O(|\JJ| \log |\JJ|)$ bit operations. Then the DFT coefficients can be found by applying Algorithm \ref{alg:Hi-DFT}, which operates in $O(|\JJ|\log |\JJ|)$ arithmetic operations.\hfill \IEEEQEDhere

\subsection{Proof of Lemma \ref{lemma:computation_shift_sample}}
The correctness of Algorithm \ref{alg:shift-sample} was discussed in the discussion preceding Lemma \ref{lemma:computation_shift_sample}. We discuss the complexity next. The algorithm needs to compute multiple (small length) Hi-DFTs, followed by solving a Vandermone system. From \cite{parker1964inverses,gohberg1997fast}, Vandermonde systems of size $\alpha$ can be solved in $c_2\alpha^2$ operations, where constant $c_2 = 6$. Thus, solving the system \eqref{eq:alg1_system} at node $v$ costs $c_2\mu_\JJ^2(v)$ operations; summing at all nodes $v$ on level $r$ incurs a complexity of \(c_2\sum_{\substack{v \text{ at level }r}} \mu_\JJ^2(v)\).

Note that finding the DFT-induced weights (or Hi-DFT) from Algorithm \ref{alg:Hi-DFT} costs $c_1 \Sizer 2^\Sizer$ operations, and this is repeated $\mu_r^\star$ times (we can pick $c_1=2$). This gives us the complexity in the statement of Lemma.

For the upper bound, note that since there are at most $2^\Sizer$ nodes at level $r$ (see Remark \ref{rem:part-homogenous-number-of-nodes}),
\[
\sum_{\substack{v \text{ at level }r}} \mu_\JJ^2(v) \leq \left(\mu^\star_r\right)^2 2^\Sizer;
\]
so that  the number of computations in Algorithm \ref{alg:shift-sample} (\texttt{SAS}) is bounded by
\[ \min_{\vec{r}} \left\{   2^\Sizer\mu^\star_r \left(c_1\Sizer + c_2\mu^\star_r\right)\right\}. 
\] \hfill \IEEEQEDhere

\subsection{Proof of Theorem \ref{thm:part-hom-Fourier-computability}}
Once given access to $\JJ$, the algorithm can first compute the truncated congruence tree $\TT_{r}(\JJ)$ and the weights $\mu_r(\JJ)$ at level $r\coloneqq r_{\text{max}}$: this requires accessing $\Sizer$ bits in the binary expansion of the elements of $\JJ$. As discussed before, this can be done in $O(k_N\Sizer) = O(k_N \log k_N)$ bit operations. We apply Algorithm \ref{alg:shift-sample}, and by Lemma \ref{lemma:computation_shift_sample}, the complexity is bounded by
\begin{align*}
2^\Sizer \alpha_2 (c_1 \Sizer + c_2 \alpha_2) &= 2^{\alpha_1} \alpha_2 k_N (c_1 \alpha_1 + c_1 \log k_N + c_2 \alpha_2)\\& = \text{ const } k_N + \text{ const }k_N \log k_N \\&= O(k_N \log k_N).
\end{align*}\hfill \IEEEQEDhere

\subsection{Proof of Corollary \ref{cor:AP}}
Let $(a,b)$ be the greatest common divisor (gcd) of $a$ and $b$. From the definition, we see that the pivots of any $\JJ \in \famJ_N$ are $(s,N), (2s,N), (3s,N), \ldots, ((k_N-1)s,N)$. Suppose $s=2^\alpha \text{(odd)}$, then $(2^\beta\text{(odd)}s, N) = 2^{\alpha+\beta}$. So the pivots are contained in $\{\alpha, \alpha+1, \ldots, \alpha+\lfloor\log k_N\rfloor\}$ there are $\log k_N + \text{const}$ pivots, thus satisfying the hypothesis of Theorem \ref{thm:part-hom-Fourier-computability}, so Algorithm \ref{alg:shift-sample} operates with $O(k_N \log k_N)$ complexity.
\hfill \IEEEQEDhere

\subsection{Proof of Lemma \ref{lem:gap-complexity}} \label{proof:gap-complexity}
Let $\JJ$ be a generalized arithmetic progression of dimension $d$, we see that
\[
\JJ-\JJ \subseteq \{a+n_1s_1+n_2s_2+\ldots+n_ds_d: -(N_j-1)\leq n_j \leq N_j-1 \}.
\]
The set on the right above has size at most $2^dN_1N_2\ldots N_d = 2^d|\JJ|$, so $|\JJ-\JJ| \leq 2^d|\JJ|$, and in fact we may generalize to \cite[Lemma~3.10]{tao2006additive}
\[
|m_1\JJ-m_2\JJ| \leq (m_1+m_2)^d|\JJ|.
\]
for any $m_1, m_2$. Note that here $m\JJ$ is the sum of all $m-$tuples with each element taken from $\JJ$, and not the dilation of $\JJ$. 

Now suppose $\JJ$ has pivots $r_1, r_2, \ldots, r_n$, in increasing order. Then by definition, we have a set $P$ of the form
\[
P = \{0,\alpha_12^{r_1}, \alpha_2 2^{r_2}, \ldots, \alpha_n2^{r_n} \}, \text{ with }\alpha_i \text{ odd };
\]
and $P \subseteq \JJ-\JJ$. First, we see that any sums constructed from distinct elements of $P\setminus \{0\}$ is distinct and nonzero. Otherwise we have an expression of the form
\begin{align*}
\alpha_{i_1}2^{r_{i_1}}+\alpha_{i_2}2^{r_{i_2}} + \ldots = 0 &\mod 2^M,\\
\alpha_{i_1}+\alpha_{i_2}2^{r_{i_2}-r_{i_1}} + \ldots = 0 &\mod 2^{M-r_{i_1}},
\end{align*}

we see that this is not possible for distinct $r_i < M$ (as is the case here). There are $2^n-1$ such sums (excluding the empty subset of $P\setminus\{0\}$), and including $0$ we thus have $2^n$ distinct elements in $nP$. So we have the following
\begin{equation}
\label{eq:gap-pivot-bound}
2^n \leq |nP| \leq |n(\JJ-\JJ)| \leq n^d2^d|\JJ|.    
\end{equation}
Thus the number of pivots $n$ of $\JJ$ satisfies $1/2|\JJ|^{1/d} \leq ne^{-n\ln2/d}$. The solution is given by
\[
n \leq -\frac{d}{\ln 2}W_{-1}\left(\frac{-\ln2}{2d|\JJ|^{1/d}}\right),
\]
where $W$ is the Lambert W function \cite{corless1996lambertw}. Since we assume $|\JJ|$ is large, the argument of $W_{-1}$ is a small (negative) number and we may apply the following approximation from \cite[Equation~4.19]{corless1996lambertw}:
\[
W_{-1}(x) = \ln(-x)-\ln(-\ln(-x)) + o(1) \text{ for small }x<0.
\]
We get that for $|\JJ|$ large enough,
\begin{align*}
n&\leq d\beta_1 + d\log d + \log |\JJ| + d\log(\beta_1+\ln d + \ln |\JJ|/d)  +o(1)\\
&\leq\log|\JJ| + d\beta_1 + d\log d + d\log 2 +\log \ln |\JJ| + o(1)\\
&= \log|\JJ| + \log \log |\JJ| + O(d\log d).
\end{align*}
Here $\beta_1 = -\log \ln 2 + 1$ in the first step. In the second step, we used $\ln|\JJ|/d > \beta_1 + \ln d$ for large enough $|\JJ|$.
This bounds the number of pivots in a generalized arithmetic progression. Now applying Algorithm \ref{alg:Hi-DFT} at these pivot locations, we get a complexity of 
\begin{align*}
&c_1 n2^n = \\
 &\quad c_1 2^{O(d\log d)}|\JJ|\log|\JJ|\left(\log|\JJ| + \log\log|\JJ| + O(d\log d) \right) \\
&=O(k_N \log^2 k_N),
\end{align*}
completing the proof. It may be possible to tighten this analysis by using a better bound on the pivots than \eqref{eq:gap-pivot-bound}; for e.g., the obtained complexity is higher than $O(k_N \log k_N)$ achievable for the case $d=1$ (Corollary \ref{cor:AP}).\hfill \IEEEQEDhere

\subsection{Proof of Theorem \ref{thm:additive-structure-complexity}}
By the Freiman-Green-Ruzsa theorem  \cite{sanders2013structure, green2007freiman}, $\JJ$ is contained in a GAP of size $C'|\JJ|$ and dimension $C''$. Since $C$ is a constant, so are $C'$ and $C''$. The result follows from the application of Lemma \ref{lem:gap-complexity} (Note that the coset progressions defined in \cite{green2007freiman} are the same as GAPs for $\Z_N$, since any subgroup in the cyclic group $\Z_N$ is of the form $\{0,2^s, 2\times 2^s,3 \times 2^s,\ldots, (2^{M-s}-1)2^s\}$; which is also an arithmetic progression).\hfill\IEEEQEDhere

\subsection{Proof of Remark \ref{rem:UOE-not-balanced}}
Set $\JJ_{i0}=\cup_{j=0}^{2^i}\{2^{a_N+i} + j\}$ as a set of $2^i$ consecutive integers starting from $2^{a_N+i}$ (so clearly $\JJ_{i0}$ is elementary). We have \(\JJ = \cup_{i=0}^{a_N} \JJ_{i0} \), so $\JJ$ is of the form \eqref{eq:UoE1} (with $\eta_i =1$). Also, note that $a_N \leq \log k_N = \log(2^{a_N+1}-1) \leq a_N+1$ (and the family is UoE). Now, by considering the difference of elements from $\JJ$, we see that $\JJ$ has pivots $\{0,1,2,\ldots,2a_N-1\}$. For this family to be balanced partly homogeneous, we need to pick $\vec{r}_N \subseteq \{0,1,2\ldots,2a_N-1\}$ (to satisfy condition 3) in Theorem \ref{thm:part-hom-Fourier-computability}. We need to pick $\Sizer \leq \alpha_1 + a_N+1$ to satisfy condition 1). However, we see now that there are $a_N-1-\alpha_1$ elements of $\JJ$ that are divisible by $2^{a_N+2+\alpha_1}$ (this is the number of elements with $j=0$, $i\geq 2+\alpha_1$ in the definition of $\JJ$ above). So for $r=r_{\text{max}}+1 = \Sizer + 1$; we have
\(\mu_{r}^\star \geq a_N-1-\alpha_1 = \Omega(\log k_N)\), so condition 2) cannot be satisfied. \hfill \IEEEQEDhere

\subsection{Proof of Proposition \ref{prop:UoE-mu-r-star-bound}}
To see this first, note that for any elementary set $\JJ_i$ of size $2^i$, we have $\mu_{\JJ_i}(v) \leq 1$ for any $s> i$; and for $s\leq i$, \(\mu_{\JJ_i}(v)=\)number of leaves in the subtree rooted at \(v= 2^{i-s}\). Now when $\JJ$, as in Lemma \ref{lem:UoE1},
\begin{align*}
\mu_\JJ(v) &= \sum_{i=0}^{a_N} \sum_{j=0}^{\eta_i-1} \mu_{\JJ_{ij}}(v) \\&= \sum_{i=0}^{s-1}\sum_{j=0}^{\eta_i-1} \mu_{\JJ_{ij}}(v) + \sum_{i=s}^{a_N}\sum_{j=0}^{\eta_i-1} \mu_{\JJ_{ij}}(v) \\
&\leq \sum_{i=0}^{s-1} \eta_i + \sum_{i=s}^{a_N} \eta_i 2^{i-s} \leq sC + C(2^{a_N+1-s}-1).
\end{align*}\hfill \IEEEQEDhere

\subsection{Proof of Lemma \ref{lem:UoE1}}
We take $\vec{r} = \{0,1,2,\dots,r-1\}$, where $r = \lfloor\log k_N- \log \log k_N\rfloor.$ Using $x-1< \lfloor x \rfloor \leq x$, we see that $k_N/2\log k_N < 2^\Sizer \leq k_N/\log k_N$. Also, from Proposition \ref{prop:UoE-mu-r-star-bound} we have,
\begin{align*}
\mu_r^\star &\leq C(r + 2^{a_N+1-r}) \\ & \leq C \left(\log k_N + 2^{a_N+1}(2\log k_N/k_N)\right) \\ & \leq C \log k_N\left(1+2^{\alpha+1}\right) = O(\log k_N).
\end{align*}

From Lemma \ref{lemma:computation_shift_sample}, the number of computations in the Algorithm \ref{alg:shift-sample} (\texttt{SAS}) is bounded by 
\(   2^\Sizer\mu^\star_r \left(c_1\Sizer + c_2\mu^\star_r\right). \)
By substituting size($\vec{r}$) and $\mu_r^*$ values of UOE in the above expression we get:
\begin{align*}
    \text{Complexity} & \leq (k_N/\log k_N) C \log k_N\left(1+2^{\alpha+1}\right) \\ &\left(c_1 \lfloor\log k_N- \log \log k_N\rfloor + c_2 C \log k_N\left(1+2^{\alpha+1}\right)\right)\\
    & \leq C \left(1+2^{\alpha+1}\right) k_N \left(c_1+c_2 C\left(1+2^{\alpha+1}\right) \right)\log k_N \\
    &= O(k_N \log k_N).
\end{align*}
\hfill\IEEEQEDhere

\subsection{Proof of Theorem \ref{thm:UoE2}}
The proof follows very similarly to that for Lemma \ref{lem:UoE1}, so we only emphasize the key differences here. 

First, we note that all the constituent sets $\JJ_{ij}$ and the set $\JJ$ are $\vec{l}_N^{(\Sizeln-a_N)-}-$part-homogeneous. Hence we can define the height of any node in the tree $\TT(\JJ_{ij})$ as per Definition \ref{def:part-H-tree} (nodes at level $l_{a_N}+1$ are mapped to height $0$). The proofs below use the height of a node (as opposed to the level for proofs of Lemma \ref{lem:UoE1} and Proposition \ref{prop:UoE-mu-r-star-bound}).

Next, we note the following result similar to Proposition \ref{prop:UoE-mu-r-star-bound}. For an index set $\JJ$ in a UoH as in Theorem \ref{thm:UoE2}. consider a node $v$ at height $s\leq a_N$ in $\TT(\JJ)$. Then 
\begin{equation}
\label{eq:UoE-mu-r-star-bound}
\mu_\JJ(v) \leq C\left( 2^{s+1}+a_N-s-1\right).
\end{equation}

This result follows from the observation of homogeneous sets below (similar to the one for elementary sets earlier): If $\JJ_i$ is a $\vec{l}_N^{(\Sizeln - i)-}-$ homogeneous, then, we have $\mu_{\JJ_i}(v) \leq 1$ for any $s< i$; and for $s\geq i$, \(\mu_{\JJ_i}(v)=\)number of leaves in the subtree rooted at \(v= 2^{s-i}\). 
Now when $\JJ$, as in Theorem \ref{thm:UoE2},
\begin{align*}
\mu_\JJ(v) &= \sum_{i=0}^{a_N} \sum_{j=0}^{\eta_i-1} \mu_{\JJ_{ij}}(v) \\&= \sum_{i=0}^{s}\sum_{j=0}^{\eta_i-1} \mu_{\JJ_{ij}}(v) + \sum_{i=s+1}^{a_N}\sum_{j=0}^{\eta_i-1} \mu_{\JJ_{ij}}(v) \\
&\leq \sum_{i=0}^{s} \eta_i 2^{s-i}  + \sum_{i=s+1}^{a_N}  \eta_i \leq C (2^{s+1}-1) + C (a_N-s).
\end{align*}
We see that this is the same as Proposition \ref{prop:UoE-mu-r-star-bound}, with $s$ replaced by $a_N-s$.

Now suppose we take $\vec{r} = \vec{l}_N^{(\Sizeln - r)-}$ (so that $\Sizer = r)$, where $r = \lfloor\log k_N- \log \log k_N\rfloor.$ Using $x-1< \lfloor x \rfloor \leq x$, we see that $k_N/2\log k_N < 2^\Sizer \leq k_N/\log k_N$. Then for nodes at height $s = a_N - r$, we have $s \leq a_N - \log k_N + \log \log k_N \leq \alpha + \log \log k_N$. So for nodes $v$ at height $s$
\begin{align*}  
\mu_\JJ(v) & \leq C\left( 2^{s+1}+a_N-s-1\right) \\ & \leq C\left(2^{\alpha+1} \log k_N + \log k_N - \log \log k_N - 1 \right) \\ & \leq C \log k_N \left(2^{\alpha+1}  + 1 \right) = O(\log k_N).\end{align*}
From Lemma \ref{lemma:computation_shift_sample}, the number of computations in Algorithm \ref{alg:shift-sample} (\texttt{SAS}) is bounded by 
\(   2^\Sizer\mu^\star_{r_{max}+1} \left(c_1\Sizer + c_2\mu^\star_{r_{max}+1}\right). \)
By substituting size($\vec{r}$) and $\mu_{r_{max}+1}^*$ values of UOH in the above expression we get:
\begin{align*}
    \text{Complexity} & \leq (k_N/\log k_N) C \log k_N\left(1+2^{\alpha+1}\right) \\ & \left(c_1 \lfloor\log k_N- \log \log k_N\rfloor + c_2 C \log k_N\left(1+2^{\alpha+1}\right)\right)\\
    & \leq C \left(1+2^{\alpha+1}\right) k_N \left(c_1+c_2 C\left(1+2^{\alpha+1}\right) \right)\log k_N \\
    &= O(k_N \log k_N).
\end{align*} 
\hfill \IEEEQEDhere

\subsection{Proof of Lemma \ref{lem:bound_mu_r_star}}
Note that $\lambda$ is $\Omega(1)$, and so $\lambda\geq b$ for some constant $b$. Also since $\lambda$ is $O(\log k)$, we have $\lambda \leq \gamma \log k$ (for large enough $N$) for some constant $\gamma>0$.

To prove \eqref{eq:mu-r-star-ub0}, we use the concentration inequality \cite{harvey2022randomized}
\begin{equation}
\Prob \left(X \geq (1+ \delta)E(X)\right) \leq \exp \left(-\delta E(X)/3)\right) \text{ for }\delta>1,
\end{equation}

where $X$ is a binomial random variable. Recall that $\mu_v\coloneqq \mu_\JJ(v)$ is a binomial random variable, with $E(\mu_v) = \lambda$. Pick $c >2 \gamma$, and apply the above to $\mu_v$, with $\delta = c\log k/\lambda - 1 \geq c/\gamma -1 > 1$. We get
\begin{align*}
    \Prob(\mu_v \geq c \log k) \leq  \exp\left(-(c\log k -\lambda)/3\right) = e^{\lambda/3}/k^{c'/3}, 
\end{align*}
where $c'=c/\ln 2$.

By union bound, the probability that $\mu_r^\star = \max\{\mu_v\}$ is large is given by
\begin{align*}
\Prob(\mu_r^\star \geq c \log k) &\leq \sum_{v\text{ at level }r}\Prob(\mu_v \geq c\log k) \\ & \leq 2^ne^{\lambda/3\ln 2} /k^{c'/3} \leq e^{\lambda/3}/\lambda k^{c'/3-1} \\ & \leq e^{\gamma \log k/3}/bk^{c'/3-1}. 
\end{align*}
Recall that level $r$ is at height $n'$; in the above, we used that there are at most $2^n$ nodes at height $n'$ (from Remark \ref{rem:part-homogenous-number-of-nodes}).
 In particular, 
if we pick $c$ such that
\(
 c=2\gamma + 3\ln 2,
\)
the upper bound computed earlier is
\[
\Prob(\mu_r^\star \geq c \log k)  \leq 1/bk^{\gamma/3\ln 2}.
\]
Hence the above probability goes to zero as $k \rightarrow \infty$.
\hfill \IEEEQEDhere

\subsection{Proof of Theorem \ref{thm:hom_rand_subset}}

We apply Algorithm \ref{alg:shift-sample} with a suitable choice of pivots $\vec{r}$. In particular we pick $\vec{r}=\vec{l}^{(\Sizel-n)-}$, with $\Sizer = n =\lfloor\log k - \log \log k\rfloor$. This choice ensures that $\lambda
\leq \mathbf{2}\log k$.

Now using Lemma \ref{lemma:computation_shift_sample}, the complexity of Algorithm \ref{alg:shift-sample} is bounded by
\[ 2^\Sizer\mu^\star_r \left(c_1\Sizer + c_2\mu^\star_r\right). \]

Using $x-1< \lfloor x \rfloor \leq x$, we see that $k/2\log k < 2^\Sizer \leq k/\log k$. By substituting the values of $\Sizer$ and $\mu^\star_r < c \log k$ (where $c= 4 + 3\ln 2$ from Lemma \ref{lem:bound_mu_r_star}) we get:

\begin{align*}
    \text{Complexity} & \leq (k/\log k) c \log k \left(c_1 (\log k - \log \log k) + c_2 c \log k \right)\\
    & \leq c  k \left(c_1+c_2 c\right) \log k \\
    &= O(k \log k).
\end{align*} \hfill \IEEEQEDhere

\subsection{Proof Lemma \ref{lem:con_SAS}}
For this, we make the following assumptions:
\begin{enumerate}
    \item The $N-$point DFT cannot be computed faster than $N\log N$: this assumption was also made in \cite{nearlyoptimalsparse}. As the Hi-DFT is a generalization of the DFT; we assume that the Hi-DFT with pivots $\vec{r}$ cannot be computed faster than $\Sizer 2^\Sizer$. This is contained in the assumption below.
    \item We assume that any $n \times n$ Vandermonde system (as in \eqref{eq:alg1_system}) involving Fourier submatrices cannot be solved faster than $n\log n$. When $n=N$, solving such a system is equivalent to finding the DFT (which is assumed to be bounded by $N\log N$ above). 
\end{enumerate}
Now consider any $\JJ \in \famJ_N$, and a $\vec{r}$ for which Algorithm \ref{alg:shift-sample} is applied (here, $\vec{r}$ is picked such that $\JJ$ is $\vec{r}-$part-homogeneous). Under the assumptions above, the complexity of (steps 2-7) in Algorithm \ref{alg:shift-sample} is at least 
\[
\Sizer2^\Sizer \mu_r^\star + \sum_{v \text{ at level }r} \mu_\JJ(v) \log \mu_\JJ(v),
\]
with $r$ and $\mu_r^\star$ as defined for Lemma \ref{lemma:computation_shift_sample}. Let $n\leq \Sizer$ be the number of $l \in \vec{r}$ such that $l$ is a pivot of $\TT(\JJ)$. Note that by construction, there are $2^n$ nodes at level $r$ in $\TT(\JJ)$, and recall that $\sum_v \mu_\JJ(v) = |\JJ|$.
Using $\mu_r^\star \geq |\JJ|/2^n$ from \eqref{eq:hom-tree-mu-r-star}; and the convexity and monotonicity (for $x\geq 1$) of $x\log x$; we get the following lower bound
\begin{align*}
|\JJ| 2^{\Sizer-n} \Sizer + |\JJ| \log \frac{|\JJ|}{2^n} & \geq |\JJ|n + |\JJ|\log \frac{|\JJ|}{2^n}\\ & = |\JJ| \log |\JJ| \\&= \Omega(k_N \log k_N).
\end{align*}\hfill\IEEEQEDhere

\subsection{Proof of Lemma \ref{lem:con_1}}
For this, we make similar assumptions to the proof of Lemma \ref{lem:con_SAS}: we assume that the $N-$ point DFT cannot be computed faster than $N\log N$.

Since $\JJ^\star$ has all the levels as pivots, to apply Algorithm \ref{alg:shift-sample}, we need to pick $\vec{r} = \{0,1,2,\ldots,\alpha-1 \}$ for some $\alpha$.
Recall that at level $\alpha$, we compute $\mu_\alpha^*$ DFTs with complexity $\mu_\alpha^\star \alpha 2^{\alpha}$. Also, at (any) level $\alpha$, note that we will only have one system of size $\mu_\alpha^* \times \mu_\alpha^*$. All the remaining nodes have weight at most one. We see that at any level $\alpha$ we have $\mu_\alpha^* = M-\alpha$. Note the range of $\alpha$ as $0 \leq \alpha \leq M-1$.
The complexity of the Algorithm \ref{alg:shift-sample} for this family is lower bounded by \(C(M,\alpha)\coloneqq (M-\alpha) \alpha2^{\alpha}\) \(+ g(M-\alpha) \). Here $g(n)$ is the complexity of solving the systems (from \eqref{eq:submatrix-method}) of size $n$; we have $ g(n) \leq 6n^2$. Assume $g$ is monotonic, and since $g$ has at most polynomial growth, we have $\lim g(M/2)/g(M) >0$. Now we have 
\[
\frac{C(M,\alpha)}{g(M)} \geq \left(1-\frac{\alpha}{M}\right)\frac{\alpha 2^\alpha}{6M} + \frac{g (M-\alpha)}{g(M)} .
\]
We show that $C(M,\alpha)/g(M)$ is bounded away from $0$ as $M \rightarrow \infty$. For $C(M,\alpha)/g(M) \rightarrow 0$ we need either 
\begin{enumerate}
    \item $1-\alpha/M \rightarrow 0$ in which case $\alpha = \Omega(M)$. But in this case $\alpha 2^\alpha /M \geq 2^{\text{(const)} M}$ and so
    \begin{align*}
    \left(1-\frac{\alpha}{M}\right)\left(\frac{\alpha 2^\alpha}{6M}\right) &\geq \left(1-\frac{M-1}{M}\right)\left(\frac{\alpha 2^\alpha}{6M}\right)\\ & \geq \frac{2^{\text{(const)} M}}{M} \rightarrow \infty; \text{ or, }
    \end{align*}
 \item $\alpha 2^\alpha/6M \rightarrow 0$, in which case we need $\alpha 2^\alpha = o(M)$ and so $\alpha = \log (o(M))$. In particular, we have 
 \[
 \lim \frac{g(M-\alpha)}{g(M)} \geq \lim \frac{g(M/2)}{g(M)} >0.
 \]
 Thus in either case, $C(M,\alpha)/g(M)$ is bounded away from $0$, and therefore $C(M,\alpha)$, and the complexity, is $\Omega(g(M))$, completing the proof.
\end{enumerate}

\subsection{Proof of Lemma \ref{lem:conv_isolation}}
This can be seen by applying the results on the structure of idempotent $h$ in terms of frequency domain zeros \cite{siripuram2020convolution}. However, it is also possible to give simple proof using elementary techniques. Let $p_\II(x) = \sum_{i \in \II}x^i$. The hypothesis of the lemma implies that $p_\II(e^{2\pi i (j_1-j_2)/N})=0$ for $j_1\neq j_2 \in \KK$. Suppose $\KK = \{2^{l_1}, 2^{l_2},\ldots,  \}$ with $0<l_1<l_2<\ldots<M$. Then for $s < r$, $e^{2\pi i (2^{l_s}-2^{l_r})/N} = e^{2\pi i l_s n'/N}$ for some odd $n'$. Now since $p_\II$ is a polynomial with integer coefficients,
\[
p_\II(e^{2\pi i l_s n'/N}) = 0 \text{ implies }\phi_{N/2^{l_s}}(x)\ | \ p_\II(x),
\]
where $\phi_n$ is the $n^{th}$ cyclotomic polynomial \cite{isaacs2009algebra}. So $p_\II$ has a factorization of the form \[p_\II(x) = g(x) \prod_{2^{l_s} \in \KK}\phi_{N/2^{l_s}}(x),\]
with some polynomial $g(x)$ with integer coefficients. Evaluating at $x=1$, we get \(p_\II(1) \geq\phi_{N/2^{l_s}}(1) = 2^{|\KK|}  \) by using the fact that $\phi_n(1)=2$ when $n$ is any non-trivial power of $2$ \cite{isaacs2009algebra}.\hfill\IEEEQEDhere



\end{document}